\documentclass[sigconf, nonacm, balance=false]{acmart}

\setcopyright{cc}

\usepackage{xspace}
\usepackage{framed}
\usepackage{booktabs}
\usepackage{multirow}
\usepackage[inline]{enumitem}
\usepackage{cleveref}
\usepackage{graphicx}
\usepackage{nicefrac}
\usepackage{fontawesome}
\usepackage[
    lambda,
    advantage,
    operators,
    sets,
    adversary,
    landau,
    probability,
    notions,
    logic,
    ff,
    mm,
    primitives,
    events,
    complexity,
    asymptotics,
    keys
]{cryptocode}
\usepackage{algorithmicx}
\usepackage[noend]{algpseudocode}
\usepackage{subcaption}

\newcommand{\FullReturn}{\State\Return}
\newcommand{\Assert}{\State \textbf{Assert}\ }
\newenvironment{algorithmic_small}{\scriptsize\begin{algorithmic}}{\end{algorithmic}}
\AtBeginEnvironment{flalign*}{\setlength{\abovedisplayskip}{0pt}}

\usepackage{tikz}
\usetikzlibrary{calc, positioning, arrows, arrows.meta, shapes, fit}
\usepackage{pgfplots}
\pgfplotsset{compat=1.18}
\usepgfplotslibrary{statistics,groupplots,units,colormaps}

\newcommand{\prepams}{\textsc{PrePaMS}\xspace}

\definecolor{uulm}{RGB}{125,154,170}
\definecolor{uulm-akzent}{RGB}{169,162,141}
\definecolor{uulm-in}{RGB}{163,38,56}
\definecolor{uulm-med}{RGB}{38,84,124}
\definecolor{uulm-mawi}{RGB}{86,170,28}
\definecolor{uulm-nawi}{RGB}{189,96,5}

\newcommand{\Eg}{E.g.,\xspace}
\newcommand{\eg}{e.g.,\xspace}
\newcommand{\Cf}{C.f.,\xspace}
\newcommand{\cf}{c.f.,\xspace}
\newcommand{\Ie}{I.e.,\xspace}
\newcommand{\ie}{i.e.,\xspace}

\makeatletter
\newcommand{\oset}[3][0ex]{%
\mathrel{\mathop{#3}\limits^{
  \vbox to#1{\kern-2\ex@
  \hbox{$\scriptstyle#2$}\vss}}}}
\makeatother
\newcommand{\?}{\overset{?}{=}}
\newcommand{\Zq}{\mathbb{Z}_q}
\newcommand{\bit}{\{0,1\}}
\newcommand{\bits}{\bit^*}
\newcommand{\drawrandom}{\oset[.2ex]{\,\$}{\leftarrow}}
\newcommand{\Oracle}{\mathcal{O}}

\newcommand{\Organizer}{\mathsf{O}}
\newcommand{\Participant}{\mathsf{P}}
\newcommand{\Service}{\mathsf{S}}
\newcommand{\Uni}{\mathsf{U}}
\newcommand{\BB}{\mathsf{BB}}
\newcommand{\UN}{\mathsf{UN}}
\newcommand{\Honest}{\mathcal{H}}
\newcommand{\Adversary}{\mathcal{A}}

\newcommand{\Setup}{\mathsf{Setup}}
\newcommand{\KeyGen}{\mathsf{KeyGen}}
\newcommand{\Register}{\mathsf{Register}}
\newcommand{\Participate}{\mathsf{Participate}}
\newcommand{\Payout}{\mathsf{Payout}}

\newcommand{\Receive}{\mathsf{Receive}}
\newcommand{\CheckCred}{\mathsf{ChkCred}}
\newcommand{\CheckParticipation}{\mathsf{ChkPart}}
\newcommand{\CheckQualification}{\mathsf{ChkQual}}

\newcommand{\id}{\mathsf{id}}
\newcommand{\SK}{\mathsf{SK}}
\newcommand\Overline[1]{%
\mathsf{}%
\begin{tikzpicture}[baseline=(a.base)]
\node[inner xsep=0pt,inner ysep=.5pt] (a) {$#1$};
\draw[line width=.8pt] ([xshift=2pt] a.north west) -- (a.north east);
\end{tikzpicture}
}
\newcommand{\qualifier}{
    {\mathchoice{\Overline{\displaystyle\delta}}{\Overline{\delta}}{\Overline{\scriptstyle{\delta}}}{\Overline{\scriptscriptstyle{\delta}}}}
}
\newcommand{\disqualifier}{\delta}
\newcommand{\PK}{\mathsf{PK}}
\newcommand{\cred}{\mathsf{cred}}
\newcommand{\CRED}{\mathsf{CRED}}
\newcommand{\Tag}{\tau}
\newcommand{\Task}{\mathsf{T}}
\newcommand{\participation}{\mathsf{part}}
\newcommand{\answers}{\vec{\mathsf{a}}}
\newcommand{\Ledger}{\mathsf{state}}
\newcommand{\username}{\mathsf{un}}
\newcommand{\attr}{\mathsf{attr}}
\newcommand{\Padding}{\mathsf{Padding}}
\newcommand{\reward}{\mathsf{R}}
\newcommand{\credential}{\mathsf{C}}
\newcommand{\Satisfy}{\mathsf{Satisfy}}
\newcommand{\Constraints}{\mathsf{constr}}

\newcommand{\send}{\mathbf{send}}
\newcommand{\recv}{\mathbf{recv}}

\newcommand{\rtx}{\mathsf{tx}}
\newcommand{\auxil}{\mathsf{aux}}
\newcommand{\tx}{\mathsf{tx}}
\newcommand{\TX}{\mathsf{BB}}
\newcommand{\PTX}{\mathsf{NUL}}
\newcommand{\Verify}{\mathsf{Verify}}
\newcommand{\AoK}[1]{\mathsf{NIZK}[\mathcal{L}_{#1}]}
\newcommand{\Pub}{\mathsf{Pub}}
\newcommand{\Sim}{\mathsf{Sim}}
\newcommand{\DS}{\mathsf{DS}}
\newcommand{\PBS}{\mathsf{PBS}}
\newcommand{\Sign}{\mathsf{Sign}}
\newcommand{\Prove}{\mathsf{Prove}}
\newcommand{\Blind}{\mathsf{Blind}}
\newcommand{\Unblind}{\mathsf{Unblind}}
\newcommand{\db}{\mathsf{UN}}%
\newcommand{\nul}{\mathsf{nul}}
\newcommand{\NUL}{\mathcal{N}}
\newcommand{\KDF}{\mathsf{KDF}}
\newcommand{\loss}{\mathsf{loss}}

\newcommand{\stmt}{\mathsf{stmt}}
\newcommand{\wit}{\mathsf{wit}}

\newcommand{\TAG}{\mathsf{Tag}}
\newcommand{\VRF}{\mathsf{VRF}}
\newcommand{\VRFN}{{\mathsf{VRF}_\NUL}}
\newcommand{\Eval}{\mathsf{Eval}}

\newcommand{\ST}{:~~~}

\newcommand{\rpk}{\mathsf{rpk}}
\newcommand{\rsk}{\mathsf{rsk}}
\newcommand{\G}{\varmathbb{G}}
\newcommand{\cL}{\vec{c}_L}
\newcommand{\cR}{\vec{c}_R}
\newcommand{\vi}{\vec{v}_i}
\newcommand{\secret}[1]{\textcolor{uulm-in}{#1}}
\newcommand{\unused}[1]{\textcolor{black!60}{#1}}

\newcommand{\skspace}[1]{\varphi_{#1}}
\newcommand{\pkspace}[1]{\Phi_{#1}}
\newcommand{\rspace}[1]{\mathbb{R}_{#1}}

\newcommand{\subsubsectionstar}[1]{\noindent\emph{#1.}}

\begin{document}

\title[\prepams: Privacy-Preserving Participant Management System]{\texorpdfstring{\prepams: Privacy-Preserving Participant Management System\\ for Studies with Rewards and Prerequisites}{PrePaMS: Privacy-Preserving Participant Management System for Studies with Rewards and Prerequisites}}

\author{Echo Mei{\ss}ner}
\orcid{0000-0002-2937-6306}
\email{echo.meissner@uni-ulm.de}
\affiliation{%
    \department{Institute of Distributed Systems}
    \institution{Ulm University, Germany}
    \city{}
    \country{}}

\author{Frank Kargl}
\orcid{0000-0003-3800-8369}
\email{frank.kargl@uni-ulm.de}
\affiliation{%
    \department{Institute of Distributed Systems}
    \institution{Ulm University, Germany}
    \city{}
    \country{}}

\author{Benjamin Erb}
\orcid{0000-0002-5432-4989}
\email{benjamin.erb@uni-ulm.de}
\affiliation{%
    \department{Institute of Distributed Systems}
    \institution{Ulm University, Germany}
    \city{}
    \country{}}

\author{Felix Engelmann}
\orcid{0000-0001-9356-0231}
\email{fe-research@nlogn.org}
\affiliation{%
    \department{MAX-IV Laboratory}
    \institution{Lund University, Sweden}
    \city{}
    \country{}}

\begin{abstract}
Taking part in surveys, experiments, and studies is often compensated by rewards to increase the number of participants and encourage attendance.
While privacy requirements are usually considered for participation, privacy aspects of the reward procedure are mostly ignored.
To this end, we introduce \prepams, an efficient participation management system that supports prerequisite checks and participation rewards in a privacy-preserving way.
Our system organizes participations with potential (dis-)qualifying dependencies and enables secure reward payoffs. By leveraging a set of proven cryptographic primitives and mechanisms such as anonymous credentials and zero-knowledge proofs, participations are protected so that service providers and organizers cannot derive the identity of participants even within the reward process.
In this paper, we have designed and implemented a prototype of \prepams to show its effectiveness and evaluated its performance under realistic workloads.
\prepams covers the information whether subjects have participated in surveys, experiments, or studies.
When combined with other secure solutions for the actual data collection within these events, \prepams can represent a cornerstone for more privacy-preserving empirical research.
\end{abstract}


\maketitle

\section{Introduction and Motivation}
\label{sec:introduction}

Surveys are imperative in opinion research, empiric evaluations, and feedback in corporations and organizations.
Experiments and studies are also essential for most empirical and behavioral sciences, such as psychology and sociology.
In all these fields, the quality of findings heavily depends on sufficiently large numbers of participations and motivated participants.
Hence, participation is often incentivized through corresponding rewards.

Most academic study programs of empirical disciplines require students to contribute to studies for a certain number of subject hours.
Similarly, successful participation is rewarded by credit points, often obligatory for graduation~\cite{Sieber1989}.
Longitudinal studies apply repeated observations in order to explore developments over time.
Hence, rewards are often tied to continual survey participation~\cite{Singer2018}.
Employee surveys provide insights into workplace characteristics and organizational conditions.
Here, companies often provide certain incentives to improve response rates for employee surveys~\cite{Dale2007, Baruch2008}.
Overall, rewarding participations is a common incentive strategy in many areas of science, education, and business.

During most experiments, surveys, and studies, data collection is handled with safeguards to protect the privacy of participants.
In particular, anonymization or pseudonymization of participants' responses is required due to research-ethical principles, data management guidelines, or compliance reasons.
However, anonymization and pseudonymization mechanisms \emph{during} a study often interfere with reward incentives provided \emph{after} the participation.
This issue becomes particularly evident when participants are required to take part in a series of studies (\eg longitudinal studies) or to take part in a certain number of different studies (\eg subject hours) to be eligible for a final reward.
Furthermore, the inclusion criteria of a study can interfere with anonymity, as a disclosed participation by itself may already reveal some information about the participant.

While there are several commercial solutions (\eg Sona Systems~\cite{sonasystems}) and academic tools (\eg hroot~\cite{Bock2014hroot}, ORSEE~\cite{Greiner2004online}) for participant management, none of these systems consider extensive privacy requirements for participations (\eg hiding participations from service providers) nor privacy-preserving rewarding processes (\eg collecting rewards anonymously while preventing the transfer of rewards between users).
This means that participants often have to trust the organizers and the service to keep identities and participations secret.

In this paper, we propose a novel privacy-preserving participation management system that protects the privacy of participants by breaking the link between participable tasks and the participating users while maintaining the possibility of individual, non-transferable rewards.
Apart from the system design with privacy and security proofs, we complement our contributions with a prototype implementation\footnote{\label{github}\emph{Prototype source code:} \url{https://github.com/vs-uulm/prepams/tree/pets25.1}} of the system and a publicly available test instance\footnote{\label{deployment}\emph{Public test deployment:} \url{https://vs-uulm.github.io/prepams/}} to demonstrate the feasibility.

\noindent\emph{Our Main Contributions:}
\begin{itemize*}[leftmargin=1.5em]
  \item We propose \prepams, a novel set of cryptographic protocols to enable participations in studies with rewards and prerequisites while preserving the privacy of participants.
  \item We define correctness, privacy, and security properties of our scheme and prove these properties for our construction.
  \item We provide an open-source web-based proof of concept implementation\footref{github} to showcase the practicability of our approach.
    This includes a publicly hosted test deployment\footref{deployment} where anyone can explore our system themselves.
  \item We evaluate the performance of our prototypical implementation.
    In this evaluation, we follow the Popper convention~\cite{jimenez2017popper} for reproducible evaluations.
\end{itemize*}

\noindent\emph{Technical Overview:}
We describe and apply a pairing-based, multi-show unlinkable anonymous credential scheme, which allows participants to authenticate themselves to the organizers.
To only allow a single participation per user and study, we derive participation tags from the user's credential using a verifiable random function.
To model (dis-)qualifiers (\ie the requirement of (not) having previously participated in referenced studies), we utilize non-interactive zero-knowledge proofs (NIZK) where a participant proves these prerequisites on participation without revealing any specific link.
Participation requirements based on attributes of a user's credential (e.g., age, handedness) are also proven using NIZK proofs.
Altogether, this prevents an organizer from linking multiple participations of the same participant while still allowing for prerequisites and rewards.
Participation rewards are issued as blind signatures, which a participant can re-randomize for a payout request without revealing which studies they have participated in to earn the rewards.
At the same time, a reward is bound to the original participant, so it cannot be passed on to another user.
Double spending is prevented by additionally publishing a verifiable nullifier of every reward.
\section{Requirements for Participation Management}
\label{sec:requirements}

We now illustrate the relevance of participation management as well as the associated privacy risks using the example of subject hours in academic study programs with empirical research methods.
We formalize these requirements later in \Cref{sec:formalization}.

Study programs with empirical methods often mandate a certain number of subject hours from their students~\cite{Sieber1989} earned through study participation.
These participations serve several purposes\,---\,self-experience for participating students, recruitment of subject pools for student research projects as well as sampling for actual research studies~\cite{goodwin2016research}.
As a concrete example, we consider an undergrad psychology program in which students are asked to take part in a certain amount of study participations.
The requirements for the participation management for such subject hours can be further divided into \underline{s}tudy management \underline{r}equirements (\emph{SR}) and \underline{r}eward management \underline{r}equirements (\emph{RR}) also shown in \Cref{tab:comparison}.
Study management requirements refer to administrative and organizational procedures such as a list of current ongoing studies (\emph{SR1}), but also includes the handling of conditions for participations.
In turn, reward management requirements targets the tracking of the students' study participations so that the associated credit points can be eventually rewarded once students reach the required amount (\emph{RR3}).

In the past, study management was mainly realized by using physical bulletin boards or websites listing the studies as well as arbitrary processes to handle preconditions of individual studies.
Reward management was typically implemented using log sheets (\ie participants receive a dated signature or stamp by the study organizer after participation) or stickers (\ie participants receive a sticker or a similar token after participation).

These analog reward approaches work well with \emph{RR1: analog studies}, such as lab experiments.
However, they lack inherent support for \emph{RR2: digital studies} (\eg online questionnaires, mobile-sensing studies) and required additional organizational processes for rewarding (\eg receiving a printable participation confirmation to be traded against a signature or sticker). 
Also, analog reward approaches often failed to prevent misuse by malicious students.
A common threat to mandatory subject hours is students selling rewards to their peers or helping their peers out by participating in a study in their name.
This of course contradicts the intended didactic goals, hence the need for \emph{RR5: non-transferable rewards}.
Furthermore, researchers want to \emph{RR4: prevent duplicate participations} of participants in a study, to not skew the gathered data.

Nowadays, most universities have superseded analog approaches with a web-based participation management systems that cover both study and reward management and accommodate these missing features.
One of the most prevalent systems is the commercial SaaS product ``Sona''~\cite{sonasystems}, which claims to be used by 1,500 customers across 30 countries and more than 23 million registered users.
Sona as well as alternative academic systems~\cite{Bock2014hroot,Greiner2004online} introduced additional new features that are not covered by their analog counterparts:
Some studies have prerequisites that participants have to satisfy in order to be admitted to the study.
With web-based systems, this can be done automatically even before the participation (\eg some common characteristics of participants are surveyed already by the participation management system at registration time), saving both researchers and participants time and resources.
This includes \emph{SR4: attributed-based preconditions}, such that only participants of a certain age are \emph{qualified} or participants with specific medical preconditions are \emph{disqualified}.
For some studies \emph{SR5: participation-based preconditions} are also necessary, either to exclude participants of a similar experiment or previous iteration to mitigate biases or to allow for longitudinal studies in which the same participants take part in a sequence of studies over time.
For lab experiments, participants often have to schedule sessions beforehand, which can be simplified by a digital system featuring \eg a calendar-based \emph{SR6: session scheduling}.

Some systems further enable researchers to query the database of potential participants when designing a study, to verify if the participant pool theoretically contains a suitable sample size based on their planned prerequisites.
Similar to preconditions this can be divided into \emph{SR2: attribute-based prescreening} and \emph{SR3: participation-based prescreening} or a combination of both.

Protecting the personal data of participants \emph{within} studies is not only a research-ethical mandate~\cite{Folkman2000}, it is also often grounded in legal requirements (\eg GDPR in European countries).
Mechanisms such as pseudonymization, anonymization, and data aggregation can help mask a participant's identity \emph{within the data set} of that study~\cite{gollwitzer2020data}.
Nevertheless, the mere fact that an individual has actually taken part in a certain study represents some kind of metadata and can already leak information about that individual.
Hence, \underline{p}rivacy \underline{r}equirements (\emph{PR}) also apply to participation management.

Depending on the prerequisites of a study, this can involve personal attributes such as ethnicity or a certain native language of the participant, but also physical (\eg dexterity, pre-existing diseases) or mental preconditions, such as currently ongoing clinical treatments, and other rather sensitive characteristics.
Furthermore, participation in a certain study can also imply previous participations (\eg longitudinal study designs with multiple measurement dates) or hint to the non-participation in a former study (\eg follow-up studies that disallow participation of subjects from previous trials).
That said, an attacker with knowledge of the sequence of participations of an individual and contextual information about the corresponding studies might already be able to create a sensitive profile of a target\,---\,completely independent from the actual (and inaccessible) response data collected in these studies.
As many people involved in the organizational process for studies have potential access to both the participation data (\eg a research assistant who is signing the participation confirmations) and the contextual study information (\eg inclusion and exclusion criteria listed on the university-internal study management system), we argue that this is a realistic threat that cannot be neglected.
Given a sequence of participations of an individual, their known identity can then be linked to a set of potentially sensitive attributes deduced from these participations.

A sticker-based analog reward management\,---\,if implemented correctly\,---\,provides an inherent privacy level similar to fiat currency for money.
Participants can participate anonymously, so that organizers are not able to link them to prior participations and the rewarded stickers are also not linked to the study they were rewarded for (\emph{PR1: anonymous participation}).
Similarly, when a student obtained the mandated subject hours they can exchange a full set of stickers for their course credit without revealing their participation history (\emph{PR2: anonymous rewarding}).
Log sheets on the other hand disclose the full prior participation history during participation to all study personal handling the sheet, and also when eventually handing in the completed sheet to exchange the hours for the final course credit.
The existing digital systems all store the individual participations in a database, which is depending on the system accessible to all researchers or just administrators.
Even if limited to just administrators of a system, this central database is still vulnerable to sensitive data leaks, \eg through a software vulnerabilities or insider attacks.

In this paper, we present a cryptographic scheme that provides the same requirements and security properties as current web-based participation management systems, while providing similar privacy guarantees to a sticker-based analog approach.
\Cref{tab:comparison} shows the supported features and privacy properties of the discussed analog approaches and current digital systems compared to our proposed system \prepams.

The listed requirements are based on an analysis of the features of state-of-the-art participation management systems and common practices for rewarding.
In addition, we double-checked with contacts from our psychology department that the derived requirements are both adequate and exhaustive from their perspective and experience.

\begin{table*}[t]
\caption{Functional requirements and privacy properties of common participation management approaches and related systems compared to our proposed system \prepams.}
\begin{tabular}{@{}rrlcccccc@{}}
\toprule
& &                                                   & Log Sheets  & Sticker & hroot~\cite{Bock2014hroot} & ORSEE~\cite{Greiner2004online} & Sona~\cite{sonasystems} & \textbf{\prepams} \\ \midrule
\multirow{6}{*}{\rotatebox{90}{\emph{study management}}}
& \textbf{SR1} & \textbf{listing of studies}                       &             &          & \faCheck   & \faCheck  & \faCheck   & \faCheck  \\
& \textbf{SR2} & \textbf{attribute-based prescreening}             &             &          & (\faCheck) & \faCheck  & \faCheck   & \phantom{\textsuperscript{x}} (\faCheck) \textsuperscript{a}   \\
& \textbf{SR3} & \textbf{participation-based prescreening}         &             &          & \faClose   & \faCheck  & \faCheck   & \phantom{\textsuperscript{x}} (\faClose) \textsuperscript{b}   \\
& \textbf{SR4} & \textbf{attribute-based prerequisites}            &             &          & \faCheck   & \faCheck  & \faCheck   & \faCheck   \\
& \textbf{SR5} & \textbf{participation-based prerequisites}        &             &          & (\faCheck) & \faCheck  & \faCheck   & \faCheck   \\
& \textbf{SR6} & \textbf{session scheduling}                       &             &          & \faCheck   & \faCheck  & \faCheck   & \phantom{\textsuperscript{x}} (\faClose) \textsuperscript{c} \\[1em]
\multirow{5}{*}{\rotatebox{90}{\emph{rewarding}}}
& \textbf{RR1} & \textbf{analog studies}                           & \faCheck    & \faCheck & \faCheck   & \faCheck  & \faCheck   & \faCheck   \\
& \textbf{RR2} & \textbf{digital studies}                          & \phantom{\textsuperscript{x}} \faClose\ \textsuperscript{d}    & \phantom{\textsuperscript{x}} \faClose\ \textsuperscript{d} & \faCheck   & \faCheck  & \faCheck   & \faCheck   \\
& \textbf{RR3} & \textbf{rewarding}                                & \faCheck    & \faCheck & \faClose   & \faCheck  & \faCheck   & \faCheck   \\
& \textbf{RR4} & \textbf{duplicate participation prevention}       & \faCheck    & \faClose & \faCheck   & \faCheck  & \faCheck   & \faCheck   \\
& \textbf{RR5} & \textbf{non-transferable rewards}                & \faCheck    & \faClose & \faClose   & \faCheck  & \phantom{\textsuperscript{x}} (\faCheck) \textsuperscript{e} & \faCheck   \\[1em]
\multirow{2}{*}{\rotatebox{90}{\parbox{.8cm}{\emph{privacy}}}}
& \textbf{PR1} & \textbf{anonymous participation}                 & \faClose    & \faCheck & \faClose   & \faClose  & \phantom{\textsuperscript{x}} \faClose\ \textsuperscript{f}   & \phantom{\textsuperscript{x}} \faCheck \textsuperscript{g}  \\
& \textbf{PR2} & \textbf{anonymous rewarding}                     & \faClose    & \faCheck & \faClose   & \faClose  & \phantom{\textsuperscript{x}} \faClose\ \textsuperscript{f}   & \phantom{\textsuperscript{x}} \faCheck  \textsuperscript{g}   \\
\bottomrule\\[-4pt]
\multicolumn{9}{p{.97\textwidth}}{
    \footnotesize
    \emph{Remarks:}
    ~\textsuperscript{a)} Attribute-based prescreening is currently not implemented, but possible if the service keeps track of which attributes are issued during registration.
    ~\textsuperscript{b)} Participation-based prescreening has intrinsic privacy issues, if centrally maintained. However, an organizer knows the number of participations of their own previous studies and can use this information for most scenarios.
    ~\textsuperscript{c)} Session scheduling does not come with strict privacy requirements and can be implemented, \eg using the participation tag as an anonymous identity.
    ~\textsuperscript{d)} Rewarding can only be implemented by using additional organizational processes, such as a subsequent conversion of digital participation confirmations into physical stickers or signatures.
    ~\textsuperscript{e)} Although credits in Sona are rewarded to a persistent pseudonym of a user, it is not verified if the pseudonym belongs to the participant.
    ~\textsuperscript{f)} Rewarding and participation uses a persistent pseudonym, which can be linked across studies and sometimes to a participant's identity.
    ~\textsuperscript{g)} \prepams also supports pseudonyms to track participants over subsequent participations for longitudinal studies. The pseudonym remains unlinkable to unrelated studies.
}
\end{tabular}
\label{tab:comparison}
\end{table*}
\section{Formalization}
\label{sec:formalization}

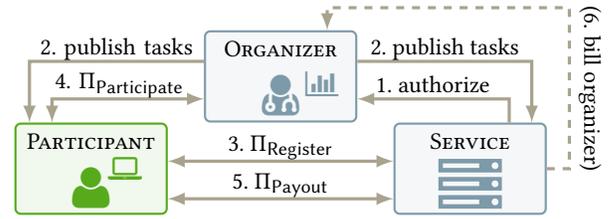
\begin{figure}
    \centering
\begin{tikzpicture}[
    label/.style={
        text width=1.9cm,
        align=center,
    },
    entity/.style={
        thick,
        text width=1.8cm,
        align=center,
        minimum height=1.2cm,
        rounded corners=2pt,
        draw=#1,
        fill=#1!8
    },
    arrow/.style={
        latex-latex,
        very thick,
        draw=uulm-akzent
    }
]
    \node[entity=uulm-mawi] (participant) {};
    \node[label, below=1pt of participant.north] {\textsc{Participant}};
    \node[xshift=.45cm,yshift=-.0cm] at (participant) {\color{uulm-mawi}\LARGE\faLaptop};
    \node[yshift=-.25cm] at (participant) {\color{uulm-mawi}\fontsize{18}{18}\faUser};

    \node[entity=uulm, right=3cm of participant] (prepams) {};
    \node[label, below=1pt of prepams.north] {\textsc{\textsc{Service}}};
    \node[yshift=-.20cm] at (prepams) {\color{uulm}\fontsize{24}{24}\faServer};

    \coordinate (midpoint) at ($(participant.north)!.5!(prepams.north)$);

    \node[entity=uulm, above=.0cm of midpoint] (organizer) {};
    \node[label, below=1pt of organizer.north] {\textsc{\textsc{Organizer}}};
    \node[xshift=.55cm,yshift=-.1cm] at (organizer) {\color{uulm}\Large\faBarChart};
    \node[yshift=-.25cm] at (organizer) {\color{uulm}\fontsize{18}{18}\faUserMd};

    \draw[arrow] ([yshift=.1cm]participant.east) -- node[above=-3pt] {3. $\Pi_\Register$} ([yshift=.1cm] prepams.west);
    \draw[arrow] ([yshift=-.4cm]participant.east) -- node[above=-3pt] {5. $\Pi_\Payout$} ([yshift=-.4cm] prepams.west);
    \draw[arrow,latex-] ([xshift=-.2cm]prepams.north east) |- node[above=-3pt,pos=.99,anchor=south west] {2. publish tasks} ([yshift=.2cm] organizer.east);
    \draw[arrow,-latex] ([xshift=-.5cm]prepams.north east) |- node[above=-3pt,pos=.97,anchor=south west,yshift=2pt] {1. authorize} ([yshift=-.3cm] organizer.east);
    \draw[arrow] ([yshift=-.3cm] organizer.west) -| node[above=-3pt,pos=-.05,anchor=south east,text width=2.05cm] {4. $\Pi_\Participate$} ([xshift=.5cm]participant.north west);
    \draw[arrow,-latex] ([yshift=.2cm] organizer.west) -| node[above=-3pt,pos=.0,anchor=south east,text width=2.05cm] {2. publish tasks} ([xshift=.2cm]participant.north west);
    \draw[arrow,dashed,-latex] (prepams.east) -- ++(.3,0) |- node[above=-2pt, rotate=-90, pos=-.02,anchor=south east] {(6. bill organizer)} ([yshift=.3cm, xshift=-.4cm] organizer.north east) -- ++(0,-.3);
\end{tikzpicture}
     \caption{
      \emph{Simplified system overview}.
      1. organizers are authorized by the service;
      2. organizer publishes tasks;
      3. participant registers with service;
      4. participant participates in organizer's task (retrieved either via service or organizer);
      5. after multiple participations, participant requests payout of rewards from service;
      6. after task is concluded, service may bill organizer for spent funds.
    }
    \label{fig:overview}
    \Description{A diagram showing a simplified system overview for \prepams. The three parties Participant, Organizer, and Service are depicted as boxes and connected with numbered and labeled arrows that indicate the interactions between these parties.}
\end{figure}

In this section, we present a formal definition of a privacy-preserving participation protocol involving three types of parties, of which two collude:
A single participation management service ($\Service$), colluding with one or multiple study organizers ($\Organizer$), and a number of participants ($\Participant$).
In general, the protocol enables participants to participate in various tasks, which are, \eg studies in the scenario described above but may be other events with rewards, and to receive a reward after successful participation without leaking who participated in which task.
Figure~\ref{fig:overview} shows a simplified overview.

\subsection{Threat Model}

Our protocol is applicable in the following setting.
The \emph{Service} acts as a covert entity during participant registration.
Similarly, when a participant requests a payout of virtual credits for real-world rewards, it will honestly fulfill this exchange request. In the university study setting, the rewards may be European Credit Transfer System (ECTS) credits or fiat currency.
However, it may use its available information to determine which tasks a participant has participated in, gaining confidential information about the participant.
It may link this to the real identity necessary for the payout procedure, which violates the participant's privacy.
We envision that this service is run by the accounting department or administration of an institution because the goals of an accounting department/administration coincide well with the goals of the service. Regarding participation privacy, we assume that the Service colludes with the Organizers.

\emph{Organizers} in our model create and publish studies and offer participation rewards to participants.
Furthermore, organizers may be malicious in the sense of attacking the privacy of the participants by linking multiple participations or linking a participation to a participant's identity.
A task or study $\Task$ is a public, opaque object with a defined participation reward and optional prerequisites that a participant has to satisfy when participating.
Such prerequisites can be divided into \emph{qualifiers} (\ie other tasks a participant must have participated in to qualify for this task), \emph{disqualifiers} (\ie previous tasks a participant must not have participated in to qualify for this task), \emph{range constraints} (\ie the value of an attribute associated with the participants credential must be contained in a specified range), or \emph{set constraints} (\ie an attribute value must be included in a specified set) a participant must fulfill.

As the Service and Organizers collude, we take advantage to simplify our formalization by modeling them as a single service $\Service$ with a separate public append-only bulletin board $\BB$.

\emph{Participants} want to participate in tasks to earn rewards in the form of virtual credits while protecting their privacy.
Participants can also be malicious by attempting to participate in a study they do not qualify for, participating twice, receiving payment for a task without participating, or attempting to request payouts without earning virtual credits first. Another problem is fraud between participants, selling credits to other users.
It is a general limitation that users may perform a task poorly, \eg in user studies where participants enter bogus data that needs to be sanitized by statistical tests, which we consider out of scope.
De-anonymization attempts via network information and metadata are also considered out of scope.
We assume participants conceal their identity when interacting with other parties through an anonymous network, such as Tor, or similar means.

\subsection{Syntax}
In the following, we provide a definition of five
  probabilistic polynomial time algorithms and interactive protocols $(\Setup,\allowbreak\KeyGen_\Service,\allowbreak\Pi_\Register,\allowbreak\Pi_\Participate,\allowbreak\Pi_\Payout)$ that comprise a participation management protocol.
In case of invalid inputs or failures in interaction, all algorithms have the option to abort.
Our notation for lists with a specific order is $[\dots]$ to which another list is concatenated or an element is appended by $\|$.
For lists, we use, \eg $A[\text{:}n]:=[A_1,\dots,A_{n}]$ to indicate slices and subscript to index, \eg $A_1$ for the first element.
For tuples $T:=(a,b,c)$ we use $T.a$ for a named element of the tuple.
For an integer $n$, we define the sequence $[n]:=[1,\dots,n]$.
We denote a negligible polynomial in $\lambda$ as $\negl$.
We use $\mathbf{assert}$ in algorithms to check a condition and abort if it evaluates to $\mathtt{false}$. \Cref{fig:variables} in the appendix summarizes the variable names.

\begin{description}[leftmargin=0.5em,nosep]
  \item[$\Setup(\secparam,m,n)\to\pp$:]
    takes the security parameter $\secparam$, the number of attributes $m \in \mathbb{N}$ of a participant and the maximum number of payouts $n\in\mathbb{N}$. It outputs the public parameters $\pp$.
  \item[$\KeyGen_\Service() \to (\sk_\Service,\pk_\Service)$:]
    outputs the key-pair $(\sk_\Service,\pk_\Service)$ used by the service and organizers.
    $\pk_\Service$ is distributed to everyone for credential verification.
    \emph{This means all subsequent algorithms have an implicit $(\pp,\pk_\Service)$ input, where $\pp$ includes $m,n$.}
  \item[$\Pi_\Register \langle \Participant(\username,\attr),\Service(\sk_\Service)\rangle\to (\cred,\username)$:]
    is an interactive protocol between $\Participant$ and $\Service$ where a participant $\Participant$, identified by a unique username $\username$ and attributes $\attr$, receives a secret credential $\cred$ for later participation in tasks.
    The service uses its secret key $\sk_\Service$ to sign the credential and gets the username to prevent duplicate registration, and verifiers the attributes.
  \item[$\Pi_\Participate \langle \Participant(\cred, \BB, \Task), \Service(\sk_\Service)\rangle\to \rtx$:]
    is an interactive protocol between $\Participant$ and $\Service$ which allows $\Participant$ with credential $\cred$ to participate in a task $\Task$ from an Organizer.
    The eligibility of $\Participant$ to participate is dependent on all previous participations by all participants described as the list of reward transactions $\BB:=[\rtx_i]_{i=1}^{|\BB|}$ on the bulletin board.
    This protocol returns a reward transaction $\rtx$ to both parties that is appended to the bulletin board $\BB\gets\BB\|\rtx$.
    A $\rtx$ includes the performed task $\Task\in\mathbb{T}$ from a task domain, defined as a tuple $\Task:=(v,\qualifier\subset\mathbb{T},\disqualifier\subset\mathbb{T},\Constraints,\Delta)$ with a reward $v$, qualifiers $\qualifier$, disqualifiers $\disqualifier$, attribute constraints $\Constraints$, and an opaque task description $\Delta\in\{0,1\}^*$, \eg a title, external link, and a verbose description.

  \item[$\Pi_\Payout \langle \Participant(\cred,\username, \BB, v,\{\tx_i\}_{i=1}^k), \Service()\rangle\to (\{\NUL_i\}_{i=1}^n,v,\username)$:]
    is an interactive protocol to allow a participant $\Participant$ with $\cred$, belonging to username $\username$, to request a payout of earned rewards from a list of $k\leq n$ reward transactions $[\rtx_i]_{i=1}^{k}\subset\BB$ with amount $v$ from $\Service$.
    It returns payout records, also called nullifiers, $\NUL_i$ which are added to the public set $\PTX$ maintained by the service and the paid out value $v$ to user $\username$.
\end{description}
\noindent
For our correctness and security definitions, we require the following four auxiliary algorithms.

\begin{description}[leftmargin=0.5em,nosep]
  \item[$\Receive(\cred,\TX,\PTX)\to (s,R)$:] returns the total amount $s$ owned by $\cred$ and the list of reward transactions $R=[\tx_i]_{i=0}^{|R|}$ which are present in the bulletin board $\TX$ and which have not yet been paid out with a record in $\PTX$.
  \item[$\CheckCred(\cred, \pk_\Service)\to\bit$:] returns 1 if the credential $\cred$ is valid under the service's public key $\pk_\Service$.
  \item[$\CheckParticipation(\cred, \rtx)\to\bit$:] returns 1 if the credential $\cred$ was used for the participation which resulted in the transaction $\rtx$.
  \item[$\CheckQualification(\cred, \TX, \Task)\to\bit$:] returns 1 if the participant satisfies all prerequisites of $\Task$ given the previous participations registered in $\TX$ which is a time ordered list of $\rtx$.
    Meaning a participant has neither participated in any disqualifier of $\Task$ nor in the task itself, but has participated in all qualifiers of $\Task$, and fulfills the attribute constraints.
\end{description}

\begin{definition}[Correctness]
\label{def:correctness}
  A participation workflow is \emph{correct} if the following conditions apply.
  For any $\lambda \in \mathbb{N}$, any $\pp \gets \Setup(\secparam)$ and any of the services' generated key pairs $(\sk_{\Service}, \pk_{\Service}) \gets \KeyGen_\Service()$, it then holds that:

  \begin{description}[leftmargin=0em,nosep]
    \item[Honest Registration:] For any previously non-registered participant identified by username $\username$ and any attributes $\attr$, it holds that
      $(\cred,\username)\gets\Pi_\Register \langle \Participant(\username,\attr),\Service(\sk_\Service)\rangle$ implies \\$\CheckCred(\cred, \pk_\Service)=1$

    \item[Honest Participation:] With any valid, signed credential $\cred$ (\ie $\CheckCred(\cred,\allowbreak \pk_\Service)=1$), a participant who fulfills the prerequisites of a task $\Task$ at any state $\TX$ (\ie $\CheckQualification(\cred,\allowbreak\TX,\allowbreak\Task)=1$), successfully participates in $\Task$ with
      $\rtx\gets\Pi_\Participate \langle \Participant(\cred, \TX,\allowbreak \Task),\allowbreak\Service(\sk_\Service)\rangle$
      such that $\Receive(\cred, [\rtx],\emptyset).v = \Task.v$.
    \item[Honest Payout:] A participant with $\cred$, given $s,R\gets\Receive(\allowbreak\cred,\BB,\PTX)$ can get a payout of an amount $v \in \{0, \dots, s\}$ with
      $(\PTX',v)\gets\Pi_\Payout \langle \Participant(\cred,\username, \TX,\PTX, v,R), \Service()\rangle$
      such that \\
      $\Receive(\cred, \TX, \PTX\cup\PTX').s \leq s - v$.
  \end{description}
\end{definition}

We define the security of our scheme with four game-based security definitions. They all follow the logic that an adversary interacts with a game and tries to win, breaking a security property. The interaction either happens through oracles, where the adversary is instructing honest parts of the game, or a challenge to the game. For our construction, we later prove, that the existence of a probabilistic polynomial time algorithm to win the game is negligible.

\subsection{Oracles}

Before defining our game-based security and privacy properties of the scheme, we require oracles for the adversary to interact with.
Our adversaries are allowed to call all oracles in a mixed order for a polynomial number of times in $\lambda$.
This allows the adversary to instruct honest participants to perform an action without access to their secret state.
The oracles presented in \Cref{fig:games} operate as one or both parties of the interactive protocols $\Pi_\Register$, $\Pi_\Participate$, and $\Pi_\Payout$ which gives a total of nine possibilities.
To highlight the adversary controlling the other party, we denote the input of this party as a part of the stateful adversary (\eg $\adv_\Participant$ for an adversary-controlled participant).
The oracles controlling both parties of the protocols are required to allow the participant adversary to observe honest interactions and their public effects.
This is important for, \eg replay attacks. The service oracles accept any participant adversary defined tasks.
Some oracles perform bookkeeping operations maintaining two lists of oracle controlled users $\mathfrak{U}^e$ with $e\in\bit$.
The epoch $e$ is required for the service oracles to separate users into two subsets.
The oracle-provided service is thereby able to track new registrations.
For each username, $\mathfrak{U}_\username^e$ maintains a tuple of $(\cred,\PTX)$: their credential $\cred$ and spent coin nullifiers $\PTX$ from payout interactions.
The oracles also track two total amounts of rewards $\mathfrak{T}^e$ depending on when the participation was rewarded.
$\Oracle\Participant\Participate$ and $\Oracle\Participant\Payout$ is parametrized with a lock set $L$ of usernames that ban them from specific interactions after a challenge from the adversary.
This is required to prevent the users from being trivially deanonymized depending on qualifying for tasks or access to rewards.
However, it does not weaken the adversary, as a winning adversary could have prepared the situation beforehand and use it in the challenge.
Additionally, the participant adversaries get an oracle $\Oracle\mathsf{state}$ to access the public state $\TX,\PTX$.

\begin{figure*}
\begin{framed}
\begin{minipage}[t]{0.3\textwidth}
\noindent
\underline{$\Oracle\Service\Register_{\sk_\Service,e}(\username,\attr,\adv_\Participant)$:}
\begin{algorithmic_small}
    \Assert {$\username \not\in\mathfrak{U}^0 \land \username\not\in\mathfrak{U}^1$ }
        \State run service side of $\Pi_\Register\langle\adv_\Participant,\Service(\sk_\Service)\rangle$
        \State $\mathfrak{U}^e_\username \gets (\bot,\emptyset)$
\end{algorithmic_small}

\noindent
\underline{$\Oracle\Service\Participate_{\sk_\Service,e}(\adv_\Participant)$:}
\begin{algorithmic_small}
    \State run service side of $\rtx\gets\Pi_\Participate\langle\adv_\Participant,\Service(\sk_\Service)\rangle$
    \State $\TX \gets \TX \| \rtx$
    \State $\mathfrak{T}^{e} \gets \mathfrak{T}^e+\rtx.\Task.v$
\end{algorithmic_small}

\noindent
\underline{$\Oracle\Service\Payout(\adv_\Participant)$:}
\begin{algorithmic_small}
    \State run $\Service$ side of $\PTX',v,\username\gets\Pi_\Payout\langle\adv_\Participant,\Service()\rangle$
    \If{$\username\in\mathfrak{U}^0$}
      \State $\mathfrak{T}^0 \gets \mathfrak{T}^0-v$
    \ElsIf{$\username\in\mathfrak{U}^1$}
     \State $\mathfrak{T}^1 \gets \mathfrak{T}^1-v$
     \EndIf
    \State $\PTX \gets \PTX \cup \PTX'$
\end{algorithmic_small}

\end{minipage}
\begin{minipage}[t]{0.33\textwidth}
\noindent
\underline{$\Oracle\Register_{\sk_\Service}(\username,\attr)$:}
\begin{algorithmic_small}
    \Assert {$\username \not\in\mathfrak{U}^0 \land \username\not\in\mathfrak{U}^1$ }
        \State run $\cred\gets\Pi_\Register\langle\Participant(\username,\attr,\pk_\Service),\Service(\sk_\Service)\rangle$
        \State $\mathfrak{U}^0_\username \gets (\cred,\emptyset)$
\end{algorithmic_small}

\noindent
\underline{$\Oracle\Participate_{\sk_\Service}(\Task,\username)$:}
\begin{algorithmic_small}
    \State run $\rtx\gets\Pi_\Participate\langle\Participant(\mathfrak{U}^0_\username.\cred,\TX,\Task),\Service(\sk_\Service)\rangle$
    \State $\TX \gets \TX \| \rtx$
\end{algorithmic_small}

\noindent
\underline{$\Oracle\Payout(\username,\TX,v)$:}
\begin{algorithmic_small}
    \State $(\cred,\PTX') \gets \mathfrak{U}^0_\username.\cred$
    \State $a,R\gets\Receive(\cred,\TX,\PTX\cup\PTX')$
    \State run $\PTX'',v,\username\gets\Pi_\Payout\langle\Participant(\cred,\TX,\PTX,v,R),\Service()\rangle$
    \State $\mathfrak{U}^0_\username.\PTX \gets \mathfrak{U}^0_\username.\PTX \cup \PTX''$
    \State $\PTX \gets \PTX \cup \PTX''$
\end{algorithmic_small}

\noindent
\underline{$\Oracle\mathsf{state}()$:}
\begin{algorithmic_small}
    \FullReturn $\TX,\PTX$
\end{algorithmic_small}

\end{minipage}
\begin{minipage}[t]{0.35\textwidth}
\noindent
\underline{$\Oracle\Participant\Register(\username,\attr,\pk_\Service,\adv_\Service)$:}
\begin{algorithmic_small}
    \Assert {$\username \not\in\mathfrak{U}^0 \land \username\not\in\mathfrak{U}^1$ }
        \State run $\Participant$ side of $\cred\gets\Pi_\Register\langle\Participant(\username,\attr,\pk_\Service),\adv_\Service\rangle$
        \State $\mathfrak{U}^0_\username \gets (\cred,\emptyset)$
\end{algorithmic_small}

\noindent
\underline{$\Oracle\Participant\Participate_L(\TX,\Task,\username,\adv_\Service)$:}
\begin{algorithmic_small}
    \If {$\username \in L$}
     \Assert {$\forall u\in L:\CheckQualification(\mathfrak{U}^0_u.\cred,\BB,\Task) = 1$}
    \EndIf
    \Assert {$\CheckQualification(\mathfrak{U}^0_\username.\cred,\BB,\Task) = 1$}
    \State run $\Participant$ side of $\rtx\gets\Pi_\Participate\langle\Participant(\mathfrak{U}^0_\username.\cred,\TX,\Task),\adv_\Service\rangle$
\end{algorithmic_small}

\noindent
\underline{$\Oracle\Participant\Payout_L(\username,\TX,v,\adv_\Service)$:}
\begin{algorithmic_small}
    \If {$\username \in L$}

    \Assert $\forall u\in L: \Receive(\mathfrak{U}^0_u.\cred,\TX,\PTX).a \geq v$
    \EndIf
    \State $\cred\gets \mathfrak{U}^0_\username.\cred$
    \State $a,R\gets\Receive(\cred,\TX,\PTX\cup\mathfrak{U}^0_\username.\PTX)$
    \State run $\Participant$ of $\PTX',v,\username\gets\Pi_\Payout\langle\Participant(\cred,\TX,\PTX,v,R),\adv_\Service\rangle$
    \State $\mathfrak{U}^0_\username.\PTX \gets \mathfrak{U}^0_\username.\PTX \cup \PTX'$
\end{algorithmic_small}

\end{minipage}

\tikz{\draw[dashed] (0,0) -- (\textwidth,0);}

\begin{minipage}[t]{0.30\textwidth}
\noindent
\underline{$\mathsf{PartSec}(\secparam)$:}
\begin{algorithmic_small}
    \State $\pp\gets\Setup(\secparam)$
    \State $\mathfrak{U}^0\gets\emptyset,\mathfrak{T}^0\gets0,\TX\gets[],\PTX\gets\emptyset$
    \State $\sk_\Service,\pk_\Service\gets\KeyGen_\Service(\pp)$
    \State $\Oracle:=\{\Oracle\Service\Register_{\sk_\Service,0},\allowbreak\Oracle\Service\Participate_{\sk_\Service,0},
    \allowbreak\Oracle\Service\Payout,\allowbreak\Oracle\Register_{\sk_\Service},\allowbreak\Oracle\Participate_{\sk_\Service},\allowbreak\Oracle\Payout,\allowbreak\Oracle\mathsf{state}\}$
    \State $\cred \gets\adv^{\Oracle}(\pp,\pk_\Service)$
    \Assert{$\CheckCred(\cred,\pk_\Service)=1$}
    \For{$t \in [|\TX|]$}
      \If{$\CheckQualification(\cred,\TX[\text{:}t],\TX_t.\Task)=0$\newline
          \text{\quad\quad\quad} $\land~~\CheckParticipation(\cred,\TX_t) = 1$} \FullReturn $\mathtt{win}$
      \EndIf
    \EndFor

\end{algorithmic_small}

\end{minipage}
\begin{minipage}[t]{0.33\textwidth}

\noindent
\underline{$\mathsf{Balance}(\secparam)$:}
\begin{algorithmic_small}
    \State $\pp\gets\Setup(\secparam)$
    \State $\mathfrak{U}^0\gets\emptyset,\mathfrak{U}^1\gets\emptyset, \mathfrak{T}^0\gets0, \mathfrak{T}^1\gets0, \TX\gets[],\PTX\gets\emptyset$
    \State $\sk_\Service,\pk_\Service\gets\KeyGen_\Service(\pp)$
    \State $\Oracle(e):=\{\Oracle\Service\Register_{\sk_\Service,e},\allowbreak\Oracle\Service\Participate_{\sk_\Service,e},\allowbreak\Oracle\Service\Payout,\allowbreak
    \Oracle\Register_{\sk_\Service},\allowbreak\Oracle\Participate_{\sk_\Service},\allowbreak\Oracle\Payout,\Oracle\mathsf{state}\}$
    \State $\auxil\gets \adv_0^{\Oracle(0)}(\pp,\pk_\Service)$
    \If{$\mathfrak{T}^0<0$}
    \Return $\mathtt{win}$
    \EndIf
    \State $\adv_1^{\Oracle(1)}(\pp,\pk_\Service,\auxil)$
    \If{$\mathfrak{T}^1<0 \lor \mathfrak{T}^0+\mathfrak{T}^1<0$}
    \Return $\mathtt{win}$
    \EndIf
\end{algorithmic_small}

\end{minipage}
\begin{minipage}[t]{0.33\textwidth}

\noindent
{\underline{$\mathsf{PartPriv}_b(\secparam)$:}}
\begin{algorithmic_small}
    \State $\pp\gets\Setup(\secparam)$
    \State $\mathfrak{U}^0\gets\emptyset,\mathfrak{T}^0\gets0,\TX\gets[],\PTX\gets\emptyset$
    \State $\Oracle(L):=\{\Oracle\Participant\Register,\allowbreak\Oracle\Participant\Participate_L,\allowbreak\Oracle\Participant\Payout_L\}$
    \State $(\TX',\username_0,\username_1,\Task,\adv_\Service,\auxil)\gets\adv_0^{\Oracle(\emptyset)}(\pp)$
    \Assert $\BB\subseteq\BB'$
    \Assert {$\forall i\in\bit: \CheckQualification(\mathfrak{U}^0_{\username_i}.\cred,\TX',\Task)=1$}
    \State run $\Participant$ side of $\rtx\gets\Pi_\Participate\langle\Participant(\mathfrak{U}^0_{\username_b}.\cred,\TX',\Task),\adv_\Service\rangle$
    \State $b'\gets\adv_1^{\Oracle(\{\username_0,\username_1\})}(\pp,\rtx,\auxil)$
    \If{$b=b'$}
      \Return $\mathtt{win}$
    \EndIf
\end{algorithmic_small}

\end{minipage}

\end{framed}
\vspace{-.8em}
\caption{Oracles with Security and Privacy Games.}
\label{fig:games}
\Description{A listing of pseudocode algorithms separated in two areas. The top area shows the 10 oracles that are available to adversaries for our game-based security and privacy definitions. The bottom area shows three pseudocode definition of our security and privacy games.}
\end{figure*}

\subsection{Participation Security}

Our first property is participation security, which ensures that participants can only participate in studies for which they qualify (checked by $\CheckQualification$).  We capture this property by defining a security game with oracles from above. The game $\mathsf{PartSec}$ below sets up the service with a generic organizer and then allows the adversary $\adv$ oracle access to register users and participate with them in studies. The oracle $\Oracle\Service\Register$ only checks that each username is registered once and $\Oracle\Service\Participate$ tracks every participation by appending the reward transaction to the list $\TX$. The adversary also has the power to observe honest users interacting with the service through $\Oracle\Register$ and $\Oracle\Participate$. There is only one epoch $e=0$. For completeness the adversary also has access to payout oracles, however they do not help in any way.
We note that a single generic organizer in this scenario has equivalent power as multiple ones. They all perform the same verification and may be controlled by a single entity.

The adversary is then challenged to present a credential $\cred$ which at any point in the list of participations successfully participated by having a reward transaction matching the credential but was not allowed to do so due to missing qualifications.

\begin{definition}[Participation Security]
\label{def:partsec}
 A participation scheme is secure if for all $\lambda\in\mathbb{N}$, all ppt adversaries $\adv$ and the game $\mathsf{PartSec}$ in \Cref{fig:games} it holds that:
 $\Pr[\mathsf{PartSec}(\secparam)=\mathtt{win}]\leq\negl$
\end{definition}

\subsection{Balance}

In addition to assuring participation security, we require that participants can only claim as much rewards as they have rightfully earned. This balance property is captured in the $\mathsf{Balance}$ game. Similarly to $\mathsf{PartSec}$, a service and organizer is set up and the adversary has oracle access to register users, participate in studies, and request a payout through $\Oracle\Service\Payout$. The adversary may observe honest interactions through the oracles controlling both parties.
The anonymous nature of the participations requires special attention to prevent users from sharing rewards. All credentials maintained by the adversary are shared and are used by the adversary to participate and get rewards. Together with the fact that the oracle cannot keep track of balances of anonymous participants, we split the adversarial actions into two epochs.
During both, the adversary can register users, participate with them, and get payouts as well as observe fully honest users in their actions.
In epoch $0$, all adversarially registered users are tracked in $\mathfrak{U}^0$ and the total reward in $\mathfrak{T}^0$. If the adversary manages to get more paid out than was rewarded ($\mathfrak{T}^0 < 0$) it already wins. This prevents, e.g. stealing funds from honest users. In addtion, the adversary gets a second chance to win, if they are able to move coins between participants.

To test that, the game switches the epoch to $e=1$ and now registers all new users in $\mathfrak{U}^1$ and all rewards are tracked in $\mathfrak{T}^1$. As the participation is anonymous, this means that also rewards to participations from users in $\mathfrak{U}^0$ are still tracked by $\mathfrak{T}^1$. If, after the interaction, more is paid out than was rewarded in epoch $1$ $(\mathfrak{T}^1<0)$ the adversary wins. The identity of the users is revealed at payout, which allows subtracting from the matching $\mathfrak{T}^e$ to when this user was registered. The adversary $\adv_1$, getting secret state from $A_0$ through $\auxil$, also wins if the sum of both epochs is negative $(\mathfrak{T}^0+\mathfrak{T}^1<0)$. It is possible to have $\mathfrak{T}^0<0$ in the following scenario: During epoch $0$, the adversary registers $\username$ but does not participate in any task $(\mathfrak{T}^0=0)$. Then in epoch $1$, the adversary asks $\username$ to participate in a study with reward $v$ $(\mathfrak{T}^1=v)$. Then $\username$ should get a payout of $v$. As $\username\in\mathfrak{U}^0$, the deduction is from $\mathfrak{T}^0$, i.e. $\mathfrak{T}^0=-v, \mathfrak{T}^1=v$. Their sum is still non negative. This property also captures phishing attacks for rewards of users, which cannot be paid out by the person who stole the credential. To win our game, the adversary must find a way to get a credential paid out to a ``new'' username which was earned by an ``old'' participation.

\begin{definition}[Balance]
\label{def:balance}
 A participation scheme is balanced, if for all $\lambda\in\mathbb{N}$ and all ppt adversaries $(\adv_0,\adv_1)$, it holds that
 $\Pr[\mathsf{Balance}(\secparam)=\mathtt{win}]\leq\negl$
 with $\mathsf{Balance}$ from \Cref{fig:games}.
\end{definition}

\subsection{Participation Privacy}

Next to the security properties protecting the service from malicious users, we define the privacy property which protects the users' anonymity from a malicious service. The participation privacy assures that as the service and the organizers collude they are still unable to track a user from the registration with their real identity to participations or link between participations or to payouts.
In our game-based definition, the adversary controls the service and the organizers. It is given oracle access to register participants and make them participate in a given study without access to their secret state ($\cred$). The adversary has no use of the double secret oracles (e.g. $\Oracle\Participate$) as all participants may be malicious. In the second invocation of the adversary (connected by $\auxil$), they have limited access to $\Oracle\Participant\Participate$ and $\Oracle\Participant\Payout$ for the users $\username_0$ and $\username_1$. This prevents trivial cases to get $b'$ from the participants' possibility to qualify for a study or access to the reward. Together with the Balance property, no rewards can be shared between the users to discriminate. As usernames are unique, we use them to identify the users. Once the adversary prepared the users, it provides a bulletin board $\TX'$ (which is a superset of oracle participations, as otherwise it is easy to make the same user participate twice) and a Task $\Task$ to fulfill by one of the two users $\username_0$ or $\username_1$.
Both users need to have valid credentials registered by the $\Oracle\Participant\Register$ oracle and fulfill the prerequisites ($\CheckQualification(\cred,\TX,\Task)=1$) to participate. This ensures that the adversary gets no advantage from an aborted participation.
The game proceeds to participate with the credential of the user specified by the bit $b$.
The interactive adversary $\adv_1$, with access to restricted oracles, needs to efficiently distinguish between the $b=0$ and $b=1$ game.

\begin{definition}[Participation Privacy]
\label{def:partpriv}
 A participation scheme has private participations, if for all $\lambda\in\mathbb{N}$ and all ppt adversaries $(\adv_0,\adv_1)$ and the game $\mathsf{PartPriv}$ defined in \Cref{fig:games}, it holds that
 $|\Pr[\mathsf{PartPriv}_0(\secparam)=\mathtt{win}]-\Pr[\allowbreak\mathsf{PartPriv}_1(\secparam)=\mathtt{win}]|\leq\negl$.
\end{definition}

\emph{Remark:} If a series of tasks are part of a longitudinal study, a possible requirement is to link participants over a longer period of time. This is modeled as weak participation privacy, where the organizer can link a previous participation to the current one. However, anonymity regarding any other task is still maintained.

\begin{corollary}[Fixed Participant State]
\label{cor:fixedstate}
 From the syntax of our formalization, it is visible that the participants get a credential through the $\Pi_\Register$ protocol, and thereafter, the credential $\cred$ is never updated. Such a property is desirable as users can back up their credential after registration and then in case of a recovery derive all secret state from this credential in combination with public information.
 One allowed exception is the need for users to store previously disclosed payout nullifiers $\PTX'$. This can be circumvented if the service maintains a public append only log of $\PTX$.
\end{corollary}

The validation of our formalization with regards to the log sheets and stickers is deferred to \Cref{appx:validation}.
\section[PrePaMS]{\prepams}
\label{sec:prepams}
In this section we present our protocol \prepams, which realizes a privacy-preserving participant management with the properties defined above.

We present the complete construction of the \prepams scheme in \Cref{fig:construction}.
It utilizes multiple building blocks, which are described in detail in \Cref{sec:defprelim}:
\begin{itemize*}[leftmargin=1.5em, nosep]
 \item a partially blind signature scheme  $\PBS = (\allowbreak\Setup,\allowbreak \KeyGen,\allowbreak \Blind,\allowbreak \Sign,\allowbreak \Unblind,\allowbreak \Verify)$ with randomness space $\rspace{\PBS}$ and an efficient NIZK for $\mathcal{L}_\PBS$ to show correct blinding.
  $\Blind$ allows a user to blind a secret message, which is then $\Sign$-ed with the signer only knowing part of the message. The user then $\Unblind$-s the result and is able to prove a valid signature with $\Verify$.
    As we need this building block twice, we instantiate two, domain-separated, versions called $\PBS\credential$ and $\PBS\reward$,
 \item a labeled verifiable random function $\VRF = (\Setup,\allowbreak \KeyGen,\allowbreak \Eval)$ where $\KeyGen$ generates a public private key pair and $\Eval$ takes a label from the domain of tasks $\mathbb{T}$ in addition to the secret key from $\skspace{\VRF}$ and outputs a deterministic, uniformly random, verifiable tag,
 \item zero-knowledge non-interactive arguments of knowledge $\AoK{} = (\Setup,\allowbreak \Prove,\allowbreak \Verify,\allowbreak \Sim)$ to prove various NP languages $\mathcal{L}$,
 \item a key derivation function $\KDF:\skspace{\VRF} \times \mathbb{T} \to \mathbb{S} \times \rspace{\PBS}$, which takes a secret key from $\skspace{\VRF}$ and a task from $\mathbb{T}$ to deterministically generate a nullifier and a $\PBS$ blinding randomness.
\end{itemize*}

We explicitly indicate exchanged messages in the construction of interactive protocols with blocking $\send$ and $\recv$ calls.

\begin{figure*}[t]
    \centering
    \begin{minipage}{.99\textwidth}
        \begin{framed}
            \begin{minipage}[t]{0.34\textwidth}
                \begin{center}
                    \vspace{-.5em}
                    \textsc{\small General Algorithms}
                    \vspace{.4em}
                \end{center}

                {$\underline{\Setup(\secparam,m,n)}$ }
                \begin{algorithmic_small}
                    \State $\db \gets \emptyset, \TX \gets [], \PTX\gets\emptyset$
                    \State $\pp_\VRF \gets \VRF.\Setup(\secparam)$
                    \State $\pp_\KDF \gets \KDF.\Setup(\secparam)$
                    \State $\pp_{\PBS\credential} \gets \PBS\credential.\Setup(\secparam,1,m+1)$
                    \State $\pp_{\PBS\reward} \gets \PBS\reward.\Setup(\secparam,2,1)$
                    \State $\pp_\Participate \gets \AoK{\Participate}.\Setup(\secparam)$
                    \State $\pp_\Payout \gets \AoK{\Payout}.\Setup(\secparam)$
                    \State $\pp:= (n, \pp_\VRF, \pp_\KDF, \pp_{\PBS\credential}, \pp_{\PBS\reward},\pp_\Participate,\allowbreak \pp_\Payout)$
                    \FullReturn $\pp$
                \end{algorithmic_small}

                \bigskip

                $\underline{\CheckCred(\cred, \pk_\Service)}$
                \begin{algorithmic_small}
                    \State $(\sk, \attr',\username', \sigma) \gets \cred$
                    \FullReturn {$\PBS\credential.\Verify(\pk_{\Service,\credential}, \sk, \attr'\|\username', \sigma)$}
                \end{algorithmic_small}

                \medskip

                $\underline{\CheckParticipation(\cred, \rtx)}$
                \begin{algorithmic_small}
                    \FullReturn {$\rtx.\Tag = \VRF.\Eval(\cred.\sk, \rtx.\Task)$}
                \end{algorithmic_small}

                \medskip

                $\underline{\CheckQualification(\cred, \TX, \Task)}$
                \begin{algorithmic_small}
                    \State $\tau \gets \VRF.\Eval(\cred.\sk, \Task)$
                    \State $Q \gets \{\VRF.\Eval(\cred.\sk, \Task.\qualifier_i)\}_{i=1}^{|\Task.\qualifier|}$
                    \State $D \gets \{\VRF.\Eval(\cred.\sk, \Task.\disqualifier_i)\}_{i=1}^{|\Task.\disqualifier|}$

                    \State $X := \{\TX_i.\Tag\}_{i=1}^{|TX|}$ \Comment set of rewarded tags in $\TX$

                    \Assert {$\tau \not\in X$} \Comment already participated in $\Task$
                    \Assert {$X \cap Q = Q$} \Comment qualifier not satisfied
                    \Assert {$X \cap D = \emptyset$} \Comment participated in disqualifier
                    \Assert {$\Satisfy(\cred.\attr, \Task.\Constraints) = 1$} \Comment attributes fulfill constraints
                \end{algorithmic_small}

                \medskip

                $\underline{\Receive(\cred, \TX, \PTX)}$
                \begin{algorithmic_small}
                    \State $a \gets 0,R\gets\emptyset$

                    \ForAll{$\rtx \in \TX$}
                        \If{$\CheckParticipation(\cred,\rtx)=1$}
                            \State $(\NUL,\rho) \gets \KDF(\cred.sk, \rtx.\Task)$
                            \If{$\PBS\reward.\Verify(\pk_{\Service,\reward}, (\NUL,\cred.\username), (\rtx.\Task.v), \rtx.r) = 1$}
                                \If{$\NUL \notin \PTX$}
                                    \State $a \gets a + \rtx.\Task.v$
                                    \State $R\gets R\cup\{\rtx\}$
                                \EndIf
                            \EndIf
                        \EndIf
                    \EndFor

                    \FullReturn $(a, R)$
                \end{algorithmic_small}
            \end{minipage}\hfill\begin{minipage}[t]{0.28\textwidth}
                \begin{center}
                    \vspace{-.5em}
                    \textsc{\small Service Algorithms}
                    \vspace{.4em}
                \end{center}

                $\underline{\KeyGen_\Service()}$
                \begin{algorithmic_small}
                    \State $(\sk_{\Service,\credential},\pk_{\Service,\credential}) \gets \PBS\credential.\KeyGen()$
                    \State $(\sk_{\Service,\reward},\pk_{\Service,\reward}) \gets \PBS\reward.\KeyGen()$
                    \State $\sk_\Service \gets (\sk_{\Service,\credential},\sk_{\Service,\reward})$
                    \State $\pk_\Service \gets (\pk_{\Service,\credential},\pk_{\Service,\reward})$
                    \FullReturn $(\sk_\Service,\pk_\Service)$
                \end{algorithmic_small}

                \bigskip

                $\underline{\Pi_{\Register,\Service}(\sk_\Service)}$
                \begin{algorithmic_small}
                    \State $(\username, \attr, \alpha, \pi) \gets \recv$
                    \Assert $\AoK{\PBS\credential}.\Verify(\pi, (\alpha)) = 1$
                    \Assert {$\username \not\in \db$}
                    \Assert {$\attr$ correct}
                    \State $\sigma' \gets \PBS\credential.\Sign(\sk_{\Service,\credential}, \alpha, \attr\|\username)$
                    \State $\db \gets \db \cup \{ \username \}$

                    \State $\send(\sigma')$

                \end{algorithmic_small}

                \bigskip

                $\underline{\Pi_{\Participate, \Service}(\sk_\Service)}$
                \begin{algorithmic_small}
                    \State $(\pi, \Task, \Tag,r') \gets \recv.\Participant$
                    \State $\stmt:=(\pk_\Service, \TX, \Task, \Tag, r')$
                    \Assert {$\AoK{\Participate}.\Verify(\pi, \stmt) = 1$}
                    \State $r \gets \PBS\reward.\Sign(\sk_{\Service,\reward}, r', (\Task.v))$
                    \State $\TX \gets \TX \| (\Task, r, \Tag)$
                \end{algorithmic_small}

                \bigskip

                $\underline{\Pi_{\Payout, \Service}(\TX,\sk_\Service)}$
                \begin{algorithmic_small}
                    \State $(\{ r'_i ,\pi'_i\}_{i=1}^n) \gets \recv$
                    \ForAll{$i \in n$} \Comment{sign padding coins}
                        \Assert {$\AoK{\PBS\reward}.\Verify(\pi'_i, r'_i) = 1$}
                        \State $v_i \gets 0$
                        \State $r''_i \gets \PBS\reward.\Sign(\sk_{\Service,\reward}, r'_i, (v_i))$
                    \EndFor
                    \State $\send(\{ r''_i \}_{i=1}^n)$

                    \State $(\pi, \{ \NUL_i \}_{i=1}^n, \username, v) \gets \recv$
                    \Assert {$\AoK{\Payout}.\Verify(\pi, (\{\NUL_i\}_{i=1}^n,\username,v)) = 1$}
                    \Assert {$\PTX \cap \{\NUL_i\}_{i=1}^n = \emptyset $}

                    \State $\PTX \gets \PTX \cup \{\NUL_i\}_{i=1}^n$
                    \State  offline payment of amount $v$ to $\username$
                \end{algorithmic_small}
            \end{minipage}\hfill\begin{minipage}[t]{0.34\textwidth}
                \begin{center}
                    \vspace{-.5em}
                    \textsc{\small Participant Algorithms}
                    \vspace{.4em}
                \end{center}

                $\underline{\Pi_{\Register,\Participant}(\username,\attr{\color{gray},\pk_\Service})}$
                \begin{algorithmic_small}
                    \State $\sk \gets \VRF.\KeyGen()$
                    \State $\rho\drawrandom\rspace{\PBS\credential}$
                    \State $\alpha \gets \PBS\credential.\Blind((\sk),\rho)$
                    \State $\stmt := (\alpha)$
                    \State $\wit := (\sk,\rho)$
                    \State $\pi \gets \AoK{\PBS\credential}.\Prove(\stmt, \wit)$
                    \State $\send(\username, \attr, \alpha, \pi)$ \Comment{with external auth token}
                    \State $\sigma' \gets \recv$
                    \State $\sigma \gets \PBS\credential.\Unblind(\pk_{\Service,\credential}, \rho, \sigma')$
                    \State $\cred := (\sk,\attr,\username, \sigma)$
                    \FullReturn $\cred$
                \end{algorithmic_small}

                \bigskip

                $\underline{\Pi_{\Participate, \Participant}(\cred, \TX, \Task)}$
                \begin{algorithmic_small}
                    \State $\Tag \gets \VRF.\Eval(\cred.\sk, \Task)$
                    \State $(\NUL,\rho) \gets \KDF(\cred.\sk,\Task)$
                    \State $r' \gets \PBS\reward.\Blind((\NUL,\cred.\username), \rho)$
                    \State $\stmt := (\pk_\Service, \TX, \Task, \Tag, r')$
                    \State $\wit := (\cred, \NUL, \rho)$
                    \State $\pi \gets \AoK{\Participate}.\Prove(\stmt, \wit)$
                    \State $\send(\pi, \Task, \Tag,r')$
                \end{algorithmic_small}

                \bigskip

                $\underline{\Pi_{\Payout, \Participant}(\cred, \TX,\PTX, v,\{\rtx_i\}_{i=1}^{k})}$
                \begin{algorithmic_small}
                    \Assert $k\leq n$
                    \Assert {$\Receive(\cred, [\rtx_i]_{i=1}^k,\PTX).v \geq v$}

                    \ForAll{$i \in [n]$} \Comment{prepare padding tx}
                        \State $(\NUL_i,\rho_i)\drawrandom\rspace{\KDF}$
                        \State $r'_i \gets \PBS\reward.\Blind((\NUL_i,\cred.\username), \rho_i)$
                        \State $\pi'_i \gets \AoK{\PBS\reward}.\Prove((r'_i),(\NUL_i,\cred.\username,\rho_i))$
                    \EndFor

                    \State $\send(\{( r'_i ,\pi'_i)\}_{i=1}^n, )$
                    \State $(\{ r''_i \}_{i=1}^n)\gets\recv$

                    \ForAll{$i \in [n]$} \Comment{prepare spendable tx}
                        \If {$i\in [k]$}
                        \State $(\NUL_i,\rho_i) \gets \KDF(\cred.sk, \rtx_{i}.\Task)$
                        \State $r_i \gets \PBS\reward.\Unblind(\pk_{\Service,\reward}, \rho_i, \rtx_{i}.r)$
                        \State $v_i \gets \rtx_{i}.\Task.v$
                        \Else
                        \State $r_i \gets \PBS\reward.\Unblind(\pk_{\Service,\reward}, \rho_i, r''_i)$
                        \State $v_i \gets 0$
                        \EndIf
                    \EndFor

                    \State $\stmt := (\{\NUL_i\}_{i=1}^n, \cred.\username,v)$
                    \State $\wit := (\{r_i, v_i\}_{i=1}^n)$

                    \State $\pi \gets \AoK{\Payout}.\Prove(\stmt, \wit)$
                    \State $\send(\pi, \{ \NUL_i \}_{i=1}^n,\cred.\username, v)$
                \end{algorithmic_small}
            \end{minipage}
        \end{framed}
    \end{minipage}
    \vspace{-.8em}
    \caption{The \prepams Construction.}
    \label{fig:construction}
    \Description{A listing of pseudocode algorithms separated into three columns: general algorithms, service algorithms, and participant algorithms. Each column contains a pseudocode definition of the algorithms and protocols supported by the respective \prepams parties.}
\end{figure*}

Given these building blocks, we present an overview of how we combine them: To register participants, we use the $\PBS\credential$ scheme to blind-sign a secret key of the $\VRF$ and the users attributes $\attr$, one of which is the user's identity $\username$. The $\PBS\credential$ signature is the credential and for each participation the participant has to create a $\VRF$ tag $\Tag$ for a specific task $\Task$ and prove correctness with a NIZK. Along with the participation, the participant sends a partially blinded coin to the organizer using $\PBS\reward.\Blind$. The coin consists of the identity $\username$ and a nullifier $\NUL$, sometimes called serial number. Once the task is fulfilled, the organizer together with the service signs coin with $\PBS\reward.\Sign$ and adds the amount of reward the task provides $\Task.v$ to get the coin, resulting in $r$ and creates a reward transaction $\rtx:=(\Task,r,\Tag)$ which is appended to the service's public log $\TX$. On payout, the participant reveals the nullifiers $\NUL_i$ and identity $\username$ and generates a NIZK which proves they are correct and the earned reward is greater than the payout. Double spending is prevented by the service which only accepts a nullifier once. Balance per user is assured by only paying out the amount to the identity of the coins. Due to the re-randomization of the coin, it cannot link it to the transaction where it was issued.

\subsection{Detailed Protocol Description}
To achieve payout privacy, the participant must not disclose their number of participations. To hide the real number of participations, we assume a system parameter $n$ denoting the maximum number of tasks to get paid out in one payout.
The $\Setup$ takes a security parameter $\secparam$, the number of attributes $m$, which is increased by one to accommodate the identity, and maximum input size for payouts $n$.
It initializes all building blocks and databases $\UN,\TX,\PTX$, and outputs a set of public parameters $\pp$.
After the initial setup and key generation of the service, participants can register (\cf $\Pi_\Participate$) to retrieve an anonymous credential $\cred$.
For this they run $\VRF.\KeyGen$ to pick a random secret key $\sk$ for later use in the verifiable random function, blind it with randomness $\rho\in\rspace{\PBS\credential}$ with $\sk$ being the blinded message and $\attr$ (including $\username$) the public part, using the partially blind signature scheme $\PBS\credential.\Blind$. They send the blinded key $\alpha$ with their identity $\username$ and an argument of knowledge $\pi$ of a correct blinding ($\mathcal{L}_{\PBS\credential}$) to the service.
The service verifies this argument $\AoK{\PBS\credential}.\allowbreak\Verify(\pi,(\alpha,\allowbreak\attr))=1$ and if it holds, the attributes are correct and the identity has not been registered yet (\ie $\username \notin \db$ where $\db$ is a set of registered users), it continues to sign the blinded seed $\alpha$ using its secret key $\sk_\Service$.
The identity $\username$ is added to the set of registered users $\UN$ and the signature $\sigma'$ is send back to the participant, who is then able to unblind the signature with $\PBS\credential.\Unblind$ and store it as part of its secret credential $\cred$.
The validity of the credential $\cred$ is checked by using $\PBS\credential.\Verify$.

On participation, the participant $\Participant$ provides a NIZK of a valid credential $\cred$ that is eligible to participate in the chosen task $\Task$.
For this construction, the argument shows four properties (\cf $\CheckQualification$):
\begin{enumerate*}
    \item $\Participant$ has not yet participated in $\Task$,
    \item $\cred$ satisfies the attributes required by $\Task$,
    \item $\Participant$ has participated in all qualifier of $\Task$,
    \item $\Participant$ has not yet participated in any disqualifier of $\Task$.
\end{enumerate*}
In our concrete instantiation, we decided to support range (from lower $l$ to upper $u$) and set constraints (in $V$), such that attributes $\attr$ satisfy a set of $\Constraints$ defined as:\\
$
\Satisfy(\attr,\Constraints) = \bigwedge_{c_i\in\Constraints} \begin{cases}
                               \attr_j\in[l,u] \text{ if } (l,u,j)\gets c_i \\
                               \attr_j\in V \text{ if } (V,j)\gets c_i
                              \end{cases}
$

To track previous participations, we utilize a verifiable random function $\VRF$ to derive a participation tag $\Tag$ for the label of a task, while allowing a participant to prove the correctness of the tag via the service's signature on the VRF seed.
This combination of anonymous credentials and verifiable random functions was previously proposed by Hohenberger et al. \cite{Hohenberger2014anonize} in their ANONIZE scheme.
Similar to ANONIZE, a participant re-randomizes their signature and then provides a zero knowledge proof that the participation tag $\Tag$ corresponding to the task $\Task$ is computed correctly.
Verifying that the participant has not yet participated in $\Task$ is as simple as a lookup that the tag $\Tag$ is not part of any reward transaction in $\TX$.

While ANONIZE only allows participation predicates based on attributes of the credential, we additionally enable qualifiers and disqualifiers based on previous participations by computing the relevant tags for all qualifiers and disqualifiers.
To not link this participation to any previous or future participations that include these tags, we construct further NIZKs to satisfy the qualifiers, disqualifiers and attribute constraints without revealing the corresponding tags and attributes.
For a \emph{qualifier} we derive the respective tag under the required task and show that it is derived using the same seed as $\Tag$ and that there exists a reward transaction in $\TX$ that stored this tag.
Essentially, providing a ring-signature on the set of referenced tags for this task.
For \emph{disqualifiers}, a random, secret value is picked that is used to randomize the set of referenced tags of the disqualifying task in $\TX$.
We use a homomorphic randomization and show the correctness of the randomization in the NIZK.
The participant then computes their own tag for the disqualifying task, again showing that the same secret key is used, and applies the same randomization as for the set of disqualifying tags.
The randomized set and the participant's own tag is then included in the argument to allow a verifier to check that it is not included, without linking a future participation in the disqualifying task to the current participation.
For an \emph{attribute constraint}, we prove that an intermediate blinding commitment holds the same attributes as the signed credential and then use a range or set membership bulletproof to show that a given attribute satisfies the constraint.
The NIZK shows knowledge of a valid credential being used to generate a valid participation and shows fulfillment of all preconditions: $\mathcal{L}_\Participate :=$
\begin{equation*}
     \left\{
        \begin{array}{l}
            \left(
                \pk_\Service,
                \TX,
                \Task,\Tag,r'
            \right) \mid \exists (\cred,\NUL,\rho)\ST \\
            \quad r'=\PBS.\Blind((\NUL,\cred.\username),\rho) \land \CheckCred(\cred, \pk_\Service) = 1 \\
            \quad \land ~~ \CheckParticipation(\cred, (T,\cdot,\Tag)) = 1  \land  \CheckQualification(\cred, \TX, \Task) = 1
        \end{array}
    \right\}
\end{equation*}
To be able to later claim a reward, the participant derives a nullifier $\NUL$ and a blinding factor $\rho$ for this participation based on their secret key $\cred.\sk$ and the task $\Task$. The participant uses $\PBS\reward$ to blind $\NUL$ and $\username$ to $r'$ and sends the blinded coin $r'$ together with their participation tag $\Tag$ and a NIZK of $\mathcal{L}_\Participate$ to the organizer of this task.
The organizer then verifies the argument $\AoK{\mathcal{L}_\Participate}.\Verify$ and if successful signs the coin of reward $v$ using the service's secret key $\sk_\Service$. In the implementation, we provide organizers access to the service signing key through an API which also keeps track of billing the organizers.
The reward transaction $\rtx$ includes the participation tag $\Tag$, the signed coin $r$, and the task $\Task$ (\ie $\rtx := (\Task,r,  \Tag)$).

After participating in one or more tasks, a participant can request a payout of earned rewards using $\Pi_\Payout$.
In this process, the participant has to disclose their identity $\username$ to receive payment outside the system. Anonymous payouts would make the balance property invalid, as rewards could be shared between participants.
In order to not link their identity to previous participations, the reward transactions in $\TX$ that the participant wants to spend are kept private.
A NIZK of $\mathcal{L}_\Payout$ proves knowledge of correct nullifiers $\NUL_i$ and identity $\username$ for the coins $r_i$ of reward transactions $\rtx_{i}$, with rewards $v_i$ that sum up to greater or equal to the payout amount $v$. This is required because a unique reward value could otherwise be used to link participations to payouts. The identity $\username$ has to be the same in all coins and equal to the payout recipient.
To also keep the number of transactions that are spent private, the participant is expected to pad the spent coins with up to $n$ fake coins $r_1,\dots,r_n$ with a fixed reward of zero, a random nullifier $\NUL_i$ and their identity $\username$. The participant generates the new coins $r'_i$, blinds them, and sends them to the service along with a proof $\pi'_i$ of $\mathcal{L}_{\PBS\reward}$ to show they have zero value. The service verifies the proofs $\pi'_i$, signs these coins with $\PBS\reward.\Sign$, and sends them back.
The padding is necessary because the proof size $|\pi|$ is proportional to the number of rewards paid out.
If the participant wants to use less than $n$ inputs, the remaining inputs are from the padded coins.

To prevent users from getting one reward paid out multiple times, the nullifier $\NUL$ is published to the service. The deterministic nullifier is detectable if the same coin is used twice.
Again, showing the correctness of the coins is performed with a NIZK. $\mathcal{L}_{\Payout}:=$
\begin{equation*}
    \left\{
        \begin{array}{l}
            \left(
                \{\NUL_i\}_{i=1}^n,\username,
                v
            \right) \mid \exists (\{(r_i,v_i)\}_{i=1}^{n})\ST \\
             \forall i\in [n]:
\PBS\reward.\Verify(\pk_\Service,(\NUL_i,\username),v_i,r_i) = 1 \land \sum_{i=1}^{n} v_i \geq v
        \end{array}
    \right\}
\end{equation*}
The service verifies the argument of knowledge $\pi$ and checks that none of the nullifier have been recorded before in $\PTX$ and all identites match. Finally, it issues the payment outside of the system and adds the new nullifiers $\NUL$ to the database $\PTX\gets\PTX\cup\NUL$.
The detailed constructions for the NIZKs are presented in \Cref{appx:languages}.
The analysis of security and privacy properties of the construction is presented in \Cref{appx:analysis}.
\section{Prototype}
\label{sec:implementation}

After showing that our scheme is cryptographically secure, we demonstrate the practicality of our approach with a proof of concept implementation.
Our motivation for this prototype is to provide a realistic deployment for the use case of managing psychological study participations at a university, as introduced in \Cref{sec:requirements}.
Hence, we implemented our \prepams{} system using modern web technologies (see \Cref{fig:prototype}) to enable usage by participants on a variety of devices including mobile devices without the need to install a native application.

\subsection{Implementation}
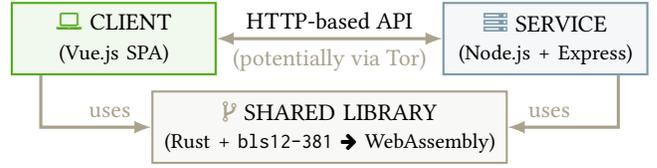
\begin{figure}[t]
    \centering
\begin{tikzpicture}[
    label/.style={
        text width=2.4cm,
        align=center,
    },
    entity/.style={
        thick,
        text width=2.5cm,
        align=center,
        inner ysep=1pt,
        minimum height=.95cm,
        draw=#1,
        fill=#1!8
    },
    arrow/.style={
        latex-latex,
        very thick,
        draw=uulm-akzent
    }
]
    \node[entity=uulm-mawi] (client) {};
    \node[label, below=1pt of client.north] {{\color{uulm-mawi}\faLaptop}~~\textsc{CLIENT}};
    \node[label, above=1pt of client.south] {\small (Vue.js SPA)};
    \node[xshift=.45cm,yshift=-.0cm] at (client) {};

    \node[entity=uulm, right=3cm of client] (server) {};
    \node[label, below=1pt of server.north] {{\color{uulm}\faServer}~~\textsc{\textsc{SERVICE}}};
    \node[label, above=1pt of server.south] {\small (Node.js + Express)};

    \coordinate (midpoint) at ($(client.south)!.5!(server.south)$);

    \node[entity=uulm-akzent, text width=4.5cm, below=.2cm of midpoint] (library) {};
    \node[label, text width=4.5cm, below=1pt of library.north] {{\color{uulm-akzent}\faCodeFork}~~\textsc{\textsc{SHARED LIBRARY}}};
    \node[label, text width=4.5cm, above=1pt of library.south] {\small (Rust + \texttt{bls12-381}~~{\footnotesize\faArrowRight}~~WebAssembly)};

    \draw[arrow,-latex] ([xshift=.4cm] client.south west) |- node[above,pos=.95,anchor=south east] {\color{uulm-akzent}uses} (library.west);
    \draw[arrow,-latex] ([xshift=-.4cm] server.south east) |- node[above,pos=.95,anchor=south west] {\color{uulm-akzent}uses} (library.east);

    \draw[arrow] (client) -- node[above] {HTTP-based API} node[below] {\color{uulm-akzent} (potentially via Tor)} (server);
\end{tikzpicture}
     \vspace{-2em}
    \caption{Graphical overview of the prototype architecture.}
    \label{fig:architecture}
    \Description{A diagram showing the components of the \prepams prototype and their relation to each other. The Client component is build with the SPA framework Vue.js and interacts with the Service component, which is built using Node.js and the Express web application framework, using an HTTP-based API which is potentially accessed via Tor. Both components additionally use a Shared Library developed in Rust utilizing the bls12-381 library and compiled to WebAssembly.}
\end{figure}

Although modern browsers natively support some high-level cryptographic operations using the Web Cryptography API~\cite{Watson17WCA}, this is insufficient for pairing-based cryptography or zero-knowledge proofs.
While this cryptography could be implemented purely in JavaScript, we opted to implement the cryptographic components of our \prepams scheme in Rust and compile it to WebAssembly.
This results in better performance and type safety, in contrast to a pure JavaScript implementation, additionally following the paradigm of a small auditable cryptography core.
We chose the BLS12-381 pairing-friendly elliptic curve that is, for example, used by the privacy-focused cryptocurrency Zcash with an estimated security level of 123 Bit \cite{wang2023}.
The main client-side application is implemented as a single page application using the Vue.js framework, and the service application is built with Node.js and the \texttt{Express} web application framework.
This allows client and service to share the same WebAssembly library for cryptographic operations.
The prototype architecture is shown in \Cref{fig:architecture}.

For the proof of concept, we implemented a basic authentication system that accepts any username that is not yet registered.
In a production deployment, this is meant to be replaced by an institutional authentication provider (\eg through OAuth, SAML).
Using the Fixed Participant State (\Cref{cor:fixedstate}), we can derive the secret keys from a user's password, which is never shared with the service.
This allows users to reconstruct any private local state when switching devices without the need to store sensitive user state in a central database (e.g., the service).

\begin{figure*}[t]
    \centering
    \begin{subfigure}[b]{0.59\textwidth}\centering
        \includegraphics[scale=.0595]{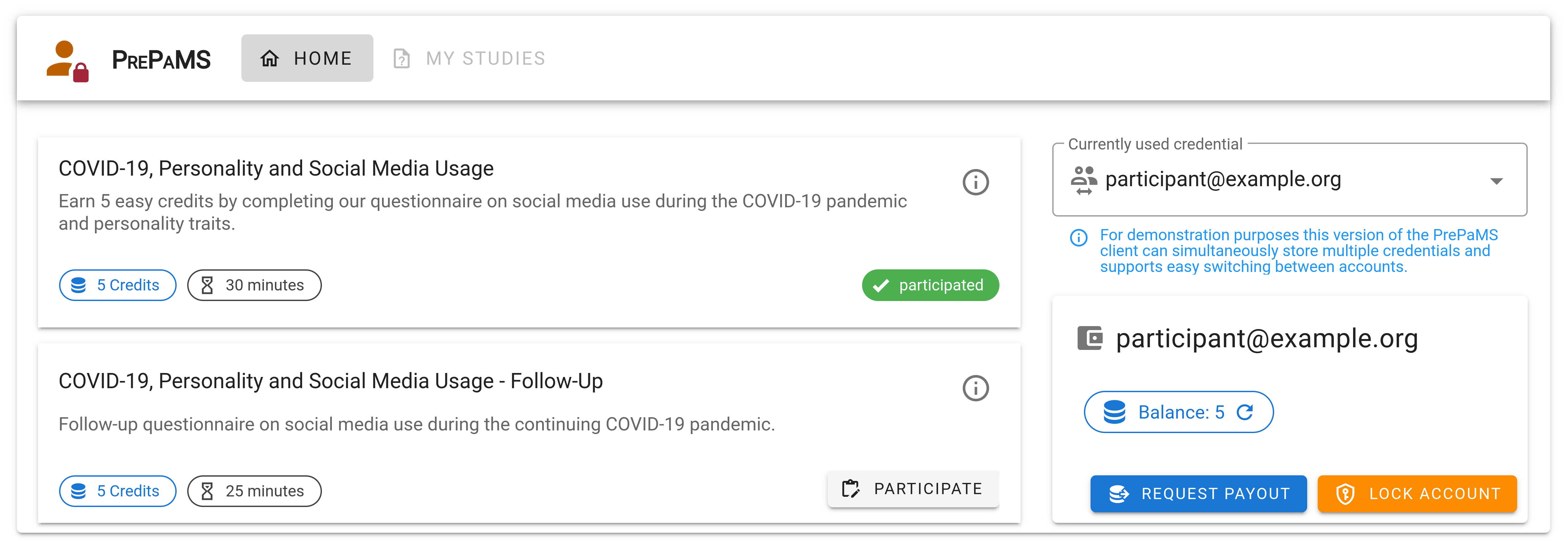}
    \end{subfigure}
    \hfill
    \begin{subfigure}[b]{0.40\textwidth}\centering
        \includegraphics[scale=.0595]{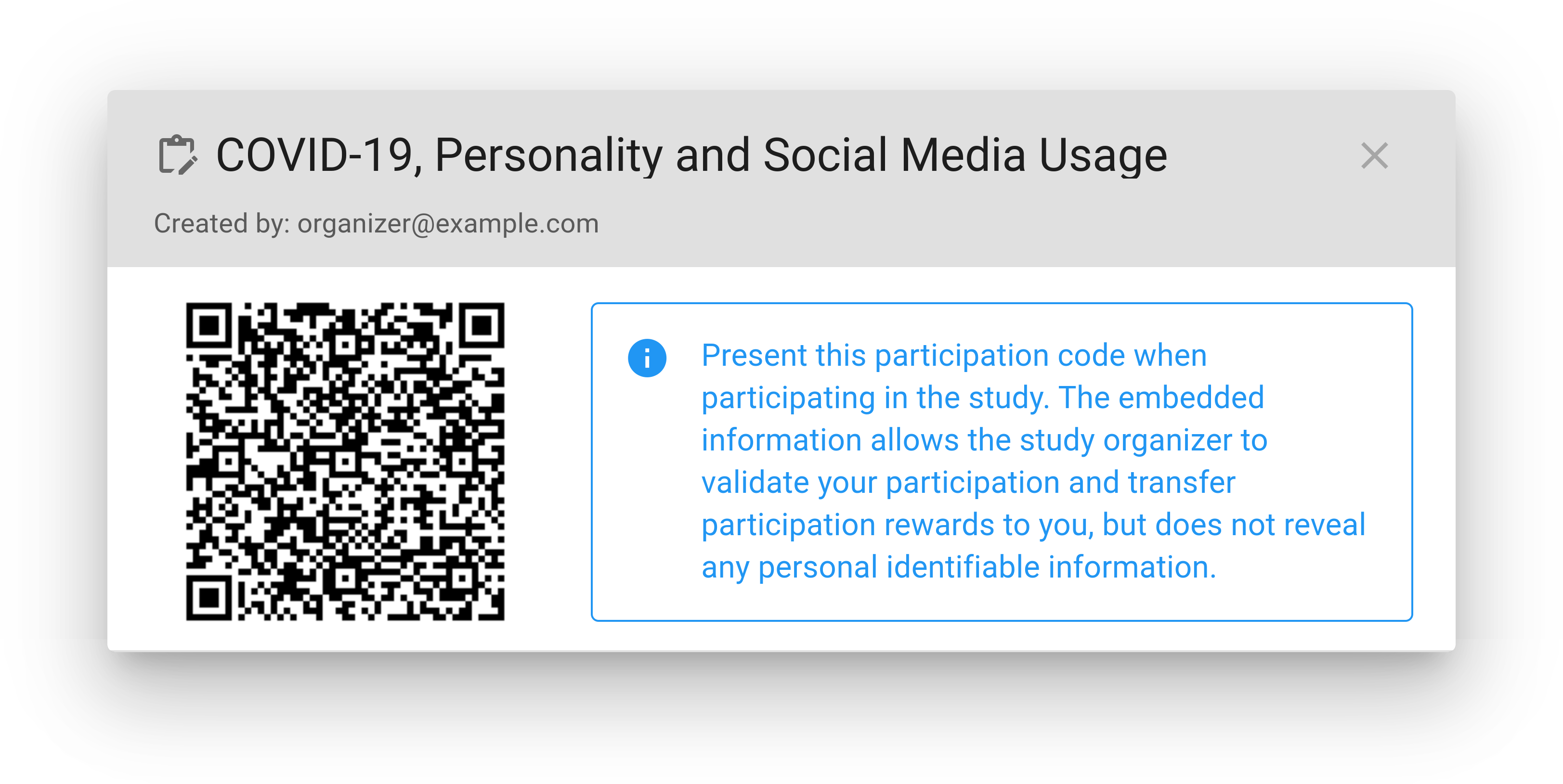}
    \end{subfigure}

    \vspace{-.7em}

    \caption{
        The system shows an overview of active studies, where a participant may participate in, and an overview of the participant's wallet with the option to request a payout of previously earned rewards.
        When participating in an offline study, the participant has to bring along the participation QR code, which can be scanned by the study organizer to confirm the participation and transfer the rewards without gaining additional information about the participant.
        For web-based studies, the participation data is directly sent to the organizer's survey system using an HTTP POST request.
        A public demo deployment is available\footref{deployment} to interactively explore the prototype.
    }
    \label{fig:prototype}
    \vspace{-.8em}
\end{figure*}

In online-only studies, such as web-based surveys, the messages of the $\Pi_\Participate$ protocol are easily submitted directly from the browser to the organizer-managed survey system.
However, in-person studies enable more complex setups and human interaction with the participant.
For this, we allow participation requests to be uploaded to the service in encrypted form.
The corresponding secret to open the encrypted participation is then given to the organizer out of band (\eg presenting a QR code during an offline lab experiment, \cf, \Cref{fig:prototype}).

After the initial registration, participants are not required to have an authenticated session with the service.
Instead, the blindly-signed $\VRF$ secret key is used as an anonymous credential to authenticate participations and payouts.
This theoretically achieves the desired privacy guarantees.
However, in practice, the service may use side channel information to obtain additional information about the participants.
To not disclose the specific studies a participant is interested in, we use the private information retrieval~\cite{chor1998private} pattern when accessing study details by requesting all public data or a random subset from the service.
Metadata, such as the client IP address of users interacting with the system, is accumulated by the service operator and potentially used to de-anonymize participants.
As addressed in \Cref{sec:formalization}, this is mitigated by utilizing anonymous communication technologies such as Tor.

Even though all private state of users is recoverable from public information, a local cache is beneficial for performance reasons.
This sensitive user state is only stored by the user's browser and not shared with the service.
To further protect the state, it is locally encrypted with the user's password.
If a user is inactive for some time, the unencrypted state is discarded, and the user will be prompted for their password.
This is particularly useful when users access the service from shared devices, such as a computer lab on campus, and may forget to log out afterward.

During registration, a recovery QR code is generated, allowing users to restore their account in case they forget their password.
In the future, the authentication and recovery process could be further improved using Web Authentication Credentials~\cite{w3c2022webauthn}, also commonly known as Passkeys.
This API enables credential generation and synchronization based on established authentication patterns, such as a device's biometric capabilities.
At the time of writing, the API does not allow the export of key material for security reasons and can not be used to derive suitable key material for \prepams.
But, for example, the proposed pseudo-random function extension~\cite[§ 10.1.4.]{w3c2022webauthn}, could potentially be used to derive \prepams credentials.

In our prototype, the service also provides a partially trusted bulletin board with append-only semantics.
In practice, the service could be operated by an institution's accounting department or comparable entity.
Alternatively, an append-only transparency log system, similar to certificate transparency~\cite{RFC6962}, can strengthen these guarantees.

In total, we implemented the \prepams client in 2356 lines of JavaScript code, the service in 569 lines of JavaScript code, performing little business logic but mostly storing and serving data, and the shared crypto library in 4240 lines of Rust code.
Client-side entities and the service backend are decoupled and interact via an HTTP-based API.
The prototype implementation of \prepams is available under an open-source license\footref{github}.
Additionally, we deployed a live test instance that is publicly available\footref{deployment}.

\subsection{Performance Evaluation}

With our performance evaluation, we assess two main aspects of our scheme and the associated proof of concept implementation:
\begin{enumerate*}[label=(\roman*)]
\item the computational impact of prerequisite complexity on performance and
\item processing times when users interact with the system under realistic conditions
\end{enumerate*}.
For the first aspect, we conducted a set of microbenchmarks to explore the influence of each parameter on the performance of the protocols.
The results of these microbenchmarks matched the expectations based on the amount of curve operations (c.f., \Cref{appx:efficiency}).
For the second aspect, we run benchmarks within an actual browser using a WebAssembly compilation of our library.

We have released all evaluation artifacts\footnote{\emph{Evaluation artifacts:} \url{https://github.com/vs-uulm/prepams/pets25.1/main/evaluation}} under an open-source license, following the Popper convention~\cite{jimenez2017popper} for reproducibility.
This features an automated, docker-based setup that executes the experiments in a remote-controlled headless browser using Puppeteer\footnote{\url{https://github.com/puppeteer/puppeteer}}.
The evaluation was conducted on a desktop computer with an Intel Core i7-7700 (quad-core with SMT; 3.60 GHz) CPU and 32 GB RAM, running Ubuntu 22.04 LTS (GNU/Linux 5.4.0) with a headless Chrome 108.
Additionally, we included a semi-automated build of the evaluation\footnote{\emph{Browser benchmark suite:} \url{https://vs-uulm.github.io/prepams/eval/}} to be used on any device featuring a modern browser.
We used this artifact to evaluate the performance on different end-user devices, more precisely, an Android smartphone (Pixel 6 w/ Android 14 and Chrome 121) and an Apple tablet (3\textsuperscript{rd} Gen iPad Pro 11'' w/ iPadOS 17.2 and Safari 172).

To put our system under load, we created a parametric workload generator.
The generator outputs a chronological sequence of valid interactions with the system (\ie registrations, participations, payouts) that can be replayed for reproducible evaluation runs.
We defined suitable defaults and ranges for each parameter (\ie distributions for each condition type) based on previous joint projects with psychology researchers to develop well-grounded workloads.
The replay phase is preceded by an initialization phase to bootstrap a set amount of studies with randomized preconditions.

\subsubsection{Computational Impact of Prerequisite Complexity}

In this evaluation, we wanted to assess the usage of varying amounts of different prerequisites in studies.
We wanted to check whether the computational costs are reasonable enough to put our approach into practice.

\noindent\emph{Workload.}
We generated a corresponding series of workloads for each type of prerequisite (\ie qualifier, disqualifier, attribute range constraint, set constraint).
Each series contains a single study with a progressive increase in prerequisites, reaching a maximum of 10 per type.
For qualifier and disqualifier workloads, we created additional studies necessary to reflect the dependencies of the target study.
$N=100$ random participants that satisfy the specified range and set constraints are created.
If necessary, every participant first participates in all $n$ qualifiers, or a set of $n$ dummy participants participates in all $n$ disqualifiers.
Every participant then participates in the target study, recording the time it takes for the operation.

Before execution of the workload, a set of $50$ additional participants follow the same steps---as warm-up and to fill the bulletin board.
Any measurements of this warm-up phase are discarded.

\begin{figure}[h]
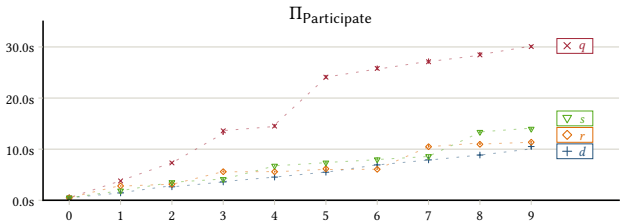

    \centering

     \vspace{-2em}
    \caption{
        Plot of measured median execution times in seconds of the participation protocol based on a synthetic workload with either \textcolor{uulm-in}{qualifier ($\times$, $q$)}, \textcolor{uulm-med}{disqualifier ($+$, $d$)}, \textcolor{uulm-nawi}{range constraints ($\diamond$, $r$)}, or \textcolor{uulm-mawi}{set constraints ($\triangledown$, $s$)} varied from $n \in [0..10]$ and all other parameters pinned to $0$.
    }
    \label{fig:scaling}
    \vspace{-1em}
\end{figure}

\subsubsectionstar{Results}
The results (see \Cref{fig:scaling}) indicate a mostly linear scaling with an increasing number of (dis-)qualifiers with an exponential offset in irregular intervals.

\subsubsectionstar{Discussion}
This step-wise increase in computation time can be accounted for by the vector width of the inner product proof being padded to the next power of two.
Each prerequisite type requires a different amount of group elements to be encoded in the inner product proof, yielding a different scaling behavior.
Quantitative empirical studies require a minimum sample size to achieve statistically significant results.
However, combining numerous prerequisites can make it difficult to recruit enough participants and thus impossible to achieve these sample sizes.
For this reason, the number of prerequisites for psychological studies is usually in the lower single-digit range, which is also the cardinality our evaluation is grounded on.
The results indicate that the scheme is also applicable for studies with non-trivial requirements that include multiple prerequisites.

\subsubsection{Performance of User Operations under Realistic Conditions}

In the second evaluation, we assessed the performance to be expected when users interact with our system.
More precisely, we measured the execution times of three crucial operations (\ie register, participate, and payout) based on actual user devices.

\subsubsectionstar{Workload}
We synthesized a workload that matches our running example based on an analysis of real studies and their typical requirements.
We limited the number of preconditions, \ie (dis-)qualifier and attribute constraints, for a given study to at most 10 per type.

For the evaluation, we generated a workload with $N=1,000$ participants, $M=1,000$ participations, and $O=100$ payout operations.
Before execution, an additional set of $N=200$ participants are registered and perform $M=200$ participations and $O=20$ payouts to fill the bulletin board and as a warm-up.
Again, all warm-up measurements are discarded.

The resulting measurements provide an indication of the real-world performance of our scheme.
In addition, other parameters can be easily explored with the open-source prototype and automated parametrized evaluation workflow.

\begin{figure}[h]
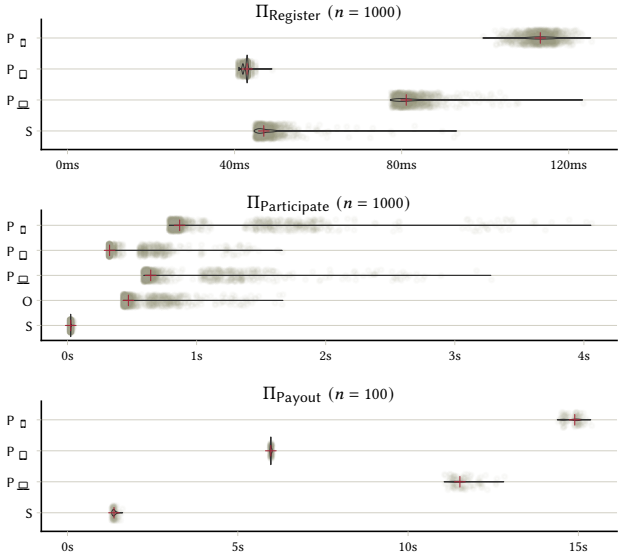

    \centering
    \let\opgfimage=\pgfimage
    \renewcommand*{\pgfimage}[2][]{\opgfimage[#1]{results/#2}}

     \let\pgfimage=\opgfimage
    \vspace{-2em}
    \caption{
        Combined violin/jitter plots of measured execution times (in seconds) of our \prepams proof of concept implementation based on a synthetic workload with $N=1,000$ participants, $M=1,000$ participations, and $O=100$ payouts.
        The individual protocols are segmented by the role of the executing party ($\Participant$: Participant, $\Organizer$: Organizer, $\Service$ Service) and partially replicated across different device types (\ie ~\faLaptop~desktop, \faTablet~tablet, and \faMobile~smartphone).
    }
    \Description{Three violin/jitter plots of measured execution times in seconds for each \prepams protocol (Register, Participate, and Payout). The y axis of each plot indicates the party that executed the protocol and optionally the type of device measured, with the x axis showing the distribution of execution time.}
    \label{fig:evaluation}
    \vspace{-1em}
\end{figure}

\subsubsectionstar{Results}
The execution times of the register, participate, and payout procedures are depicted in \Cref{fig:evaluation}.
All participant measurements were executed separately using three different types of end-user devices (\ie ~\faLaptop~desktop, \faTablet~tablet, and \faMobile~smartphone).

Across devices the mean execution time of the registration procedure was $71.5 ms$ ($\sigma=28.6 ms$).
The cost of the participation procedures highly depends on the number of prerequisites of the respective study.
For our workload, this resulted in a mean participation time of $624.0 ms$ ($\sigma=529.0 ms$) with a mean proof size of $7.1 kB$ ($\sigma=2.6 kB$).
The execution time of the payout protocol is constant, depending only on the fixed number of maximum inputs, in our case, 10 study rewards.
This resulted in a mean execution time of $8.4 s$ ($\sigma=5.2 s$) with a mean payload size of $18.8 kB$ ($\sigma=4.1 B$).

The average execution times on the service side for registration was $48.1 ms$ ($\sigma=4.25 ms$), rewards issuance took an average of $26.0 ms$ ($\sigma=2.84 ms$), and payout validation was performed in $1.4 s$ ($\sigma=61.3 ms$) on average.

\subsubsectionstar{Discussion}
The results indicate that the registration and most participation procedures are in the range of the typically accepted application response times of $100ms$ to $1s$ \cite{nielsen1994usability}.
Depending on the number of prerequisites some participation proofs may take up to $4.1 s$ to compute.
However, we argue that this is still very acceptable in practice because the time-consuming procedures are limited to less frequent user interactions.
The payout operation shows the highest execution times due to its costly NIZK.
Nevertheless, it is the least frequently issued operation in the system (\eg once in a lifetime for a student account when claiming credits for subject hours), which makes an average of $14.8$ seconds on smartphones a justifiable time.
Additionally, progress bars are displayed during the expensive operations to indicate progress and bridge the waiting time.
The execution time of the participation procedures scales with the amount of referenced (dis-)qualifiers.
However, this is inherently bounded by the lifespan of a study.

The execution time of the service side is generally faster than the client-side.
Participations, as the most commonly invoked operation, take only $26.0 ms$ ($\sigma=2.84 ms$), which would enable $38.46$ requests per second to the service with a single thread.
Depending on the amount of users active in the systems, a multi-core CPU can be leveraged to process multiple participations simultaneously.
In terms of our running example, this should allow a moderately sized university to offer this service to all students with a single server using commodity hardware.

\subsection{Prototype Limitations}
While our \prepams system enables privacy-enhanced participation management with rewards, we are aware of the following technical limitations of our web-based prototype.

Web applications have the benefit of being easily accessible without any prior installation and are supported in a wide variety of devices.
However, they introduce additional attack vectors that local applications are not susceptible to.
The client-side application logic\,---\,including critical code of cryptographic protocols\,---\,is fetched from the platform provider.
Although transport layer security prevents other parties from modifying this code, this layer of protection does not prevent a rogue service provider from including a backdoor in the client-side code.
Theoretically, the client application can be hosted separately from the service, allowing it to be hosted by a somewhat trusted external party.
For example, a public GitHub repository with a continuous deployment on GitHub pages could mitigate some attacks.
WAIT~\cite{Meissner2021} showcases an alternative approach to verify web applications against a public transparency log, similar to certificate transparency~\cite{RFC6962}.
Applying such an approach could mitigate more targeted attacks and enable detection of attacks by the platform provider.

As noted before, a service provider can also use available metadata, such as the client IP address of users, to potentially track and de-anonymize participants.
When a participant is accessing the service via the institution's own network infrastructure, this metadata may already link the participant's identity (\eg WiFi network with RADIUS-based authentication).
This can be addressed using anonymous communication technologies, such as Tor. %
In certain circumstances, trusting a network operator might also be enough, especially when legal regulations like GDPR mandate data protection compliance.
\section{Related Work}
\label{sec:related-work}

ANONIZE~\cite{Hohenberger2014anonize} is an anonymous survey system where a survey organizer can invite a set of participants to a survey based on their identity.
Participation in a survey does not reveal their identity, and only permitted participants are accepted.
\prepams builds on the combination of partially blind signatures and verifiable random functions used in ANONIZE and extends it with support for a privacy-preserving reward procedure and dynamic prerequisites/exclusion criteria dependent on previous participations.

ZebraLancer~\cite{lu2018zebralancer} and zkCrowd~\cite{zhu2020zkcrowd} are anonymous crowd-sens-ing systems utilizing smart contracts on a blockchain to exchange crowd-sourced data with rewards.
These works are motivated by similar requirements as \prepams, \ie to survey data in exchange for rewards.
However, both systems only focus on rewards and do not support prerequisites and exclusion criteria of other surveys.

BLAC~\cite{tsang2007blacklistable}, EPID~\cite{brickell2007enhanced}, PEREA~\cite{tsang2008perea}, and FAUST~\cite{lofgren2011faust} are anonymous credential schemes with blocklists where access to a service can be revoked for misbehaving users while maintaining privacy.
PE(AR)2~\cite{ying2012pear}, BLACR~\cite{au2012blacr}, PERM~\cite{au2012perm}, EXBLACR~\cite{wang2014exblacr}, \cite{Nakanishi2020efficient}, and \cite{chow2023scored} extend the binary exclusion of the previous systems by a reputation-based scoring where only users with score greater than a defined threshold gain access to the service.
DAC~\cite{garman2014decentralized} and DBLACR~\cite{yang2019decentralized} extend this with a decentralized registration.
FARB~\cite{xi2014farb} and Arbra~\cite{xi2014arbra} are centralized systems which allow more complex thresholds based on scores in different categories.
Although these systems may be adapted to support dependencies on previous participations, they do not support the exclusion criteria supported by \prepams.

CLARC~\cite{bemmann2018fully} was built for reviewing services after a verified purchase.
It is based on anonymous credentials, which include signed attributes.
Users can prove to service providers which attributes they hold (or a complex combination), allowing for flexible access control structures.
The user gets a single-show token to rate the service if used correctly.
Any double use to sway the opinion will make the reviews linkable through the equal token.

Untraceable Payments by David Chaum~\cite{Chaum1983} introduced a payment scheme, also based on blind signatures, in which a central bank issues and accepts electronic coins.
Users can receive coins and transfer them to merchants, who can then withdraw received coins from the bank without the bank learning the previous owner's identity.
The bank can check for double-spending during withdrawal by comparing the coin to a list of spent coins.
Electronic cash~\cite{Chaum1990} and follow-up works~\cite{Camenisch2005,Belenkiy2009} introduced offline double-spending protection, where a merchant can accept payments without involving the bank.
The scheme guarantees that this payment will either be honored by the bank or that the identity of the double spender will be revealed.
This does not fit well with our participation protocol because the roles are reversed.
Suppose an organizer double-spends a coin to two participants.
In that case, the later payout operation will reveal the identity of the double-spender (\ie the other participant or the organizer) and hence reveal the link between the participant and the study, which violates our privacy properties.

In summary, we acknowledge that there are multiple previous contributions solving adjacent or generalized problems, but all miss some functionality or trade performance for generality unnecessesary for our requirements. Our PrePaMS construction took inspiration from many of them.
\section{Conclusion}
\label{sec:conclusion}

In this paper, we have introduced \prepams, a privacy-preserving participation management system that also takes rewarding into account.
We derived the requirements and features of such a system based on an analysis of existing systems as well as the involved parties, their motivations, and potential attackers.
It supports analog and digital studies with single participations or follow up studies with pseudonymous participants.
We then specified correctness, security, and anonymity properties for \prepams.
Furthermore, we proposed a concrete instantiation of our \prepams scheme using the partially blind signature scheme by \cite{Hohenberger2014anonize} and an anonymous payment system, which provably provides the security and anonymity properties of our formalization.

We implemented a proof-of-concept prototype\footref{github} of \prepams and deployed a publicly available demo installation\footref{deployment}.
In our technical evaluation, we measured the performance impact of processing overhead and cryptographic operations and showed that the \prepams instantiation provides reasonable performance results.

We consider the current \prepams implementation a core building block for privacy-preserving participation management systems that could be extended and adapted for specific use cases.
The credential management process can be simplified using proposed extensions of the Web Authentication API~\cite{w3c2022webauthn}, utilizing FIDO Passkeys or other already available hardware authentication tokens.
For instance, smart cards are widely available in many relevant scenarios\,---\,\eg student identity cards at universities or employee identity cards in companies\,---\,and could be further incorporated into the system design as a source of key material.

\prepams only protects the participation privacy, but it does not take into account the actual study and the corresponding subject data (\eg survey responses in an online study).
Orthogonal solutions exist that target privacy-enhanced processes for data collection and analysis in empirical research, \eg by extending the methodological practice of pre-registered studies with trusted computing concepts \cite{meissner2021peqes} or by enabling statistical analyses without revealing participant data to researchers using secure multiparty computation \cite{lapets2018}.
Integrating \prepams for participation management with a privacy-enhancing study platform facilitates the development of a comprehensive privacy-preserving study management system that covers the entire process of conducting quantitative empirical research studies while rewarding participants and upholding complete data privacy.

\subsubsectionstar{Outlook}
One drawback of the presented implementation is that participants have to trust the organizer to pay after legitimate participation.
Participants may contact the service provider and use the transcript of the interaction with an organizer to dispute the transaction, which harms their privacy.
\emph{Contingent payments}~\cite{contingent2020}, where an organizer has to commit the transaction before retrieving the full response of a participant, could address this issue.
For example, a two-phase protocol where a participant first commits a chunked encryption of their response, followed by opening a single uniformly random chunk to show the validity of their submission.
If the decryption looks good, they proceed with the financial exchange so that spending the reward reveals the encryption keys to the organizer.
However, this approach only works for online submissions and not necessarily for offline lab experiments.
Nevertheless, here, a participant still has the option to not leave without a signature on the reward transaction.

Overall, \prepams contributes to making the scientific process more privacy friendly while maintaining all features and properties that scientists and study participants are used to in today's practice.
 
\begin{acks}
 This work was partially supported by an internal research grant from Ulm University (\emph{ProTrainU/21-22/PePER}).
 Part of this work by Felix Engelmann was done while at the ITU Copenhagen.
\end{acks}

\bibliographystyle{ACM-Reference-Format}

\begin{thebibliography}{48}

%
%
%
%
%
%
%
%
%
%
%
%
%
%
%
%

\ifx \showCODEN    \undefined \def \showCODEN     #1{\unskip}     \fi
\ifx \showDOI      \undefined \def \showDOI       #1{#1}\fi
\ifx \showISBNx    \undefined \def \showISBNx     #1{\unskip}     \fi
\ifx \showISBNxiii \undefined \def \showISBNxiii  #1{\unskip}     \fi
\ifx \showISSN     \undefined \def \showISSN      #1{\unskip}     \fi
\ifx \showLCCN     \undefined \def \showLCCN      #1{\unskip}     \fi
\ifx \shownote     \undefined \def \shownote      #1{#1}          \fi
\ifx \showarticletitle \undefined \def \showarticletitle #1{#1}   \fi
\ifx \showURL      \undefined \def \showURL       {\relax}        \fi
%
%
\providecommand\bibfield[2]{#2}
\providecommand\bibinfo[2]{#2}
\providecommand\natexlab[1]{#1}
\providecommand\showeprint[2][]{arXiv:#2}

\bibitem[Attema et~al\mbox{.}(2021)]%
        {ACR20}
\bibfield{author}{\bibinfo{person}{Thomas Attema}, \bibinfo{person}{Ronald
  Cramer}, {and} \bibinfo{person}{Matthieu Rambaud}.}
  \bibinfo{year}{2021}\natexlab{}.
\newblock \showarticletitle{Compressed $\varSigma$-Protocols for~Bilinear Group
  Arithmetic Circuits and~Application to~Logarithmic Transparent Threshold
  Signatures}. In \bibinfo{booktitle}{\emph{Advances in Cryptology -- ASIACRYPT
  2021}}, \bibfield{editor}{\bibinfo{person}{Mehdi Tibouchi} {and}
  \bibinfo{person}{Huaxiong Wang}} (Eds.). \bibinfo{publisher}{Springer
  International Publishing}, \bibinfo{address}{Cham},
  \bibinfo{pages}{526--556}.
\newblock
\showISBNx{978-3-030-92068-5}


\bibitem[Au and Kapadia(2012)]%
        {au2012perm}
\bibfield{author}{\bibinfo{person}{Man~Ho Au} {and} \bibinfo{person}{Apu
  Kapadia}.} \bibinfo{year}{2012}\natexlab{}.
\newblock \showarticletitle{PERM: practical reputation-based blacklisting
  without TTPS}. In \bibinfo{booktitle}{\emph{Proceedings of the 2012 ACM
  Conference on Computer and Communications Security}} (Raleigh, North
  Carolina, USA) \emph{(\bibinfo{series}{CCS '12})}.
  \bibinfo{publisher}{Association for Computing Machinery},
  \bibinfo{address}{New York, NY, USA}, \bibinfo{pages}{929–940}.
\newblock
\showISBNx{9781450316514}
\urldef\tempurl%
\url{https://doi.org/10.1145/2382196.2382294}
\showDOI{\tempurl}


\bibitem[Au et~al\mbox{.}(2012)]%
        {au2012blacr}
\bibfield{author}{\bibinfo{person}{Man~Ho Au}, \bibinfo{person}{Apu Kapadia},
  {and} \bibinfo{person}{Willy Susilo}.} \bibinfo{year}{2012}\natexlab{}.
\newblock \showarticletitle{{BLACR:} TTP-Free Blacklistable Anonymous
  Credentials with Reputation}. In \bibinfo{booktitle}{\emph{19th Annual
  Network and Distributed System Security Symposium, {NDSS} 2012, San Diego,
  California, USA, February 5-8, 2012}}. \bibinfo{publisher}{The Internet
  Society}.
\newblock
\urldef\tempurl%
\url{https://www.ndss-symposium.org/ndss2012/blacr-ttp-free-blacklistable-anonymous-credentials-reputation}
\showURL{%
\tempurl}


\bibitem[Barreto et~al\mbox{.}(2003)]%
        {barreto2003constructing}
\bibfield{author}{\bibinfo{person}{Paulo S. L.~M. Barreto},
  \bibinfo{person}{Ben Lynn}, {and} \bibinfo{person}{Michael Scott}.}
  \bibinfo{year}{2003}\natexlab{}.
\newblock \showarticletitle{Constructing Elliptic Curves with Prescribed
  Embedding Degrees}. In \bibinfo{booktitle}{\emph{Security in Communication
  Networks}}, \bibfield{editor}{\bibinfo{person}{Stelvio Cimato},
  \bibinfo{person}{Giuseppe Persiano}, {and} \bibinfo{person}{Clemente Galdi}}
  (Eds.). \bibinfo{publisher}{Springer Berlin Heidelberg},
  \bibinfo{address}{Berlin, Heidelberg}, \bibinfo{pages}{257--267}.
\newblock
\showISBNx{978-3-540-36413-9}


\bibitem[Baruch and Holtom(2008)]%
        {Baruch2008}
\bibfield{author}{\bibinfo{person}{Yehuda Baruch} {and}
  \bibinfo{person}{Brooks~C. Holtom}.} \bibinfo{year}{2008}\natexlab{}.
\newblock \showarticletitle{Survey response rate levels and trends in
  organizational research}.
\newblock \bibinfo{journal}{\emph{{Human Relations}}} \bibinfo{volume}{61},
  \bibinfo{number}{8} (\bibinfo{year}{2008}), \bibinfo{pages}{1139--1160}.
\newblock
\urldef\tempurl%
\url{https://doi.org/10.1177/0018726708094863}
\showDOI{\tempurl}


\bibitem[Belenkiy et~al\mbox{.}(2009)]%
        {Belenkiy2009}
\bibfield{author}{\bibinfo{person}{Mira Belenkiy}, \bibinfo{person}{Melissa
  Chase}, \bibinfo{person}{Markulf Kohlweiss}, {and} \bibinfo{person}{Anna
  Lysyanskaya}.} \bibinfo{year}{2009}\natexlab{}.
\newblock \showarticletitle{Compact E-Cash and Simulatable VRFs Revisited}. In
  \bibinfo{booktitle}{\emph{Pairing-Based Cryptography -- Pairing 2009}},
  \bibfield{editor}{\bibinfo{person}{Hovav Shacham} {and}
  \bibinfo{person}{Brent Waters}} (Eds.). \bibinfo{publisher}{Springer Berlin
  Heidelberg}, \bibinfo{address}{Berlin, Heidelberg},
  \bibinfo{pages}{114--131}.
\newblock
\urldef\tempurl%
\url{https://doi.org/10.1007/978-3-642-03298-1\_9}
\showDOI{\tempurl}


\bibitem[Bemmann et~al\mbox{.}(2018)]%
        {bemmann2018fully}
\bibfield{author}{\bibinfo{person}{Kai Bemmann}, \bibinfo{person}{Johannes
  Bl\"{o}mer}, \bibinfo{person}{Jan Bobolz}, \bibinfo{person}{Henrik
  Br\"{o}cher}, \bibinfo{person}{Denis Diemert}, \bibinfo{person}{Fabian
  Eidens}, \bibinfo{person}{Lukas Eilers}, \bibinfo{person}{Jan Haltermann},
  \bibinfo{person}{Jakob Juhnke}, \bibinfo{person}{Burhan Otour},
  \bibinfo{person}{Laurens Porzenheim}, \bibinfo{person}{Simon Pukrop},
  \bibinfo{person}{Erik Schilling}, \bibinfo{person}{Michael Schlichtig}, {and}
  \bibinfo{person}{Marcel Stienemeier}.} \bibinfo{year}{2018}\natexlab{}.
\newblock \showarticletitle{Fully-Featured Anonymous Credentials with
  Reputation System}. In \bibinfo{booktitle}{\emph{Proceedings of the 13th
  International Conference on Availability, Reliability and Security}}
  (Hamburg, Germany) \emph{(\bibinfo{series}{ARES 2018})}.
  \bibinfo{publisher}{Association for Computing Machinery},
  \bibinfo{address}{New York, NY, USA}, Article \bibinfo{articleno}{42},
  \bibinfo{numpages}{10}~pages.
\newblock
\showISBNx{9781450364485}
\urldef\tempurl%
\url{https://doi.org/10.1145/3230833.3234517}
\showDOI{\tempurl}


\bibitem[Bock et~al\mbox{.}(2014)]%
        {Bock2014hroot}
\bibfield{author}{\bibinfo{person}{Olaf Bock}, \bibinfo{person}{Ingmar Baetge},
  {and} \bibinfo{person}{Andreas Nicklisch}.} \bibinfo{year}{2014}\natexlab{}.
\newblock \showarticletitle{hroot: Hamburg Registration and Organization Online
  Tool}.
\newblock \bibinfo{journal}{\emph{European Economic Review}}
  \bibinfo{volume}{71} (\bibinfo{year}{2014}), \bibinfo{pages}{117--120}.
\newblock
\showISSN{0014-2921}
\urldef\tempurl%
\url{https://doi.org/10.1016/j.euroecorev.2014.07.003}
\showDOI{\tempurl}


\bibitem[Brickell and Li(2007)]%
        {brickell2007enhanced}
\bibfield{author}{\bibinfo{person}{Ernie Brickell} {and}
  \bibinfo{person}{Jiangtao Li}.} \bibinfo{year}{2007}\natexlab{}.
\newblock \showarticletitle{Enhanced privacy id: a direct anonymous attestation
  scheme with enhanced revocation capabilities}. In
  \bibinfo{booktitle}{\emph{Proceedings of the 2007 ACM Workshop on Privacy in
  Electronic Society}} (Alexandria, Virginia, USA) \emph{(\bibinfo{series}{WPES
  '07})}. \bibinfo{publisher}{Association for Computing Machinery},
  \bibinfo{address}{New York, NY, USA}, \bibinfo{pages}{21–30}.
\newblock
\showISBNx{9781595938831}
\urldef\tempurl%
\url{https://doi.org/10.1145/1314333.1314337}
\showDOI{\tempurl}


\bibitem[B{\"{u}}nz et~al\mbox{.}(2018)]%
        {buenz2018}
\bibfield{author}{\bibinfo{person}{Benedikt B{\"{u}}nz},
  \bibinfo{person}{Jonathan Bootle}, \bibinfo{person}{Dan Boneh},
  \bibinfo{person}{Andrew Poelstra}, \bibinfo{person}{Pieter Wuille}, {and}
  \bibinfo{person}{Gregory Maxwell}.} \bibinfo{year}{2018}\natexlab{}.
\newblock \showarticletitle{Bulletproofs: Short Proofs for Confidential
  Transactions and More}. In \bibinfo{booktitle}{\emph{2018 {IEEE} Symposium on
  Security and Privacy, {SP} 2018, Proceedings, 21-23 May 2018, San Francisco,
  California, {USA}}}. \bibinfo{publisher}{{IEEE} Computer Society},
  \bibinfo{address}{Washington, D.C., USA}, \bibinfo{pages}{315--334}.
\newblock
\urldef\tempurl%
\url{https://doi.org/10.1109/SP.2018.00020}
\showDOI{\tempurl}


\bibitem[Camenisch et~al\mbox{.}(2005)]%
        {Camenisch2005}
\bibfield{author}{\bibinfo{person}{Jan Camenisch}, \bibinfo{person}{Susan
  Hohenberger}, {and} \bibinfo{person}{Anna Lysyanskaya}.}
  \bibinfo{year}{2005}\natexlab{}.
\newblock \showarticletitle{Compact E-Cash}. In
  \bibinfo{booktitle}{\emph{Advances in Cryptology -- EUROCRYPT 2005}},
  \bibfield{editor}{\bibinfo{person}{Ronald Cramer}} (Ed.).
  \bibinfo{publisher}{Springer Berlin Heidelberg}, \bibinfo{address}{Berlin,
  Heidelberg}, \bibinfo{pages}{302--321}.
\newblock
\urldef\tempurl%
\url{https://doi.org/10.1007/11426639\_18}
\showDOI{\tempurl}


\bibitem[Chaum(1983)]%
        {Chaum1983}
\bibfield{author}{\bibinfo{person}{David Chaum}.}
  \bibinfo{year}{1983}\natexlab{}.
\newblock \showarticletitle{Blind Signatures for Untraceable Payments}. In
  \bibinfo{booktitle}{\emph{Advances in Cryptology}},
  \bibfield{editor}{\bibinfo{person}{David Chaum}, \bibinfo{person}{Ronald~L.
  Rivest}, {and} \bibinfo{person}{Alan~T. Sherman}} (Eds.).
  \bibinfo{publisher}{Springer US}, \bibinfo{address}{Boston, MA},
  \bibinfo{pages}{199--203}.
\newblock
\urldef\tempurl%
\url{https://doi.org/10.1007/978-1-4757-0602-4\_18}
\showDOI{\tempurl}


\bibitem[Chaum et~al\mbox{.}(1990)]%
        {Chaum1990}
\bibfield{author}{\bibinfo{person}{David Chaum}, \bibinfo{person}{Amos Fiat},
  {and} \bibinfo{person}{Moni Naor}.} \bibinfo{year}{1990}\natexlab{}.
\newblock \showarticletitle{Untraceable Electronic Cash}. In
  \bibinfo{booktitle}{\emph{Advances in Cryptology --- CRYPTO' 88}},
  \bibfield{editor}{\bibinfo{person}{Shafi Goldwasser}} (Ed.).
  \bibinfo{publisher}{Springer New York}, \bibinfo{address}{New York, NY},
  \bibinfo{pages}{319--327}.
\newblock
\urldef\tempurl%
\url{https://doi.org/10.1007/0-387-34799-2\_25}
\showDOI{\tempurl}


\bibitem[Chor et~al\mbox{.}(1998)]%
        {chor1998private}
\bibfield{author}{\bibinfo{person}{Benny Chor}, \bibinfo{person}{Eyal
  Kushilevitz}, \bibinfo{person}{Oded Goldreich}, {and} \bibinfo{person}{Madhu
  Sudan}.} \bibinfo{year}{1998}\natexlab{}.
\newblock \showarticletitle{Private information retrieval}.
\newblock \bibinfo{journal}{\emph{J. ACM}} \bibinfo{volume}{45},
  \bibinfo{number}{6} (\bibinfo{date}{nov} \bibinfo{year}{1998}),
  \bibinfo{pages}{965–981}.
\newblock
\showISSN{0004-5411}
\urldef\tempurl%
\url{https://doi.org/10.1145/293347.293350}
\showDOI{\tempurl}


\bibitem[Chow et~al\mbox{.}(2023)]%
        {chow2023scored}
\bibfield{author}{\bibinfo{person}{Sherman S.~M. Chow}, \bibinfo{person}{Jack
  P.~K. Ma}, {and} \bibinfo{person}{Tsz~Hon Yuen}.}
  \bibinfo{year}{2023}\natexlab{}.
\newblock \showarticletitle{Scored Anonymous Credentials}. In
  \bibinfo{booktitle}{\emph{Applied Cryptography and Network Security}},
  \bibfield{editor}{\bibinfo{person}{Mehdi Tibouchi} {and}
  \bibinfo{person}{XiaoFeng Wang}} (Eds.). \bibinfo{publisher}{Springer Nature
  Switzerland}, \bibinfo{address}{Cham}, \bibinfo{pages}{484--515}.
\newblock
\showISBNx{978-3-031-33491-7}


\bibitem[Dodis and Yampolskiy(2005)]%
        {dodis2005verifiable}
\bibfield{author}{\bibinfo{person}{Yevgeniy Dodis} {and}
  \bibinfo{person}{Aleksandr Yampolskiy}.} \bibinfo{year}{2005}\natexlab{}.
\newblock \showarticletitle{A Verifiable Random Function with Short Proofs and
  Keys}. In \bibinfo{booktitle}{\emph{Public Key Cryptography - PKC 2005}},
  \bibfield{editor}{\bibinfo{person}{Serge Vaudenay}} (Ed.).
  \bibinfo{publisher}{Springer Berlin Heidelberg}, \bibinfo{address}{Berlin,
  Heidelberg}, \bibinfo{pages}{416--431}.
\newblock
\showISBNx{978-3-540-30580-4}
\urldef\tempurl%
\url{https://doi.org/10.1007/978-3-540-30580-4\_28}
\showDOI{\tempurl}


\bibitem[Folkman(2000)]%
        {Folkman2000}
\bibfield{author}{\bibinfo{person}{Susan Folkman}.}
  \bibinfo{year}{2000}\natexlab{}.
\newblock \showarticletitle{Privacy and confidentiality}.
\newblock In \bibinfo{booktitle}{\emph{Ethics in Research With Human
  Participants}}, \bibfield{editor}{\bibinfo{person}{Bruce~D Sales} {and}
  \bibinfo{person}{Susan Folkman}} (Eds.). \bibinfo{publisher}{American
  Psychological Association}, \bibinfo{address}{Washington, DC, USA},
  \bibinfo{pages}{49--57}.
\newblock


\bibitem[Garman et~al\mbox{.}(2014)]%
        {garman2014decentralized}
\bibfield{author}{\bibinfo{person}{Christina Garman}, \bibinfo{person}{Matthew
  Green}, {and} \bibinfo{person}{Ian Miers}.} \bibinfo{year}{2014}\natexlab{}.
\newblock \showarticletitle{Decentralized Anonymous Credentials}. In
  \bibinfo{booktitle}{\emph{21st Annual Network and Distributed System Security
  Symposium, {NDSS} 2014, San Diego, California, USA, February 23-26, 2014}}.
  \bibinfo{publisher}{The Internet Society}.
\newblock
\urldef\tempurl%
\url{https://www.ndss-symposium.org/ndss2014/decentralized-anonymous-credentials}
\showURL{%
\tempurl}


\bibitem[Gollwitzer et~al\mbox{.}(2020)]%
        {gollwitzer2020data}
\bibfield{author}{\bibinfo{person}{Mario Gollwitzer}, \bibinfo{person}{Andrea
  Abele-Brehm}, \bibinfo{person}{Christian Fiebach}, \bibinfo{person}{Roland
  Ramthun}, \bibinfo{person}{Anne~M Scheel}, \bibinfo{person}{Felix~D
  Sch\"{o}nbrodt}, {and} \bibinfo{person}{Ulf Steinberg}.}
  \bibinfo{year}{2020}\natexlab{}.
\newblock \bibinfo{title}{Data Management and Data Sharing in Psychological
  Science: Revision of the DGPs Recommendations}.
\newblock
\newblock
\urldef\tempurl%
\url{https://doi.org/10.31234/osf.io/24ncs}
\showDOI{\tempurl}


\bibitem[Goodwin and Goodwin(2016)]%
        {goodwin2016research}
\bibfield{author}{\bibinfo{person}{C.J. Goodwin} {and} \bibinfo{person}{K.A.
  Goodwin}.} \bibinfo{year}{2016}\natexlab{}.
\newblock \bibinfo{booktitle}{\emph{Research In Psychology Methods and
  Design}}.
\newblock \bibinfo{publisher}{Wiley}, \bibinfo{address}{Hoboken, NJ, USA}.
\newblock
\showISBNx{978-1-119-33044-8}
\showLCCN{2016036350}


\bibitem[Greiner(2004)]%
        {Greiner2004online}
\bibfield{author}{\bibinfo{person}{Ben Greiner}.}
  \bibinfo{year}{2004}\natexlab{}.
\newblock \bibinfo{booktitle}{\emph{The Online Recruitment System ORSEE 2.0 - A
  Guide for the Organization of Experiments in Economics}}.
\newblock \bibinfo{type}{Working Paper Series in Economics}~10.
  \bibinfo{institution}{University of Cologne, Department of Economics}.
\newblock
\urldef\tempurl%
\url{https://EconPapers.repec.org/RePEc:kls:series:0010}
\showURL{%
\tempurl}


\bibitem[Hohenberger et~al\mbox{.}(2014)]%
        {Hohenberger2014anonize}
\bibfield{author}{\bibinfo{person}{Susan Hohenberger}, \bibinfo{person}{Steven
  Myers}, \bibinfo{person}{Rafael Pass}, {and} \bibinfo{person}{abhi shelat}.}
  \bibinfo{year}{2014}\natexlab{}.
\newblock \showarticletitle{ANONIZE: A Large-Scale Anonymous Survey System}. In
  \bibinfo{booktitle}{\emph{2014 IEEE Symposium on Security and Privacy}}.
  \bibinfo{publisher}{IEEE}, \bibinfo{address}{Washington, D.C., USA},
  \bibinfo{pages}{375--389}.
\newblock
\urldef\tempurl%
\url{https://doi.org/10.1109/SP.2014.31}
\showDOI{\tempurl}


\bibitem[Jimenez et~al\mbox{.}(2017)]%
        {jimenez2017popper}
\bibfield{author}{\bibinfo{person}{Ivo Jimenez}, \bibinfo{person}{Michael
  Sevilla}, \bibinfo{person}{Noah Watkins}, \bibinfo{person}{Carlos Maltzahn},
  \bibinfo{person}{Jay Lofstead}, \bibinfo{person}{Kathryn Mohror},
  \bibinfo{person}{Andrea Arpaci-Dusseau}, {and} \bibinfo{person}{Remzi
  Arpaci-Dusseau}.} \bibinfo{year}{2017}\natexlab{}.
\newblock \showarticletitle{{The Popper Convention: Making Reproducible Systems
  Evaluation Practical}}. In \bibinfo{booktitle}{\emph{2017 IEEE International
  Parallel and Distributed Processing Symposium Workshops (IPDPSW)}}. IEEE,
  \bibinfo{publisher}{{IEEE} Computer Society}, \bibinfo{address}{Washington,
  D.C., USA}, \bibinfo{pages}{1561--1570}.
\newblock
\urldef\tempurl%
\url{https://doi.org/10.1109/IPDPSW.2017.157}
\showDOI{\tempurl}


\bibitem[Jones et~al\mbox{.}(2022)]%
        {w3c2022webauthn}
\bibfield{author}{\bibinfo{person}{Michael~B. Jones}, \bibinfo{person}{Akshay
  Kumar}, {and} \bibinfo{person}{Emil Lundberg}.}
  \bibinfo{year}{2022}\natexlab{}.
\newblock \bibinfo{booktitle}{\emph{{Web Authentication: An API for accessing
  Public Key Credential}}}.
\newblock \bibinfo{type}{Editor's Draft}. \bibinfo{institution}{W3C}.
\newblock
\urldef\tempurl%
\url{https://w3c.github.io/webauthn/}
\showURL{%
\tempurl}


\bibitem[Lai et~al\mbox{.}(2019)]%
        {lai19Succint}
\bibfield{author}{\bibinfo{person}{Russell W.~F. Lai}, \bibinfo{person}{Giulio
  Malavolta}, {and} \bibinfo{person}{Viktoria Ronge}.}
  \bibinfo{year}{2019}\natexlab{}.
\newblock \showarticletitle{Succinct Arguments for Bilinear Group Arithmetic:
  Practical Structure-Preserving Cryptography}. In
  \bibinfo{booktitle}{\emph{Proceedings of the 2019 ACM SIGSAC Conference on
  Computer and Communications Security}} (London, United Kingdom)
  \emph{(\bibinfo{series}{CCS '19})}. \bibinfo{publisher}{Association for
  Computing Machinery}, \bibinfo{address}{New York, NY, USA},
  \bibinfo{pages}{2057–2074}.
\newblock
\showISBNx{9781450367479}
\urldef\tempurl%
\url{https://doi.org/10.1145/3319535.3354262}
\showDOI{\tempurl}


\bibitem[Lapets et~al\mbox{.}(2018)]%
        {lapets2018}
\bibfield{author}{\bibinfo{person}{Andrei Lapets}, \bibinfo{person}{Frederick
  Jansen}, \bibinfo{person}{Kinan~Dak Albab}, \bibinfo{person}{Rawane Issa},
  \bibinfo{person}{Lucy Qin}, \bibinfo{person}{Mayank Varia}, {and}
  \bibinfo{person}{Azer Bestavros}.} \bibinfo{year}{2018}\natexlab{}.
\newblock \showarticletitle{Accessible Privacy-Preserving Web-Based Data
  Analysis for Assessing and Addressing Economic Inequalities}. In
  \bibinfo{booktitle}{\emph{Proceedings of the 1st ACM SIGCAS Conference on
  Computing and Sustainable Societies}} (Menlo Park and San Jose, CA, USA)
  \emph{(\bibinfo{series}{COMPASS '18})}. \bibinfo{publisher}{Association for
  Computing Machinery}, \bibinfo{address}{New York, NY, USA}, Article
  \bibinfo{articleno}{48}, \bibinfo{numpages}{5}~pages.
\newblock
\showISBNx{9781450358163}
\urldef\tempurl%
\url{https://doi.org/10.1145/3209811.3212701}
\showDOI{\tempurl}


\bibitem[Laurie et~al\mbox{.}(2013)]%
        {RFC6962}
\bibfield{author}{\bibinfo{person}{B. Laurie}, \bibinfo{person}{A. Langley},
  {and} \bibinfo{person}{E. Kasper}.} \bibinfo{year}{2013}\natexlab{}.
\newblock \bibinfo{booktitle}{\emph{Certificate Transparency}}.
\newblock \bibinfo{type}{RFC} 6962. \bibinfo{institution}{Internet Engineering
  Task Force}.
\newblock
\showISSN{2070-1721}


\bibitem[Lofgren and Hopper(2011)]%
        {lofgren2011faust}
\bibfield{author}{\bibinfo{person}{Peter Lofgren} {and}
  \bibinfo{person}{Nicholas Hopper}.} \bibinfo{year}{2011}\natexlab{}.
\newblock \showarticletitle{FAUST: Efficient, TTP-Free Abuse Prevention by
  Anonymous Whitelisting}. In \bibinfo{booktitle}{\emph{Proceedings of the 10th
  Annual ACM Workshop on Privacy in the Electronic Society}} (Chicago,
  Illinois, USA) \emph{(\bibinfo{series}{WPES '11})}.
  \bibinfo{publisher}{Association for Computing Machinery},
  \bibinfo{address}{New York, NY, USA}, \bibinfo{pages}{125–130}.
\newblock
\showISBNx{9781450310024}
\urldef\tempurl%
\url{https://doi.org/10.1145/2046556.2046572}
\showDOI{\tempurl}


\bibitem[Lu et~al\mbox{.}(2018)]%
        {lu2018zebralancer}
\bibfield{author}{\bibinfo{person}{Yuan Lu}, \bibinfo{person}{Qiang Tang},
  {and} \bibinfo{person}{Guiling Wang}.} \bibinfo{year}{2018}\natexlab{}.
\newblock \showarticletitle{ZebraLancer: Private and Anonymous Crowdsourcing
  System atop Open Blockchain}. In \bibinfo{booktitle}{\emph{2018 IEEE 38th
  International Conference on Distributed Computing Systems (ICDCS)}}.
  \bibinfo{publisher}{{IEEE} Computer Society}, \bibinfo{address}{Washington,
  D.C., USA}, \bibinfo{pages}{853--865}.
\newblock
\urldef\tempurl%
\url{https://doi.org/10.1109/ICDCS.2018.00087}
\showDOI{\tempurl}


\bibitem[Mei\ss{}ner et~al\mbox{.}(2021a)]%
        {meissner2021peqes}
\bibfield{author}{\bibinfo{person}{Echo Mei\ss{}ner}, \bibinfo{person}{Felix
  Engelmann}, \bibinfo{person}{Frank Kargl}, {and} \bibinfo{person}{Benjamin
  Erb}.} \bibinfo{year}{2021}\natexlab{a}.
\newblock \showarticletitle{PeQES: A Platform for Privacy-Enhanced Quantitative
  Empirical Studies}. In \bibinfo{booktitle}{\emph{Proceedings of the 36th
  Annual ACM Symposium on Applied Computing}} (Virtual Event, Republic of
  Korea) \emph{(\bibinfo{series}{SAC '21})}. \bibinfo{publisher}{Association
  for Computing Machinery}, \bibinfo{address}{New York, NY, USA},
  \bibinfo{pages}{1226–1234}.
\newblock
\showISBNx{9781450381048}
\urldef\tempurl%
\url{https://doi.org/10.1145/3412841.3441997}
\showDOI{\tempurl}


\bibitem[Mei\ss{}ner et~al\mbox{.}(2021b)]%
        {Meissner2021}
\bibfield{author}{\bibinfo{person}{Echo Mei\ss{}ner}, \bibinfo{person}{Frank
  Kargl}, {and} \bibinfo{person}{Benjamin Erb}.}
  \bibinfo{year}{2021}\natexlab{b}.
\newblock \showarticletitle{WAIT: Protecting the Integrity of Web Applications
  with Binary-Equivalent Transparency}. In \bibinfo{booktitle}{\emph{The 36th
  ACM/SIGAPP Symposium on Applied Computing}} (Virtual Event, Republic of
  Korea) \emph{(\bibinfo{series}{SAC '21})}. \bibinfo{publisher}{ACM},
  \bibinfo{address}{New York, NY, USA}, \bibinfo{pages}{1950--1953}.
\newblock
\showISBNx{978-1-4503-8104-8/21/03}
\urldef\tempurl%
\url{https://doi.org/10.1145/3412841.3442143}
\showDOI{\tempurl}


\bibitem[Nakanishi and Kanatani(2020)]%
        {Nakanishi2020efficient}
\bibfield{author}{\bibinfo{person}{Toru Nakanishi} {and}
  \bibinfo{person}{Takeshi Kanatani}.} \bibinfo{year}{2020}\natexlab{}.
\newblock \showarticletitle{Efficient blacklistable anonymous credential system
  with reputation using a pairing-based accumulator}.
\newblock \bibinfo{journal}{\emph{IET Information Security}}
  \bibinfo{volume}{14}, \bibinfo{number}{6} (\bibinfo{year}{2020}),
  \bibinfo{pages}{613--624}.
\newblock
\urldef\tempurl%
\url{https://doi.org/10.1049/iet-ifs.2018.5505}
\showDOI{\tempurl}
\showeprint{https://ietresearch.onlinelibrary.wiley.com/doi/pdf/10.1049/iet-ifs.2018.5505}


\bibitem[Nguyen et~al\mbox{.}(2020)]%
        {contingent2020}
\bibfield{author}{\bibinfo{person}{Ky Nguyen}, \bibinfo{person}{Miguel
  Ambrona}, {and} \bibinfo{person}{Masayuki Abe}.}
  \bibinfo{year}{2020}\natexlab{}.
\newblock \showarticletitle{WI is Almost Enough: Contingent Payment All Over
  Again}. In \bibinfo{booktitle}{\emph{Proceedings of the 2020 ACM SIGSAC
  Conference on Computer and Communications Security}} (Virtual Event, USA)
  \emph{(\bibinfo{series}{CCS '20})}. \bibinfo{publisher}{Association for
  Computing Machinery}, \bibinfo{address}{New York, NY, USA},
  \bibinfo{pages}{641–656}.
\newblock
\showISBNx{9781450370899}
\urldef\tempurl%
\url{https://doi.org/10.1145/3372297.3417888}
\showDOI{\tempurl}


\bibitem[Nielsen(1994)]%
        {nielsen1994usability}
\bibfield{author}{\bibinfo{person}{Jakob Nielsen}.}
  \bibinfo{year}{1994}\natexlab{}.
\newblock \bibinfo{booktitle}{\emph{Usability Engineering}}.
\newblock \bibinfo{publisher}{Morgan Kaufmann Publishers Inc.},
  \bibinfo{address}{San Francisco, CA, USA}.
\newblock
\showISBNx{9780080520292}


\bibitem[Rose et~al\mbox{.}(2007)]%
        {Dale2007}
\bibfield{author}{\bibinfo{person}{Dale~S. Rose}, \bibinfo{person}{Stuart~D.
  Sidle}, {and} \bibinfo{person}{Kristin~H. Griffith}.}
  \bibinfo{year}{2007}\natexlab{}.
\newblock \showarticletitle{A Penny for Your Thoughts: Monetary Incentives
  Improve Response Rates for Company-Sponsored Employee Surveys}.
\newblock \bibinfo{journal}{\emph{{Organizational Research Methods}}}
  \bibinfo{volume}{10}, \bibinfo{number}{2} (\bibinfo{year}{2007}),
  \bibinfo{pages}{225--240}.
\newblock
\urldef\tempurl%
\url{https://doi.org/10.1177/1094428106294687}
\showDOI{\tempurl}


\bibitem[Sieber and Saks(1989)]%
        {Sieber1989}
\bibfield{author}{\bibinfo{person}{Joan~E. Sieber} {and}
  \bibinfo{person}{Michael~J. Saks}.} \bibinfo{year}{1989}\natexlab{}.
\newblock \showarticletitle{A census of subject pool characteristics and
  policies.}
\newblock \bibinfo{journal}{\emph{{American Psychologist}}}
  \bibinfo{volume}{44}, \bibinfo{number}{7} (\bibinfo{year}{1989}),
  \bibinfo{pages}{1053--1061}.
\newblock
\urldef\tempurl%
\url{https://doi.org/10.1037/0003-066X.44.7.1053}
\showDOI{\tempurl}


\bibitem[Singer(2018)]%
        {Singer2018}
\bibfield{author}{\bibinfo{person}{Eleanor Singer}.}
  \bibinfo{year}{2018}\natexlab{}.
\newblock \showarticletitle{The Use and Effects of Incentives in Surveys}.
\newblock In \bibinfo{booktitle}{\emph{The Palgrave Handbook of Survey
  Research}}. \bibinfo{publisher}{Springer}, \bibinfo{address}{Berlin,
  Heidelberg}, \bibinfo{pages}{63--70}.
\newblock
\showISBNx{978-3-319-54395-6}
\urldef\tempurl%
\url{https://doi.org/10.1007/978-3-319-54395-6\_9}
\showDOI{\tempurl}


\bibitem[Sona~Systems(2024)]%
        {sonasystems}
\bibfield{author}{\bibinfo{person}{Ltd. Sona~Systems}.}
  \bibinfo{year}{2002--2024}\natexlab{}.
\newblock \bibinfo{title}{{Sona Systems: Cloud-based Participant Management
  Software}}.
\newblock \bibinfo{howpublished}{\url{https://www.sona-systems.com/}}.
\newblock


\bibitem[Tsang et~al\mbox{.}(2007)]%
        {tsang2007blacklistable}
\bibfield{author}{\bibinfo{person}{Patrick~P. Tsang}, \bibinfo{person}{Man~Ho
  Au}, \bibinfo{person}{Apu Kapadia}, {and} \bibinfo{person}{Sean~W. Smith}.}
  \bibinfo{year}{2007}\natexlab{}.
\newblock \showarticletitle{Blacklistable anonymous credentials: blocking
  misbehaving users without ttps}. In \bibinfo{booktitle}{\emph{Proceedings of
  the 14th ACM Conference on Computer and Communications Security}}
  (Alexandria, Virginia, USA) \emph{(\bibinfo{series}{CCS '07})}.
  \bibinfo{publisher}{Association for Computing Machinery},
  \bibinfo{address}{New York, NY, USA}, \bibinfo{pages}{72–81}.
\newblock
\showISBNx{9781595937032}
\urldef\tempurl%
\url{https://doi.org/10.1145/1315245.1315256}
\showDOI{\tempurl}


\bibitem[Tsang et~al\mbox{.}(2008)]%
        {tsang2008perea}
\bibfield{author}{\bibinfo{person}{Patrick~P. Tsang}, \bibinfo{person}{Man~Ho
  Au}, \bibinfo{person}{Apu Kapadia}, {and} \bibinfo{person}{Sean~W. Smith}.}
  \bibinfo{year}{2008}\natexlab{}.
\newblock \showarticletitle{PEREA: towards practical TTP-free revocation in
  anonymous authentication}. In \bibinfo{booktitle}{\emph{Proceedings of the
  15th ACM Conference on Computer and Communications Security}} (Alexandria,
  Virginia, USA) \emph{(\bibinfo{series}{CCS '08})}.
  \bibinfo{publisher}{Association for Computing Machinery},
  \bibinfo{address}{New York, NY, USA}, \bibinfo{pages}{333–344}.
\newblock
\showISBNx{9781595938107}
\urldef\tempurl%
\url{https://doi.org/10.1145/1455770.1455813}
\showDOI{\tempurl}


\bibitem[Wang et~al\mbox{.}(2014)]%
        {wang2014exblacr}
\bibfield{author}{\bibinfo{person}{Weijin Wang}, \bibinfo{person}{Dengguo
  Feng}, \bibinfo{person}{Yu Qin}, \bibinfo{person}{Jianxiong Shao},
  \bibinfo{person}{Li Xi}, {and} \bibinfo{person}{Xiaobo Chu}.}
  \bibinfo{year}{2014}\natexlab{}.
\newblock \showarticletitle{ExBLACR: Extending BLACR System}. In
  \bibinfo{booktitle}{\emph{Information Security and Privacy}},
  \bibfield{editor}{\bibinfo{person}{Willy Susilo} {and}
  \bibinfo{person}{Yi~Mu}} (Eds.). \bibinfo{publisher}{Springer International
  Publishing}, \bibinfo{address}{Cham}, \bibinfo{pages}{397--412}.
\newblock
\showISBNx{978-3-319-08344-5}


\bibitem[Wang et~al\mbox{.}(2023)]%
        {wang2023}
\bibfield{author}{\bibinfo{person}{Xiaofeng Wang}, \bibinfo{person}{Peng
  Zheng}, {and} \bibinfo{person}{Qianqian Xing}.}
  \bibinfo{year}{2023}\natexlab{}.
\newblock \showarticletitle{Security Analysis of Pairing-based Cryptography}.
\newblock \bibinfo{journal}{\emph{CoRR}}  \bibinfo{volume}{abs/2309.04693}
  (\bibinfo{year}{2023}), \bibinfo{numpages}{23}~pages.
\newblock
\urldef\tempurl%
\url{https://doi.org/10.48550/ARXIV.2309.04693}
\showDOI{\tempurl}
\showeprint[arXiv]{2309.04693}


\bibitem[Watson(2017)]%
        {Watson17WCA}
\bibfield{author}{\bibinfo{person}{Mark Watson}.}
  \bibinfo{year}{2017}\natexlab{}.
\newblock \bibinfo{booktitle}{\emph{Web Cryptography {API}}}.
\newblock \bibinfo{type}{{W3C} Recommendation}. \bibinfo{institution}{W3C}.
\newblock
\newblock
\shownote{https://www.w3.org/TR/2017/REC-WebCryptoAPI-20170126/}.


\bibitem[Xi and Feng(2014)]%
        {xi2014farb}
\bibfield{author}{\bibinfo{person}{Li Xi} {and} \bibinfo{person}{Dengguo
  Feng}.} \bibinfo{year}{2014}\natexlab{}.
\newblock \showarticletitle{FARB: Fast Anonymous Reputation-Based Blacklisting
  without TTPs}. In \bibinfo{booktitle}{\emph{Proceedings of the 13th Workshop
  on Privacy in the Electronic Society}} (Scottsdale, Arizona, USA)
  \emph{(\bibinfo{series}{WPES '14})}. \bibinfo{publisher}{ACM},
  \bibinfo{address}{New York, NY, USA}, \bibinfo{pages}{139–148}.
\newblock
\showISBNx{9781450331487}
\urldef\tempurl%
\url{https://doi.org/10.1145/2665943.2665947}
\showDOI{\tempurl}


\bibitem[Xi et~al\mbox{.}(2014)]%
        {xi2014arbra}
\bibfield{author}{\bibinfo{person}{Li Xi}, \bibinfo{person}{Jianxiong Shao},
  \bibinfo{person}{Kang Yang}, {and} \bibinfo{person}{Dengguo Feng}.}
  \bibinfo{year}{2014}\natexlab{}.
\newblock \showarticletitle{ARBRA: Anonymous Reputation-Based Revocation with
  Efficient Authentication}. In \bibinfo{booktitle}{\emph{Information
  Security}}, \bibfield{editor}{\bibinfo{person}{Sherman S.~M. Chow},
  \bibinfo{person}{Jan Camenisch}, \bibinfo{person}{Lucas C.~K. Hui}, {and}
  \bibinfo{person}{Siu~Ming Yiu}} (Eds.). \bibinfo{publisher}{Springer
  International Publishing}, \bibinfo{address}{Cham}, \bibinfo{pages}{33--53}.
\newblock
\showISBNx{978-3-319-13257-0}


\bibitem[Yang et~al\mbox{.}(2019)]%
        {yang2019decentralized}
\bibfield{author}{\bibinfo{person}{Rupeng Yang}, \bibinfo{person}{Man~Ho Au},
  \bibinfo{person}{Qiuliang Xu}, {and} \bibinfo{person}{Zuoxia Yu}.}
  \bibinfo{year}{2019}\natexlab{}.
\newblock \showarticletitle{Decentralized blacklistable anonymous credentials
  with reputation}.
\newblock \bibinfo{journal}{\emph{Comput. Secur.}}  \bibinfo{volume}{85}
  (\bibinfo{year}{2019}), \bibinfo{pages}{353--371}.
\newblock
\urldef\tempurl%
\url{https://doi.org/10.1016/J.COSE.2019.05.009}
\showDOI{\tempurl}


\bibitem[Yu et~al\mbox{.}(2012)]%
        {ying2012pear}
\bibfield{author}{\bibinfo{person}{Kin~Ying Yu}, \bibinfo{person}{Tsz~Hon
  Yuen}, \bibinfo{person}{Sherman S.~M. Chow}, \bibinfo{person}{Siu~Ming Yiu},
  {and} \bibinfo{person}{Lucas C.~K. Hui}.} \bibinfo{year}{2012}\natexlab{}.
\newblock \showarticletitle{PE(AR)2: Privacy-Enhanced Anonymous Authentication
  with Reputation and Revocation}. In \bibinfo{booktitle}{\emph{Computer
  Security -- ESORICS 2012}}, \bibfield{editor}{\bibinfo{person}{Sara Foresti},
  \bibinfo{person}{Moti Yung}, {and} \bibinfo{person}{Fabio Martinelli}}
  (Eds.). \bibinfo{publisher}{Springer Berlin Heidelberg},
  \bibinfo{address}{Berlin, Heidelberg}, \bibinfo{pages}{679--696}.
\newblock
\urldef\tempurl%
\url{https://doi.org/10.1007/978-3-642-33167-1\_39}
\showDOI{\tempurl}


\bibitem[Zhu et~al\mbox{.}(2020)]%
        {zhu2020zkcrowd}
\bibfield{author}{\bibinfo{person}{Saide Zhu}, \bibinfo{person}{Zhipeng Cai},
  \bibinfo{person}{Huafu Hu}, \bibinfo{person}{Yingshu Li}, {and}
  \bibinfo{person}{Wei Li}.} \bibinfo{year}{2020}\natexlab{}.
\newblock \showarticletitle{zkCrowd: A Hybrid Blockchain-Based Crowdsourcing
  Platform}.
\newblock \bibinfo{journal}{\emph{IEEE Transactions on Industrial Informatics}}
  \bibinfo{volume}{16}, \bibinfo{number}{6} (\bibinfo{year}{2020}),
  \bibinfo{pages}{4196--4205}.
\newblock
\urldef\tempurl%
\url{https://doi.org/10.1109/TII.2019.2941735}
\showDOI{\tempurl}


\end{thebibliography}

\appendix

\section{Notation}
In the following, we present a concrete group-level instantiation of our scheme components.
Let $\G$ and $\G_2$ be paring-friendly cyclic groups of prime order $q$ with generators $g$, $g_2$ where the discrete log assumption holds.
Let $\G_T$ be a cyclic group of prime order $q$ with an efficient bilinear non-degenerate mapping $e: \G \times \G_2 \to \G_T$, with generator $g_T := e(g, g_2)$.
For our construction $\Setup$ produces a description of a Barreto, Lynn, Scott curve\cite{barreto2003constructing}.
We follow the common multiplicative notation for group operations (\eg $x,y \in \G: x \cdot y$ and $a\in\Zq,x \in \G: x^a$).
Let $\langle{}\vec a, \vec b\rangle{} = \sum_{i=1}^{n} a_i \cdot b_i$ denote the inner product of two vectors $a,b \in \Zq^n$.
For a vector of group elements $\vec x$ and a vector of scalars $\vec x$ we denote the element wise exponentiation as $\vec{a}^{\circ \vec x} = (a_1^{x_1}, a_2^{x_2}, \dots, a_n^{x_n})$.
The sum of all elements of a vector is abbreviated as $\sum \vec{a} = \sum_{a\in\vec{a}} a$, analogously the product of all elements of a vector is denoted as $\prod\vec{a}$.

There exists a cryptographic hash function $\id:\mathbb{T} \to \Zq$ that maps tasks to a unqiue scalar.
We also refer to the output of this function as the identifier of a study.
Further, there exist cryptographic hash functions $H_1: \mathsf{X} \to \ZZ_q$, $H_2: \mathsf{X} \to \rspace{\PBS}$ that map arbitraty length inputs $\mathsf{X}$ to scalars in the respective domains.

\Cref{fig:variables} provides an overview of variables used in our formalization and construction.

\begin{table}
\caption{Variable names and usage.}
\label{fig:variables}
\begin{tabular}{l|l}
$\username$& participant identifier\\
$\pk_\Service$ & service public key\\
$\attr$ & attributes of a participant, e.g. age\\
$\cred$ & anonymous credential\\
$\tx$ & reward transaction \\
$\BB$ & bulletin board, list of $\tx$\\
$\Task$ & task from domain $\mathbb{T}$ \\
$\nul$ & reward nullifier\\
$\NUL$ & set of nullifiers\\
$\bar\delta$ & set of qualifiers\\
$\delta$ & set of disqualifiers\\
$\mathfrak{U}$& set of users managed by the oracles\\
$\mathfrak{T}$& bookkeeping of amount held by adversary\\
$\auxil$& channel for interactive adversary\\
$\UN$ & set of service registered usernames\\
\end{tabular}
\end{table}
\section{Model Validation}
\label{appx:validation}

To show that our formalization captures the rewarding and privacy properties of existing systems, we show that log sheets and stickers are partial instantiations of our model. These two decentralized systems are the only interesting, as the ones with a central database simply rely on a trusted server.

The log sheet system relies on unforgeable and unclonable (manual) signatures and uses the following construction:

\begin{description}[leftmargin=0.5em,nosep]
  \item[$\Setup(\secparam,m,n)\to\pp$:]
    Create a sheet design with space for $m$ attributes and $n$ lines for signatures. This is $\pp$
  \item[$\KeyGen_\Service() \to (\sk_\Service,\pk_\Service)$:]
    Create an unforgeable signature ($\sk_\Service$) and publish how it looks like ($\pk_\Service$).
  \item[$\Pi_\Register \langle \Participant(\username,\attr),\Service(\sk_\Service)\rangle\to (\cred,\username)$:]
    The participant identifies themselves ($\username$) and receive a log sheet ($\cred$) from the service with with their attributes $\attr$ filled in and signed with $\sk_\Service$. The service keeps track of the identity.
  \item[$\Pi_\Participate \langle \Participant(\cred, \BB, \Task), \Service(\sk_\Service)\rangle\to \rtx$:]
    The participant shows the log sheet $\cred$ to the organizer who uses the previous signatures on $\cred$ to check the qualifications and upon successful participation of $\Task$ signs the sheet too ($\rtx$ is a line on the sheet).
  \item[$\Pi_\Payout \langle \Participant(\cred,\username, \BB, v,\{\tx_i\}_{i=1}^k), \Service()\rangle\to (\{\NUL_i\}_{i=1}^n,v,\username)$:~~]
    The\linebreak{} participant hands the log sheet $\cred$ over to the service, who keeps it and pays out the amount signed by each organizer.
  \item[$\Receive(\cred,\TX,\PTX)\to (s,R)$:] Add all plaintext rewards with signatures on the log sheet.
  \item[$\CheckCred(\cred, \pk_\Service)\to\bit$:] The sheet is authentic and has attributes signed with $\pk_\Service$.
  \item[$\CheckParticipation(\cred, \rtx)\to\bit$:] Check if the log sheet $\cred$ has the signature for $\rtx$ on it.
  \item[$\CheckQualification(\cred, \TX, \Task)\to\bit$:] Check for previous participations based on signatures on the log sheet.
\end{description}

The log sheet only fulfills the $\mathsf{PartSec}$ and $\mathsf{Balance}$ property, but obviously not the $\mathsf{PartPriv}$.
With unforgeable signatures, $\mathsf{PartSec}$ holds, because the organizers check the qualifications in plaintext and the participation for a given study requires a signature from the organizer. Similarly, to break the balance property, signatures need to be transferred between sheets, which physically cannot be done without cloning.

Complementary, the sticker based solution is more privacy friendly. It assumes unclonable, indistinguishable stickers and fits to our model as follows with an empty $\Setup$:

\begin{description}[leftmargin=0.5em,nosep]
  \item[$\KeyGen_\Service() \to (\sk_\Service,\pk_\Service)$:]
    Create an unforgeable sticker design ($\sk_\Service$) and publish how it looks like ($\pk_\Service$).
  \item[$\Pi_\Register \langle \Participant(\username,\attr),\Service(\sk_\Service)\rangle\to (\cred,\username)$:]
    The participant identifies themselves ($\username$) and receive an empty booklet ($\cred$) from the service.
  \item[$\Pi_\Participate \langle \Participant(\cred, \BB, \Task), \Service(\sk_\Service)\rangle\to \rtx$:]
    Upon successful participation of $\Task$, the participant gets a sticker ($\rtx$).
  \item[$\Pi_\Payout \langle \Participant(\cred,\username, \BB, v,\{\tx_i\}_{i=1}^k), \Service()\rangle\to (\{\NUL_i\}_{i=1}^n,v,\username)$:~~]
    The\linebreak{} participant hands the booklet with a subset of their stickers over to the service, who keeps it and pays out the number of stickers.
  \item[$\Receive(\cred,\TX,\PTX)\to (s,R)$:] Count the stickers in $\cred$.
\end{description}
$\CheckCred, \CheckParticipation$ and $\CheckQualification$ are not possible.

Given indistinguishable stickers, the system has $\mathsf{PartPriv}$. In $\Pi_\Participate$, no information is dependent on $b$. With unclonable stickers, $\mathsf{Balance}$ is only satisfied up to the first winning condition. The adversary is unable to copy any sticker and get payed out more than they earned. The second winning condition is breakable as the adversary can move stickers between users.

\section{Deferred Preliminaries}
\label{sec:defprelim}

Here we present definitions of the preliminaries used in our construction.

\subsection{Key Derivation Function}
We use a key derivation function $\KDF:\skspace{\VRF} \times \mathbb{T} \to \mathbb{S} \times \rspace{\PBS}$ which takes a secret key from $\skspace{\VRF}$ and a task from $\mathbb{T}$ to generate a nullifier $\NUL$ and a $\PBS$ blinding randomness.
It consists of two algorithms:

\begin{description}[leftmargin=1.5em]
    \item[$\Setup(\secparam)\to\pp$:] takes the security parameter $\secparam$ and outputs public parameters $\pp$.
    \item[$\KDF(s, \Task)\to(\NUL, \rho)$:] takes a secret $s\in\skspace{\VRF}$ together with a task $\Task$ and generates a nullifier $\NUL$  and $\rho$, a $\PBS$ blinding randomness.
\end{description}

As concrete instantiation, we use cryptographic hash functions $H_1$ and $H_2$, set $\mathbb{S}=\ZZ_q$, and calculate $\NUL:=H_1(s\|\id(\Task))$,  $\rho:=H_2(s\|\id(\Task))$, and output $\KDF(s,\Task):=(\NUL, \rho)$.

\subsection{VRF scheme}
A $\VRF$ enables generating unique, deterministic tags for a public key and a label.
It consists of the following three algorithms:
\begin{description}[topsep=2pt,leftmargin=1.5em]
    \item[$\Setup(\secparam)\to\pp$:] takes the security parameter $\secparam$ and outputs public parameters $\pp$.
    \item[$\KeyGen()\to(\sk,\pk)$:] generates a public private key pair $\sk,\pk$ in the domain $\skspace{\VRF}\times\pkspace{\VRF}$ with a helper function $\Pub:\skspace{\VRF}\to\pkspace{\VRF}$ to calculate the corresponding public key to a secret key.
    \item[$\Eval(\sk, x)\to\Tag$:] takes a secret key $\sk$ and a label $x\in\mathbb{T}$ and outputs a tag $\Tag$ from the domain $\chi_\VRF$.
\end{description}

The VRF scheme must satisfy the following properties:
\begin{description}[topsep=2pt,leftmargin=1.5em]
 \item[Correctness:] A honestly generated NIZK with $\mathcal{L}_\VRF$ for a tag $\Tag$ and witness $\sk$ generated by $\Eval$ is valid.
 \item[Tag Pseudorandomness:] A tag $\Tag$ generated by $\Eval$ must only be predictable for the party knowing the secret key $\sk$ and looks uniformly random to everyone else.
 \item[Tag Uniqueness:] For a given secret key $\sk$ and label $x$, there must exist one unique, valid tag $\Tag$
\end{description}

The detailed definitions of these properties together with a construction is presented in \cite{dodis2005verifiable} which is summarized as:
\begin{description}[topsep=2pt,leftmargin=1.5em]
    \item[$\Setup(\secparam)$:] returns $(g,\G,q)$.
    \item[$\KeyGen()\to(\sk,\pk)$:] chooses $\sk \in \Zq$, outputs $\sk,\pk := g^\sk$.
    \item[$\Eval(\sk, x)\to \Tag$:] computes and outputs $\Tag := g^{\frac{1}{\sk + \id(x)}}$.
\end{description}

In addition, we require an efficient NIZK for the correct evaluation.
\Ie the relation $\mathcal{L}_{\VRF}((\pk,\Tag,x) \exists (\sk) \ST \pk=\Pub(\sk) \land \Tag=\Eval(\sk,x))$.
A concrete instantiation of $\mathcal{L}_{\VRF}$ is adapted from \cite{Hohenberger2014anonize}.
The prover first picks a random blinding $b \drawrandom \Zq$ and commits on $E \gets e(g, \Tag)^b$.
The commitment is sent to the verifier, who then replies with a challenge $c \drawrandom \Zq$.
The response is computed as $z \gets b + c \cdot \sk$ and sent to the verifier.
Then the verifier verifies the proof by checking: $ E \cdot e(g,g_2)^c \cdot e(g, \Tag) ^ {-cx} = e(g_2, \Tag)^z$

\subsection{PBS scheme}

To issue anonymous credentials with signer defined attributes and use them in a multi-show unlinkable way, we need a partially blind signature scheme.
This hybrid of a regular digital signature and a blind signature, splits the message in a part that is visible to the signer and one that is hidden from the signer.
We mostly follow the ANONIZE~\cite{Hohenberger2014anonize} definition using six probabilistic polynomial time algorithms, but generalize the signature to allow more than one hidden message, and more than one visible message:

$\PBS = (\Setup, \KeyGen, \Blind, \Sign, \Verify, \Unblind)$: 
\begin{description}[topsep=2pt,leftmargin=1.5em]
    \item[$\Setup(\secparam, m,n)\to\pp$:] takes a security parameter $\secparam$, the number of hidden messages $n$, the number of visible messages $m$, and returns public parameters $\pp$.
    \item[$\KeyGen()\to(\sk,\pk)$:] generates a public private key pair to sign credentials $(\sk,\pk) \in \skspace{\PBS}\times\pkspace{\PBS}$.
    \item[$\Blind(S,\rho)\to\alpha$:] takes hidden message parts $S \in\rspace{\PBS}^n$ and a blinding factor $\rho\in\rspace{\PBS}$, and returns a blinded message $\alpha$
    \item[$\Sign(\sk, \alpha, M)\to\sigma'$:] takes a secret key $\sk$, a blinded message $\alpha$, and the public message parts $M\in\rspace{\PBS}^m$ and outputs a signature $\sigma'$.
    \item[$\Unblind(\rho, \sigma')\to\sigma$:] takes the blinding factor $\rho$ and the blinded signature $\sigma'$ and outputs the unblinded version $\sigma$.
    \item[$\Verify(\pk, S, M, \sigma)\to\bit$:]  takes the public key $\pk$, the secret and public message parts $S$, $M$, and a signature $\sigma$ on it, and outputs 1 if the signature is valid and 0 otherwise.
\end{description}

A PBS scheme satisfies the following three properties:
\begin{description}[topsep=2pt,leftmargin=1.5em]
 \item[Correctness:] Every honestly blinded, signed, and unblinded signature verifies.
 \item[Partial Blindness:] For multiple signatures that use the same public message part, a signer cannot link these signatures to the respective signing sessions.
 \item[Unforgeability:] An adversary that interacts at most $l$ times with the signer cannot produce more than $l$ valid message-signature pairs.
\end{description}

We utilize the efficient pairing-based partially blind signature scheme by \citeauthor{Hohenberger2014anonize}~\cite{Hohenberger2014anonize}.
We generalized their scheme, to allow for a fixed set of public and private messages.

\begin{description}[topsep=2pt,leftmargin=1.5em]
    \item[$\Setup(\secparam, m, n)\to\pp$:] sample random group generators $\vec{U} \in \G^m$, $\vec{V} \in \G^n$, $h, g \in \G$, $g_2 \in \G_2$, outputs $(\vec{U}, \vec{V}, h, g, g_2)$ and sets the randomness space $\rspace{\PBS}:=\Zq$.
    \item[$\KeyGen()$:] choose a secret $\sk \in \Zq$, compute $\pk \gets e(g, g_2)^{\sk}$, and output $(\sk, \pk)$.
    \item[$\Blind(\vec{S}, \rho)$:] compute $\alpha \gets \textstyle\prod\vec{V}^{\circ\vec{S}} \cdot g^\rho$ and output $\alpha$.
    \item[$\Sign(\sk, \alpha, M)$:] choose random $w \in \Zq$, compute $\sigma_1 \gets g^\sk (\alpha \cdot h \cdot \textstyle\prod\vec{U}^{\circ\vec{M}})^w$, $\sigma_2 \gets g^w$, $\sigma_3 \gets g_2^w$, and output $(\sigma_1, \sigma_2, \sigma_3)$.
    \item[$\Unblind(\rho, (\sigma_1, \sigma_2, \sigma_3))$:] compute and output $\left(\sigma_1 \cdot \sigma_2^{-\rho}, \sigma_3\right)$.
    \item[$\Verify(\pk, \vec{S}, \vec{M}, (\sigma_1, \sigma_2))$:] check if $\pk \cdot e(\prod\vec{V}^{\circ\vec{S}} \cdot \prod{\vec{U}^{\circ\vec{M}}} \cdot h, \sigma_2) = e(\sigma_1, g_2)$ and output 1, 0 otherwise.
\end{description}

Proofs for correctness, partial blindness, and unforgeability in the case $|\vec{V}| = |\vec{M}| = 1$ are described in \cite[(Theorem 11, Thereom 10, Theorem 12)]{Hohenberger2014anonize}.
Assuming the hardness of DLP in $\G$, an instantiation with $|\vec{V}| = 2$ and $|\vec{M}| = 2$ is computationally indistinguishable from an instantiation with $|\vec{V}| = |\vec{M}| = 1$ if no discrete logarithm relation is known between generators $\vec{V}_1, \vec{V}_2, \vec{M}_1, \vec{M}_2$ (\cf hiding property of Pedersen Commitment).
Per induction this holds for $|\vec{V}| > 2$ and $|\vec{M}| > 2$.

In \cite{Hohenberger2014anonize}, the authors also describe an efficient Schnorr-based NIZK $\mathcal{L}_{\PBS}$ for a correct blinding, and a language $\mathcal{L}_{\mathsf{VerPBS}}$ for verifying the signature without revealing any part of the message.
We adapated these again for a fixed set of public message parts $\vec{M}$ and private messsage parts $\vec{S}$:

\begin{flalign*}
    & \mathcal{L}_{\PBS}:= \left\{
        \begin{array}{l}
            (\alpha) \mid \exists (\vec{S}, \rho)\ST \alpha = \textstyle\prod\vec{V}^{\circ\vec{S}} \cdot g^\rho \\
        \end{array}
    \right\} &&
\end{flalign*}

The prover first picks blinding factors $\vec{B} \drawrandom \Zq^n$, $b \drawrandom \Zq$ and commits to $\gamma \gets \vec{V}^{\circ\vec{B}} \cdot g^{b}$.
The verifier receives $\alpha, \gamma$ and responds with a challenge $c \drawrandom \Zq$.
The response is computed as:
\[
    \vec{Y} \gets \left[\vec{B}_i + c \cdot \vec{S}_i \right]_{i=1}^n\hspace{2em}
    z \gets b+ c \cdot \rho
\]
The verifier then verifies $\textstyle\prod\vec{V}^{\circ\vec{Y}} \cdot g^{z} = \alpha^c \cdot \gamma$.

For the anonymous verification of a signature, we use

\begin{flalign*}
    & \mathcal{L}_{\mathsf{VerPBS}}:= \left\{~(\pk) \mid \exists (\vec{S}, \vec{M}, \sigma)\ST \PBS.\Verify(\pk, \vec{S}, \vec{M}, \sigma) = 1~\right\} &&
\end{flalign*}

The prover first re-randomizes their signature $\sigma := (\sigma_1, \sigma_2)$ by picking a new random $d \drawrandom \Zq$ and computing:
\[
    s_1 \gets \sigma_1 \cdot \left(\prod\vec{V}^{\circ\vec{S}} \prod \vec{U}^{\circ{\vec{M}}} \cdot h\right)^d \hspace{2em}
    s_2 \gets \sigma_2 \cdot g_2^d
\]

They then picks random values $\vec{B} \drawrandom \Zq^m$ and a random group element $J \drawrandom \G$ and compute a commitment as:
\[
    E \gets e(J, g_2) \cdot e(\textstyle\prod\vec{U}^{\circ\vec{B}}, s_2)^{-1} \hspace{2em}
\]

The verifier receives $(s_2, E)$ and responds with a challenge $c \drawrandom \Zq$.
The prover computes a response:
\[
    \vec{Y} \gets \left[\vec{B}_i + c \cdot \vec{M}_i \right]_{i=1}^m\hspace{2em}
    z \gets {s_1}^c \cdot J
\]

This can be verified by the verifier by checking:
\[
    E \cdot \pk^c \cdot e(h, s_2)^c = e(z, g_2) \cdot e(\textstyle\prod\vec{U}^{\circ\vec{Y}}, s_2)^{-1}
\]

\subsection{Non-Interactive Zero-Knowledge Proofs}

We use NIZKs with the following three algorithms $(\Setup,\Prove,\allowbreak\Verify)$ that satisfy simulation-extractable soundness and simulatable zero-knowledge.
As a construction, we rely on the extended bulletproofs from \cite{lai19Succint}, but once there is an efficient implementation of a more efficient proving system, we can easily upgrade to \eg compressed $\Sigma$-protocols \cite{ACR20}.

Bulletproofs\cite{buenz2018} allow a prover to convince a verifier that the inner product of two commited scalar vectors is equal to a value with logarithmic communication.
\citeauthor{lai19Succint}~\cite{lai19Succint} further describe an efficient outer protocol to prove knowledge of bilinear group arithmetic relations.
We use the non-pairing part of \cite{lai19Succint} which allows proofs for a proving system of the following relation $\mathcal{L}_{\text{LRM19}}:=$\\[-8pt]
\[\left\{
    \begin{array}{l}
\vec K\in\G^n, \{\vec v^{(')}_i\in \Zq^m, \mathsf{cls}_i\in\{\mathfrak{mul},\mathfrak{dir},\mathfrak{sum},\mathfrak{one}\},c_i\in\Zq\}_{i=1}^o  \\[4pt]
\exists (\cL,\cR)\in (\Zq^m\times\Zq^m) \ST \prod_{i=1}^{n} K_i^{\vec{c}_{L,i}} = I:=g^0\\[4pt]
\qquad\bigwedge_{i\in\{1,\dots,o\}}: \begin{cases}
    \langle\cL, \vi\rangle = c_i \text{ if } \mathsf{cls}_i = \mathfrak{dir}\\
    \langle\cL, \cR \circ \vi\rangle = c_i \text{ if } \mathsf{cls}_i = \mathfrak{mul}\\
    \langle\cL, \vi\rangle + \langle\cR, \vi'\rangle = c_i  \text{ if } \mathsf{cls}_i = \mathfrak{sum}\\
    \langle\cL - \cR - \vec{1}^m, \vi\rangle = c_i   \text{ if } \mathsf{cls}_i = \mathfrak{one}\\
    \end{cases}
\end{array}
    \right\}
\]

with $n\leq m$ and a proof size of $5 \cdot |\Zq| + (4 + 2 \cdot \lceil{}log_2(|\vec K|)\rceil{}) \cdot |\G|$.
The number of necessary point additions and scalar multiplications in relation to the length of $\vec K$ is shown in \Crefrange{eq:efficiency1}{eq:efficiency2}.\\[-8pt]
\begin{flalign}
    \label{eq:efficiency1}
    &|\mathsf{p}_\times|  &=~& -1 &+~4~|\vec K|& + 2 \cdot {\lceil{}\log_2|\vec K|\rceil{}} ~+~ &         &2^{\lceil{}\log_2|\vec K|\rceil{}} & ~ \\
    &|\mathsf{p}_+|  &=~& 7  &+~4~|\vec K|& + 2 \cdot {\lceil{}\log_2|\vec K|\rceil{}} ~+~ & 26~\cdot&2^{\lceil{}\log_2|\vec K|\rceil{}} & ~ \\
    &|\mathsf{v}_\times| &=~& 12 &+~2~|\vec K|& + 2 \cdot {\lceil{}\log_2|\vec K|\rceil{}} ~+~ &         &2^{\lceil{}\log_2|\vec K|\rceil{}} & ~ \\
    &|\mathsf{v}_+| &=~& 17 &+~2~|\vec K|& + 2 \cdot {\lceil{}\log_2|\vec K|\rceil{}} ~+~ & 21~\cdot&2^{\lceil{}\log_2|\vec K|\rceil{}} & ~
    \label{eq:efficiency2}
\end{flalign}
\section{NIZKs for our Languages}

In this section we provide a concrete construction of the languages for the \prepams scheme on top of the aformentioned preliminaries.

\label{appx:languages}
\subsection{Participation Proof}

On a high-level, the participation proof requires a participant to convince an organizer that they have a valid, signed credential $\cred$ signed by the service's public key $\pk_\Service$, provide a correct blinding for a reward bound to the user's credential, provide a correct participation tag $\Tag$ for task $\Task$ (both in the tuple $\rtx$), and meet the prerequisites for this task.

\begin{flalign*}
    & \mathcal{L}_{\Participate}:= &&\\
    & \quad\quad\left\{
        \begin{array}{l}
            \left(
                \pk_\Service,
                \TX,
                \Task,\Tag,r'
            \right) \mid \exists (\cred, \NUL, \rho)\ST \\
            \quad r'=\PBS\reward.\Blind((\NUL, \cred.\username),\rho) \\
            \quad \land ~~ \CheckCred(\cred, \pk_\Service) = 1 \\
            \quad \land ~~ \CheckParticipation(\cred, \rtx) = 1 \\
            \quad \land ~~ \CheckQualification(\cred, \TX, \rtx.\Task) = 1
        \end{array}
    \right\} &&
\end{flalign*}

For an efficient proof, we split this language into three sub languages:
$\mathcal{L}_\mathsf{CheckCredTag}$ for the anonymous credential including the participation tag, $\mathcal{L}_{\mathsf{CheckQual}}$ for the prerequisites, and $\mathcal{L}_{\mathsf{RewardBlinding}}$ for a correct reward blinding.
These proofs are joined in an AND composition by concatenation and verifying that the tag $\Tag$ in both statements is equal, thereby committing to the same secret key in both parts due to tag uniqueness.
A vector commitment $p$ to the values of the user's attributes and user identity is used to bind the values used in $\mathcal{L}_{\mathsf{CheckQual}}$ to the correct values in $\cred$ proven in $\mathcal{L}_\mathsf{CheckCredTag}$ and to the blinded reward request.
It is blinded by a random $b_p \drawrandom \Zq$ and computed as $p \gets g^{b_p} \cdot \textstyle\prod\vec{U}^{\circ\attr\|\username}$.

\begin{flalign*}
    & \mathcal{L}'_{\Participate}:= &&\\
    & \quad\quad\left\{
        \begin{array}{l}
            \left(
                \pk_\Service,
                \TX,
                \Task,\Tag,r',p
            \right) \mid \exists (\cred, \NUL, \rho, b_p)\ST \\
            \quad\left(\stmt = (\pk_\Service,\Task,\Tag,p), \wit=(\cred, b_p)\right)\\
            \qquad \in \mathcal{L}_{\mathsf{CheckCredTag}}\\
            \quad\land \left(\stmt = (\TX,\Task,\Tag,p), \wit=(\cred, b_p)\right)\\
            \qquad \in \mathcal{L}_{\mathsf{CheckQual}}\\
            \quad\land \left(\stmt=(r', p),\wit=(\cred, \NUL, \rho, b_p)\right)\\
            \quad\quad \in \mathcal{L}_{\mathsf{RewardBlinding}}
        \end{array}
    \right\} &&
\end{flalign*}

\subsubsection*{Credential and Participation Tag}

Participation tags $\Tag$ are derived from the participant's secret key $\cred.\sk$ and a task $\Task$ by computing $\Tag=\VRF.\Eval(\cred.\sk,\Task)$.
The correctness of the participation tag is shown by proving that the VRF evaluates to the provided participation tag.
The validity of the anonymous credential $\cred.\sigma$ can be expressed as the validity of $\PBS.\Verify$.
Both participation tag $\Tag$ and the commitment $p$ are used to bind the statements of the three sublanguages.
Thereby we formulate:
\begin{flalign*}
    & \mathcal{L}_\mathsf{CheckCredTag}:= &&\\
    & \quad\quad\left\{
        \begin{array}{l}
            \left(
                \pk_\Service,\Task,\Tag,p
            \right) \mid \exists (\cred, b_p)\ST \\
            \quad p = g^{b_p} \cdot \textstyle\prod\vec{U}^{\circ\cred.\attr\|\cred.\username} \\
            \quad \land~~ \VRF.\Eval(\cred.\sk,\Task) = \Tag \\
            \quad \land~~ \PBS\credential.\Verify(\pk_\Service,\cred.\sk,\cred.\sigma) = 1
        \end{array}
    \right\} &&
\end{flalign*}

Concrete NIZKs for the correctness of a VRF output and the validity of PBS signature have been given in \cref{sec:defprelim}.
The $\PBS\credential$ instantiation has a single private message part $\vec{S}=(\cred.\sk)$, which also serves as the seed for the VRF.

\begin{flalign*}
    & \mathcal{L}'_\mathsf{CheckCredTag}:= &&\\
    & \quad\quad\left\{
        \begin{array}{l}
            \left(
                \pk_\Service,\Task,\Tag,p
            \right) \mid \exists (\cred,b_p)\ST \\
            \quad p = g^{b_p} \cdot \textstyle\prod\vec{U}^{\circ\cred.\attr\|\cred.\username} \\
            \quad\land(\stmt = (\pk_\Service,\Tag,\id(\Task)), \wit=(\cred.\sk)) \in \mathcal{L}_{\VRF}\\
            \quad\land~\bigg(\begin{array}{l}
                \stmt = (\pk_\Service),\\
                \wit = \left((\cred.\sk), \cred.\attr\|\cred.\username, \cred.\sigma\right)\\
                (\stmt,\wit) \in \mathcal{L}_{\mathsf{VerPBS}}
            \end{array}\bigg)
        \end{array}
    \right\} &&
\end{flalign*}

An AND-composition of both proofs can be created, by using the same blinding value for ${S}_1$ in $\mathcal{L}_{\mathsf{VerPBS}}$ and for $\sk$ in $\mathcal{L}_{\VRF}$.
$\Tag := g^{({\sk + \id(\Task)})^{-1}}$ is the participation tag of a participant with credential $\cred$ (using its secret key $\cred.\sk$) for the task $\Task$ with unique identifier $\id(\Task)$.
The concatenated protocol is also extended by an additional vector pedersen commitment to bind it to the other statements in $\mathcal{L}_{\Participate}$.
This results in the following interactive sigma-style proof.
$\Participant$ picks $d \drawrandom \Zq$, $a_1, a_2 \drawrandom \Zq$, $\vec{B} \drawrandom \Zq^{m + 1}$, and a random group element $J \drawrandom \G$ and computes:

\begin{flalign*}
    & s_1 \gets \sigma_1 \cdot ({V}_1^{\sk} h \textstyle\prod\vec{U}^{\circ\attr\|\username})^d & (\text{re-randomized} \\
    & s_2 \gets \sigma_2 \cdot g_2^d & \text{signature}) \\[4pt]
    & E_1 \gets e(J, g_2) \cdot e({V}_1^{a_1} \cdot \textstyle\prod\vec{U}^{\circ\vec{B}}, s_2)^{-1} & (\text{commitment PBS}) \\
    & E_2 \gets e(g, \Tag)^b & (\text{commitment VRF}) \\
    & E_3 \gets g ^ {a_2} \cdot \textstyle\prod\vec{U}^{\circ\vec{B}} & (\text{commitment binding})
\end{flalign*}

The participant sends $E_1,E_2,E_3,p$ to the organizer and gets a challenge $c \drawrandom \Zq$ back from the organizer.
The last message is computed as follows:

\[
    \vec{Y} \gets \left[{B}_i + c \cdot {M}_i \right]_{i=1}^{m}
\]
\[
    z_1 \gets a_1 + c \cdot d \hspace{2em}
    z_2 \gets {s_1}^c \cdot J \hspace{2em}
    z_3 \gets a_2 + \cdot c \cdot b_p \hspace{2em}
\]

The statement is verified by the organizer with these three equations:
\[\arraycolsep=1.4pt
    \begin{array}{rl}
        E_1 \cdot {\pk_\Service}^c \cdot e(h, s_2)^c =& e(z_2, g_2) \cdot e(\textstyle\prod\vec{U}^{\circ\vec{Y}}, s_2)^{-1} \\
        E_2 \cdot e(g,g_2)^c \cdot e(g, \Tag) ^ {-c \cdot \id(\Task)} =& e(g_2, \Tag)^{z_1} \\
        E_3 \cdot p^c =& g^{z_3} \cdot \textstyle\prod\vec{U}^{\circ\vec{Y}}
    \end{array}
\]

\subsubsection*{Prerequisites}
The second part of the participation proof requires a participant to show that they qualify for the participation with $\CheckQualification$.
In our model this essentially means that they satisfy all attribute constraints, have previously participated in every qualifier task, and have not participated in any disqualifier task.
$\mathcal{L}_{\textsf{CheckQual}}$ takes all previous transactions $\TX$ and iterates over all qualifier tasks $\Task_q\in\Task.\qualifier$ and selects the set $Q\Tag$ as tags from transactions which contain $\Task_q$.
Similarly $D\Tag$ is specified for all disqualifiers.
The participation tag $\Tag$ is used to bind $\mathcal{L}_{\qualifier}$ and $\mathcal{L}_{\disqualifier}$ to the credential used in $\mathcal{L}_{\textsf{CheckCredTag}}$, whereas the pedersen commitment $p$ is used in $\mathcal{L}_{\Satisfy}$.

\begin{flalign*}
    & \mathcal{L}_{\textsf{CheckQual}}:= &&\\
    &\enskip\left\{
        \begin{array}{l}
            \left(
                \TX:=[(\Task_i,\cdot, \Tag_i)],
                \Task,\Tag,p
            \right) \mid \exists (\cred,b_p)\ST \\
            \forall \Task_q\in\Task.\qualifier:\begin{cases}
            Q\Tag:=\{\tx.\Tag | \tx \in \{\tx' \in \TX | \tx'.\Task=\Task_q \}  \}\\
                \stmt = (Q\Tag,\Tag,\Task,\Task_q),\\
                \wit=(\cred.sk)\\
            (\stmt,\wit)\in \mathcal{L}_{\qualifier}
            \end{cases}\\
            \land \forall \Task_d\in\Task.\disqualifier: \begin{cases} D\Tag:=\{\tx.\Tag | \tx \in \{\tx' \in \TX | \tx'.\Task=\Task_d \}  \}\\
               \stmt = (D\Tag,\Tag,\Task,\Task_d),\\
                \wit=(\cred.sk)\\
            (\stmt,\wit) \in \mathcal{L}_{\disqualifier}
            \end{cases}\\
            \land \forall c \in\Task.\Constraints: \begin{cases}
               \stmt = (\Tag,c,p),\\
                \wit=(\cred.\attr,\cred.username, b_p)\\
            (\stmt,\wit) \in \mathcal{L}_{\Satisfy}
            \end{cases}\\
        \end{array}
    \right\} &&
\end{flalign*}

To exploit the repetitive structure, we further divide this proof into sub languages for a single qualifier task $\Task_q$ and a single disqualifer task $\Task_d$.
To bind the conjunction of the proofs for every (dis-)qualifying task, we use the current task tag $\tau$ which assures that the participant uses the same credential secret key $\cred.\sk$ to calculate the tags correctly.
These statements also include the public set of participation tags ($Q\Tag$, $D\Tag$) from all previous participations in the (dis-)qualifier.

\subsubsection*{Qualifier}
For a qualifying task $\Task_q$, the participant computes their correct participation tag for this task $\Task_q$ and shows that $\VRF.\allowbreak\Eval(\cred.\sk,\Task_q)$ is part of the set of qualifying tags $Q\Tag$.
Additionally, it is necessary to show that the secret key of the participant $\cred.\sk$ matches the secret key used to derive the participation tag $\Tag$ for the task $\Task$ where the participant wants to qualify for.
\begin{flalign*}
    & \mathcal{L}_\qualifier:= &&\\
    & \quad\quad\left\{
        \begin{array}{l}
            \left(
                Q\Tag,
                \Tag,
                \Task,
                \Task_q
            \right) \mid \exists (\sk)\ST \\
            \quad \VRF.\Eval(\sk,\Task_q) \in Q\Tag \\
            \quad \land~~ \VRF.\Eval(\sk,\Task) = \Tag
        \end{array}
    \right\} &&
\end{flalign*}

Using the VRF construction $(\Tag=g^{\frac{1}{\sk+\id(\Task)}})$, this corresponds to a ring signature on the set of qualifying tags $Q\Tag$, where the participant has the correct $\sk$ bound by $\Tag$ to equate the discrete logarithm of element $Q\Tag_j$ at position $j$.
\begin{flalign*}
    & \mathcal{L}'_{\qualifier}:= &&\\
    & \quad\quad\left\{
        \begin{array}{l}
            \left(
                Q\Tag,
                \Tag,
                \Task,
                \Task_q
            \right) \mid \exists (\sk,j)\ST \\
            \quad \Tag = g^{\frac1{\sk+\id(\Task)}}\\
            \quad \land Q\Tag_j = g^{\frac1{\sk+\id(\Task_q)}}
        \end{array}
    \right\} &&
\end{flalign*}
This language is directly provable in $\mathcal{L}_{\mathsf{LMR19}}$.

\subsubsection*{Disqualifier}
For disqualifiers the participant also evaluates the verifiable random function for the disqualifying task $\Task_d$ and shows that the tag $\VRF.\Eval(\cred.\sk,\Task_d)$ is not in the set of disqualifying tags $D\Tag$.
It is again necessary to to also show that the same secret key $\cred.\sk$ is used to compute the disqualifying tag and the actual participation tag $\Tag$.

\begin{flalign*}
    & \mathcal{L}_\disqualifier:= &&\\
    & \quad\quad\left\{
        \begin{array}{l}
            \left(
                D\Tag,
                \Tag,
                \Task,
                \Task_d
            \right) \mid \exists (\sk)\ST \\
            \quad \VRF.\Eval(\sk,\Task_d) \not\in D\Tag\\
            \quad \land ~~ \VRF.\Eval(\sk,\Task) = \Tag
        \end{array}
    \right\} &&
\end{flalign*}

This is achieved by choosing a secret blinding factor $r \drawrandom \Zq$ to calculate a blinded version of the disqualifying tag $R\Tag\gets\Tag^r$ and the set of equally blinded disqualifying tags ($\forall i\in\{1,\dots,|D\Tag|\}:RD\Tag_i\gets D\Tag_i^r$).
Hence, allowing the participant to proof in zero-knowledge that the challenge set was correctly blinded (\ie exponentiated with the random factor $r$), as well as the correct computation and blinding of the disqualifying tag.
The verifier is then able to check whether the blinded disqualifying tag $R\Tag$ is equal to any blinded tag in the challenge set $RD\Tag$, without learning the participant's real tag $\VRF.\Eval(\cred.\sk,\Task_d)$ for the disqualifying task $\Task_d$.
This provides participation privacy for future participations with the same disqualifying task.

\begin{flalign*}
    & \mathcal{L}'_{\disqualifier}:= &&\\
    & \quad\quad\left\{
        \begin{array}{l}
            \left(
                D\Tag,
                RD\Tag,
                R\Tag,
                \Tag,
                \Task,
                \Task_d
            \right) \mid \exists (\sk,r)\ST \\
            \quad \Tag = g^{\frac1{\sk+\id(\Task)}}\\
            \quad \land R\Tag = g^{\frac{r}{\sk+\id(\Task_d)}}\\
            \quad \land \forall i\in[|D\Tag|] : RD\Tag_i = D\Tag_i^r
        \end{array}
    \right\} &&
\end{flalign*}

Again this language is provable in the $\mathcal{L}_\mathsf{LMR19}$ language.

\subsubsection*{Satisfy}
For attribute constraints, we differentiate between range and element constraints, which we split up into two separate languages $\mathcal{L}'_{\Satisfy,\mathsf{rng}}$ and $\mathcal{L}'_{\Satisfy,\mathsf{ele}}$.
Based on the type of constraint $c$ is either composed as $c_{\textsf{rng}} := (l, u, i)$ with lower bound $l$, upper bound $u$, and index $i$ of the referenced attribute, or as $c_{\mathsf{ele}} = (V, i)$ with set of allowed element $V$ and referenced attribute index $i$.
Only the statement of one language can be valid at any time, but for simplicity we denote the choice of sublanguage as a logical or in the following language.
Both languages use the pedersen commitment $p$ for showing equality of the attributes used in $\mathcal{L}_{\mathsf{CheckCredTag}}$.

\begin{flalign*}
    & \mathcal{L}_{\Satisfy}:= &&\\
    & \quad\quad\left\{
        \begin{array}{l}
            \left(
                \Tag,c,p
            \right) \mid \exists (\attr,\username, b_p)\ST \\
            (\stmt=(c.l, c.u, c.i, p),\wit=(\attr,\username,b_p)) \in \mathcal{L}_{\Satisfy,\mathsf{rng}}\\
            \lor (\stmt=(c.V, c.i, p),\wit=(\attr,\username,b_p)) \in \mathcal{L}_{\Satisfy,\mathsf{ele}}\\
        \end{array}
    \right\} &&
\end{flalign*}

Range constraints use two binary decompositions to show that the referenced attribute value $\attr_i$ is in the interval of $[l,u]$.
This can be expressed as $v_l := \attr_i - l$ and $v_u := u - \attr_i$ being a positive integers.
\begin{flalign*}
    & \mathcal{L}'_{\Satisfy,\mathsf{rng}}:= &&\\
    & \quad\quad\left\{
        \begin{array}{l}
            \left(
                l,u,i,p
            \right) \mid \exists (\attr,\username,b_p)\ST \\
            \quad p = g^{b_p} \cdot \textstyle\prod\vec{U}^{\circ\attr\|\username} \\
            \quad \land ~~ l + v_l = \attr_i \land v_l\in\{0,\dots,2^{\lceil{}log_2|u-l|\rceil+1} - 1\}\\
            \quad \land ~~ u = \attr_i + v_u \land v_u\in\{0,\dots,2^{\lceil{}log_2|u-l|\rceil+1} - 1\}
        \end{array}
    \right\} &&
\end{flalign*}

Element constraints are essentially a ring signature similar to qualifier prerequisites, where the participant's attribute equates the discrete logarithm of the element in $V$ with index $j$.
\begin{flalign*}
    & \mathcal{L}'_{\Satisfy,\mathsf{ele}}:= &&\\
    & \quad\quad\left\{
        \begin{array}{l}
            \left(
                V,i,p
            \right) \mid \exists (\attr,\username,b_p,j)\ST \\
            \quad p = g^{b_p} \cdot \textstyle\prod\vec{U}^{\circ\attr\|\username} \\
            \quad \land V_j = \attr_i\\
        \end{array}
    \right\} &&
\end{flalign*}
Both languages can be expressed in $\mathcal{L}_{\mathsf{LMR19}}$ and are bound to $\mathcal{L}_{\mathsf{CheckCredTag}}$ using the shared commitment $p$.

\subsubsection*{Reward Blinding}
The correct blinding of a reward key can be shown as a statement of $\mathcal{L}_{\mathsf{\PBS\reward}}$ (\cf \Cref{sec:defprelim}).
Additionally, the matching to the vector pedersen $p$ is shown by an AND composition using the same challenge in the final message of the sigma protocol.

\begin{flalign*}
    & \mathcal{L}_\mathsf{RewardBlinding}:= &&\\
    & \quad\quad\left\{
        \begin{array}{l}
            \left(r', p\right) \mid \exists (\cred, \NUL, \rho, b_p)\ST \\
            \quad p = g^{b_p} \cdot \textstyle\prod\vec{U}^{\circ\cred.\attr\|\cred.\username} \\
            \quad\land \left(\stmt = (r'), \wit=((\NUL, \cred.\username), \rho)\right) \in \mathcal{L}_{\PBS\reward}\\
        \end{array}
    \right\} &&
\end{flalign*}

\subsubsection*{Efficiency}

$\mathcal{L}_{\mathsf{CheckCredTag}}$ requires a total of $13 + 5m$ point additions, $13 + 5m$ scalar multiplications, and $3$ pairing operations for the proof and $9 + 2m$ additions, $10 + 2m$ multiplications, and $6$ pairings for the verification.
The proof of $\mathcal{L}_{\mathsf{\PBS\reward}}$ requires $4$ point additions, $6$ scalar multiplications, and $0$ pairing operations.
The verification requires 3 additions, 4 multiplications, and 0 pairing operations.

The witness length of $\mathcal{L}_{\mathsf{CheckQual}}$ in $\mathcal{L}_{\mathsf{LMR19}}$ is as follows:
\begin{flalign*}
    &\enskip 10 + m & (\text{binding}) \\
    & \quad + \textstyle\sum_{i=0}^{|\qualifier|}2 + |Q\Tag_i|  & (\text{qualifier}) \\
    & \quad +~4 \cdot |\disqualifier| & (\text{disqualifier}) \\
    & \quad + \textstyle\sum_{i=0}^{|\Constraints,\mathsf{ele}|}1 + |\{e_i\}| & (\text{element constraints}) \\
    & \quad + \textstyle\sum_{i=0}^{|\Constraints,\mathsf{rng}|}2 \cdot \lceil 1 + log_2(|u_i-l_i|)\rceil & (\text{range constraints})
\end{flalign*}

The necessary operations for the full composition of $\mathcal{L}_{\Participate}$ is strongly dependent on the complexity of the study.
For better comprehensibility, we define $\kappa(\Task)$ as the nuber of group elements required to encode the prerequisites of a task $\Task$:
\begin{align*}
    \kappa(\Task) =& 10 + m + 2 |\Task.\qualifier| + 4 |\Task.\disqualifier| + |\Task.\Constraints,\mathsf{ele}| + 2 |\Task.\Constraints,\mathsf{rng}|\\
    &+ \textstyle\sum_{i=0}^{|\Task.\qualifier|}|Q\Tag_i|\\
    &+ \textstyle\sum_{i=0}^{|\Task.\Constraints,\mathsf{ele}|}|\{e_i\}|\\
    &+ \textstyle\sum_{i=0}^{|\Task.\Constraints,\mathsf{rng}|}\lceil 1 + log_2(|u_i-l_i|)\rceil
\end{align*}

Adding the necessary operations of $\mathcal{L}_\mathsf{CheckCredTag}$, $\mathcal{L}_{\mathsf{\PBS\reward}}$, and $\mathcal{L}_{\mathsf{CheckQual}}$ in $\mathcal{L}_{\mathsf{LMR19}}$, we get the following for proving $\mathsf{p}$ and verifying $\mathsf{v}$:
\begin{flalign*}
    &|\mathsf{p}_\times| &=&~18 + 5m &+~4~\kappa(\Task) + 2 \cdot {\lceil{}\log_2\kappa(\Task)\rceil{}} + &&          2^{\lceil{}\log_2\kappa(\Task)\rceil{}} \\
    &|\mathsf{p}_+|      &=&~24 + 5m &+~4~\kappa(\Task) + 2 \cdot {\lceil{}\log_2\kappa(\Task)\rceil{}} + && 26~\cdot 2^{\lceil{}\log_2\kappa(\Task)\rceil{}} \\
    &|\mathsf{p}_e|      &=&~3& \\[4pt]
    &|\mathsf{v}_\times| &=&~22 + 2m &+~2~\kappa(\Task) + 2 \cdot {\lceil{}\log_2\kappa(\Task)\rceil{}} + &&          2^{\lceil{}\log_2\kappa(\Task)\rceil{}} \\
    &|\mathsf{v}_+|      &=&~26 + 2m &+~2~\kappa(\Task) + 2 \cdot {\lceil{}\log_2\kappa(\Task)\rceil{}} + && 21~\cdot 2^{\lceil{}\log_2\kappa(\Task)\rceil{}} \\
    &|\mathsf{v}_e|      &=&~6&
\end{flalign*}

\subsection{Payout Proof}
The payout proof allows a participant to prove ownership of up to $n$ reward coins of respective value $v_i$ that sum up to at least the payout amount $v \leq \sum_{i=0}^{n} v_i$.
Every coin is a $\PBS$ signature by the service with a value $v_i$ and bound to a nullifier $\NUL_i$ and the username $\username$ of a participant, both hidden from the service when signing.
To prevent double spending, participants have to reveal the nullifier $\NUL_i$ for each spent reward and prove their correctnes.
The exact number of inputs is hidden by padding the input set to a fixed length of $n$ inputs.
Padding inputs are first requested from the service, similar to regular participations.
The validity of a payout request is shown with a statement of the following language.
\begin{flalign*}
    & \mathcal{L}_{\Payout}:= &&\\
    & \quad\quad\left\{
        \begin{array}{l}
            \left(
                \{\NUL_i\}_{i=1}^n,
                \username,
                v
            \right) \mid \exists (\{(r_i,v_i)\}_{i=1}^{n})\ST \\
            \quad \hspace*{0pt}\sum_{i=1}^{n} v_i \geq v \\
            \quad \land ~ \forall i\in [n]: \PBS\reward.\Verify(\pk_{\Service,\reward},(\NUL_i, \username),(v_i),r_i) = 1\\
        \end{array}
    \right\} &&
\end{flalign*}

Substituting the concrete instantiation from our building blocks, we get the language $\mathcal{L}'_{\Payout}$.
We allow to claim less payout than the sum of all rewards $\sum_{i=1}^{n} v_i \geq v$ which is implemented as a positive difference $v^*$.
This helps participants to hide themselves in a larger anonymity set of \eg students who all claim the same amount equal to the required subject hours for their study programme.
Again, we split this language into two sub languages, for an efficient proof.
The composition of the two sub languages is bound by adding separate pedersen commitments for each input $p_i$ to the statement each blinded by random $a_i \drawrandom \Zq$, computed as $p_i \gets g^a_i \cdot {V}_1 ^ {\NUL_i} \cdot {V}_2 ^ {\username} \cdot {U}_1 ^ {v_i}$.
\begin{flalign*}
    & \mathcal{L}'_{\Payout}:= &&\\
    & \quad\quad\left\{
        \begin{array}{l}
            \left(
                \{\NUL_i, p_i\}_{i=1}^n,
                \username,
                v
            \right) \mid \exists (\{(r_i,v_i,a_i)\}_{i=1}^{n})\ST \\
            \quad \stmt = (\{\NUL_i, p_i\}_{i=1}^n, \username, v),\\
            \quad \wit=(\{(r_i,v_i,a_i)\}_{i=1}^{n}),\\[4pt]
            \quad (\stmt, \wit) \in \mathcal{L}'_{\mathsf{PayoutSum}}\\
            \quad \land ~~ (\stmt, \wit) \in \mathcal{L}'_{\mathsf{CheckReward}}\\
        \end{array}
    \right\} &&
\end{flalign*}

The correctness of the re-randomizations is shown using statements of $\mathcal{L}'_{\mathsf{CheckReward}}$.
The range proof that that the sum of inputs and the positive difference $v^*$ is equal to the payout value $v$ is proven with a statement of $\mathcal{L}'_{\mathsf{PayoutSum}}$ using a bit decomposition of $v^*$.
Depending on the average value of rewards, the bit range $\mathfrak{B}$ of $v^*$ may be adapted.
For reference, in our prototype we selected a range of $\mathfrak{B} = 8$, to match our use case of subject hours in an empirical study programme, where typically one payout of 30 hours equals one ECTS.
\begin{flalign*}
    & \mathcal{L}'_{\mathsf{PayoutSum}} := &&\\
    & \quad\quad\left\{
        \begin{array}{l}
            \left(
                \{\NUL_i, p_i\}_{i=1}^n,
                \username,
                v
            \right) \mid \exists (\{(r_i,v_i,a_i)\}_{i=1}^{n})\ST \\
            \quad \forall i\in[n]: p_i = {\pk_{\Service,\reward}}^{a_i} \cdot V_1 ^ {\NUL_i} \cdot V_2 ^ {\username} \cdot U_1 ^ {v_i} \\
            \quad \land ~~ \sum_{i=1}^{n} v_i = v + v^* \land v^*\in\{0,\dots,2^\mathfrak{B}-1\}
        \end{array}
    \right\} &&
\end{flalign*}

The language $\mathcal{L}'_{\mathsf{PayoutSum}}$ is directly representable in $\mathcal{L}_\mathsf{LMR19}$.
$\mathcal{L}'_{\mathsf{CheckReward}}$ can be expressed as multiple iterations of $\mathcal{L}_\mathsf{VerPBS}$ bound to $\mathcal{L}'_{\mathsf{PayoutSum}}$ using individual pedersen commitments.
\begin{flalign*}
    & \mathcal{L}'_{\mathsf{CheckReward}}:= &&\\
    & \quad\quad\left\{
        \begin{array}{l}
            \left(
                \{\NUL_i, p_i\}_{i=1}^n,
                \username,
                v
            \right) \mid \exists (\{(r_i,v_i,a_i)\}_{i=1}^{n})\ST \\
            \quad \forall i\in [n]:\begin{cases}
                \stmt = (\pk_{\Service,\reward}),\\
                \wit=((\NUL_i, \username), (v_i), r_i)\\
                (\stmt, \wit) \in \mathcal{L}_{\mathsf{VerPBS}}\\
                \land~~ p_i = {\pk_{\Service,\reward}}^{a_i} \cdot {V}_1 ^ {\NUL_i} \cdot {V}_2 ^ {\username} \cdot {{U}}_1 ^ {v_i} \\
            \end{cases}\\
        \end{array}
    \right\} &&
\end{flalign*}

For better comprehensibility, we denote the set of generators of $\PBS\reward$ $({V}_1,{V}_2,{U}_1)$ as $\vec{\mathfrak{G}}$, and the set of coin messages $(\username, \NUL_i, v_i)$ as $\vec{\mathfrak{M}}_i$.
The input signatures are first re-randomized with random $d_i \in \Zq$ resulting in:
\[
    r_i = \sigma_{i,1} \cdot (\textstyle\prod \vec{\mathfrak{G}}^{\circ\vec{\mathfrak{M}}_i} \cdot h)^{d_i} \hspace{2em}
    s_i = \sigma_{i,2} \cdot g_2^{d,i}
\]
The participant creates a committment for each input by picking random values $\vec{B_i} \drawrandom \Zq^3$, $a_i \drawrandom \Zq$, $b_i \drawrandom \Zq$, and a random group element $J_i \drawrandom \G$, and computing:
\begin{flalign*}
    & {E}_i \gets e(J_i, g_2) \cdot e(\textstyle\prod\vec{\mathfrak{G}}^{\circ\vec{B_i}}, \vec{s}_i)^{-1}  & (\text{commitment PBS}) \\
    & {F}_i \gets g ^{a_i} \cdot \textstyle\prod\vec{\mathfrak{G}}^{\circ\vec{B_i}} & (\text{commitment binding}) \\
    & {p}_i \gets g ^{b_i} \cdot \textstyle\prod\vec{\mathfrak{G}}^{\circ\vec{\mathfrak{M}}_i} & (\text{binding})
\end{flalign*}

The participant sends $\vec{s}, \vec{E}, \vec{F}, \vec{p}$ to the service and gets a challenge $c\drawrandom\Zq$ back.
The response is computed as follows:
\[
    {Y}_i \gets \left[{B_i}_j + c \cdot \vec{\mathfrak{M}_i}_j \right]_{j=1}^{3} \hspace{2em}
    {z_E}_i = {{r}_i}^c \cdot J_i \hspace{2em}
    {z_F}_i = {F_i}^c \cdot b_i
\]

The statements are verified by the service by checking:
\[
    \begin{array}{rl}
        E_i \cdot {\pk_\Service}^c \cdot e(h, {s}_i)^c & = e(z_2, g_2) \cdot e(\textstyle\prod\vec{\mathfrak{G}}^{\circ\vec{Y}_i}, s_2)^{-1}\\
        {{p}_i}^c + {F_i} & = g^{{z_F}_i} \cdot \textstyle\prod\vec{\mathfrak{G}}^{\circ\vec{Y}_i}
    \end{array}
\]

\subsubsection*{Efficiency}
The $n$ iterations of $\mathcal{L}'_{\mathsf{CheckReward}}$ each require the participant to perform $19$ point additions, $16$ scalar multiplications, and $2$ pairing operations.
The verification requires $11$ additions, $11$ multiplications, and $3$ pairing operations.
$\mathcal{L}'_{\mathsf{PayoutSum}}$ has an encoded witness length of $|\mathfrak{B}| + 5 \cdot n$, hence requiring the following operations to compute:
\begin{flalign*}
    &|\mathsf{p}_\times| &=& -1 &+ 4 \mathfrak{B} + 20n + 2 \cdot {\lceil{}\log_2(\mathfrak{B} + 5n)\rceil{}} &&+~           2^{\lceil{}\log_2(\mathfrak{B} + 5n)\rceil{}} \\
    &|\mathsf{p}_+| &=& 7  &+ 4 \mathfrak{B} + 20n + 2 \cdot {\lceil{}\log_2(\mathfrak{B} + 5n)\rceil{}} &&+~  26~\cdot 2^{\lceil{}\log_2(\mathfrak{B} + 5n)\rceil{}} \\
    &|\mathsf{p}_e| &=&  &2n & \\[4pt]
    &|\mathsf{v}_\times| &=& 12 &+ 2 \mathfrak{B} + 10n + 2 \cdot {\lceil{}\log_2(\mathfrak{B} + 5n)\rceil{}} &&+~           2^{\lceil{}\log_2(\mathfrak{B} + 5n)\rceil{}} \\
    &|\mathsf{v}_+| &=& 17 &+ 2 \mathfrak{B} + 10n + 2 \cdot {\lceil{}\log_2(\mathfrak{B} + 5n)\rceil{}} &&+~  21~\cdot 2^{\lceil{}\log_2(\mathfrak{B} + 5n)\rceil{}}\\
    &|\mathsf{v}_e| &=&  &3n  &
\end{flalign*}

Using the example parameters used for our prototype ($\mathfrak{B} = 8, n = 10$), the full $\mathcal{L}_{\Payout}$ proof can be computed in $2105$ point additions, $531$ multiplications, and $20$ pairing operations.
The verification requires $1599$ additions, $314$ multiplications, and $30$ pairing operations.
\subsection{Analysis}
\label{appx:analysis}

By inspection, our \prepams{} construction is correct according to \Cref{def:correctness}.

\begin{theorem}[Participation Security]
 Given an unforgeable $\PBS\credential$ scheme, a deterministic~$\VRF$, and a simulation-extractable (SE) NIZK, \prepams satisfies the participation security defined in \Cref{def:partsec}.
\end{theorem}
\vspace{-1.5em}
\begin{proof}
We require that a credential $\cred$ is binding to a specific secret key.
This holds because of the unforgeability of the PBS scheme.
Additionally, a tag must be binding in the sense, that only a unique pair of key and task result in this tag. Our deterministic $\VRF$ scheme satisfies this.
We now show that the winning condition of $\mathsf{PartSec}$ has a negligible probability: Every reward transaction $\rtx$ in the list of transactions $\TX$ is valid because organizers validate each participation NIZK in $\Oracle\Service\Participate$ or $\Oracle\Participate$ and only append to $\TX$ if valid.
The valid participation thereby implies a valid NIZK for the language $\mathcal{L}_\Participate$.
By the extractability of the NIZK, there exists an efficient extractor $\mathcal{E}_\Participate$ which extracts the witness $\cred$ from $\pi$ for the winning $\rtx$ ($\CheckParticipation(\cred,\TX_t)=1\land\CheckQualification(\cred,\TX[\text{:}t],\TX_t.T)=0$).
According to the language, $\cred$ satisfies at least $\CheckParticipation(\cred,\rtx)=1$ and $\CheckQualification(\cred,\TX,\allowbreak\Tag)=1$.
The $\CheckQualification$ predicate together with the binding tag assures that the participant has not participated in the same task before.
As $\TX$ in the statement is exactly $\TX[\text{:}n]$ containing all previous participations up to this point, the winning probability of the adversary is reduced to the negligible probability of forging a SE NIZK without knowledge of a witness $\cred.\sk$. Thereby \prepams has participation security.
\end{proof}

\begin{theorem}[Balance]
 Given an unforgeable $\PBS\credential$ and $\PBS\reward$ scheme, simulation-extractable (SE) NIZKs and participation security (Def \ref{def:partsec}), \prepams is balanced according to \Cref{def:balance}.
\end{theorem}
\vspace{-1.5em}
\begin{proof}
In the first epoch, the adversary is tasked to get a higher payout from the service as they earned through their participations. The system requires every reward transaction $\rtx$ used for the payout to be in the list of valid transactions $\TX$.
Given participation security, the organizers only blind-sign a reward coin and add it to the $\TX$ with the correct amount for the completed task. All rewards to the adversary are also maintained by the oracle in state $\mathfrak{T}^0$. Therefore the only option to increase the adversary's balance is to steal from honest users or attack the payout protocol. Every payout in $\Oracle\Service\Payout$ includes a valid proof $\pi$ for the language $\mathcal{L}_\Payout$.
With the NIZK extractor $\mathcal{E}_\Payout$, we extract a valid witness $\{(r_i,v_i)\}_{i=1}^n$. These are the coins $r_i$ with a valid $\PBS\reward$ signature and the blinded identity $\username$ and nullifier $\NUL_i$, preventing repeated spending of the same key. The identifier prevents spending of rewards by someone else. Any change of the identity, nullifier or value by a participant would break the unforgeability of the $\PBS\reward$ signature. The extracted rewards $v_i$ to be paid out are larger than the payout amount $v$. Preventing double spending and the greater or equal constraint on the amount leaves the adversary a negligible advantage to breaking the balance property by attacking the building blocks.

In the second epoch, the adversary has full access to old credentials and their rewards. Again, as $\mathcal{E}_\Payout$ extracts the identity at payout, only payouts accumulated in $\mathfrak{T}^1$ can be paid out to users in $\mathfrak{U}^1$, staying non-negative.
The ideneity in every coin is binding due to the unforgeability of $\PBS\credential$ and $\mathcal{L}_{\Participate}$.
\end{proof}

\begin{theorem}[Participation Privacy]
 Given blinding partially blind $\PBS\credential$ and $\PBS\reward$ schemes, a pseudo-random verifiable random function $\VRF$, simulatable NIZKs and Balance, \prepams satisfies the participation privacy defined in \Cref{def:partpriv}.
\end{theorem}
\vspace{-1.5em}
\begin{proof}
 We show the participation privacy through a series of hybrids changing the experiment from $b=0$ to $b=1$.
 \begin{description}[leftmargin=0em,nosep]
  \item[Hyb1:] The first hybrid is equal to $\mathsf{PartPriv}_0(\lambda)$. The information dependent on $b$ is the communication in $\Pi_\Participate$, i.e. $(\pi_0,\allowbreak\Task_0,\Tag_0,\allowbreak r_0)$, but $\Task_0=\Task_1=\Task$ is equal for both $b$. The oracles after the challenge only perform actions, if they were possible by both $\username_0$ and $\username_1$, thereby being independnt of $b$.
  \item[Hyb2:] This hybrid uses the existence of the NIZK simulator to generate $\pi_0$ as $\pi_\mathsf{Sim}$ which is indistinguishable from the real proof and does not need a witness.
  \item[Hyb3:] As the proof $\pi_\mathsf{Sim}$ is now simulated, this hybrid uses the credential $\cred_1$ to generate a $\Tag_1$. The hybrid is indistinguishable by the pseudorandomness of the VRF.
  \item[Hyb4:] This hybrid replaces the $r_0$ with $r_1$ as both are blinding and indistinguishable.
  \item[Hyb5:] With the information about $b$ available to the adversary being $(\pi_\mathsf{Sim},\Tag_1,r_1)$, it remains to change the simulated NIZK to a real one again, using $\cred_1$ as witness. This is indistinguishable due to the simulatability of NIZKs.
 \end{description}
 Hyb5 is equal to $\mathsf{PartPriv}_1(\lambda)$, concluding that the adversary has a negligible advantage of breaking the participation privacy.
\end{proof}
\section{Efficiency}
\label{appx:efficiency}

\begin{table*}[h]
    \caption{
        Effiency of the concrete instantiation of the \prepams protocol.
        $\times$ indicates the number of scalar multiplications, $+$ the number of group element additions, $e$ the number of pairing operations, and $|.|$ the communication size defined as multiples of group elements sizes ($|\G|$, $|\G_2|$, $|\G_T|$) and scalars ($|\Zq|$).
    }
    \begin{tabular}{lcccc}
        \toprule
        & \multicolumn{3}{c}{Computation Efficiency} \\
        & $\times$ & $+$ & $e$ \\
        \midrule
        $\Pi_{\Register,\Participant}$ & $9 + m$ & $7 + m$ & $2$\\
        $\Pi_{\Register,\Service}$ & $8 + 3m$ & $4 + m$ \\[6pt]

        $\Pi_{\Participate, \Participant}$ & $18 + 5m + 4\kappa(\Task) + 2 \cdot {\lceil{}\log_2\kappa(\Task)\rceil{}} + 2^{\lceil{}\log_2\kappa(\Task)\rceil{}}$ & $24 + 5m + 4\kappa(\Task) + 2 \cdot {\lceil{}\log_2\kappa(\Task)\rceil{}} + 26~\cdot 2^{\lceil{}\log_2\kappa(\Task)\rceil{}}$ & $3$ \\
        $\Pi_{\Participate, \Service}$ & $22 + 2m + 2\kappa(\Task) + 2 \cdot {\lceil{}\log_2\kappa(\Task)\rceil{}} + 2^{\lceil{}\log_2\kappa(\Task)\rceil{}}$ & $26 + 2m + 2\kappa(\Task) + 2 \cdot {\lceil{}\log_2\kappa(\Task)\rceil{}} + 21~\cdot 2^{\lceil{}\log_2\kappa(\Task)\rceil{}}$ & $6$ \\[6pt]

        $\Pi_{\Payout, \Participant}$ & $-1 + 4 \mathfrak{B} + 20n + 2 \cdot {\lceil{}\log_2(\mathfrak{B} + 5n)\rceil{}} + 2^{\lceil{}\log_2(\mathfrak{B} + 5n)\rceil{}}$ & $7 + 4 \mathfrak{B} + 20n + 2 \cdot {\lceil{}\log_2(\mathfrak{B} + 5n)\rceil{}} + 26~\cdot 2^{\lceil{}\log_2(\mathfrak{B} + 5n)\rceil{}}$ & $2n$ \\
        $\Pi_{\Payout, \Service}$ & $12 + 2 \mathfrak{B} + 10n + 2 \cdot {\lceil{}\log_2(\mathfrak{B} + 5n)\rceil{}} + 2^{\lceil{}\log_2(\mathfrak{B} + 5n)\rceil{}}$ & $17 + 2 \mathfrak{B} + 10n + 2 \cdot {\lceil{}\log_2(\mathfrak{B} + 5n)\rceil{}} + 21~\cdot 2^{\lceil{}\log_2(\mathfrak{B} + 5n)\rceil{}}$ & $3n$ \\
        \bottomrule\\
    \end{tabular}

    \begin{tabular}{lccccccc}
        \toprule
        & \multicolumn{4}{c}{Communication Size} \\
        & $\cdot|\Zq|$ & & $\cdot|\G|$ & & $\cdot|\G_2|$ & & $\cdot|\G_T|$ \\
        \midrule
        $\Pi_{\Register,\Participant}$ & $(2 + m)$ & & $2$ & \\
        $\Pi_{\Register,\Service}$ & & & $2$ & & $1$ \\[6pt]

        $\Pi_{\Participate, \Participant}$ & $\biggl($ \parbox{5.3cm}{$15 + m + |\Task.\qualifier| + 2|\Task.\disqualifier|\\~~~ + 3|\Task.\Constraints,\mathsf{ele}| + \textstyle\sum_{c\in |\Task.\Constraints,\mathsf{rng}|} |c.V|$} $\biggr)$ & & $\biggl($ \parbox{5.3cm}{$12 + 2m + 2 \lceil{}log_2(|\vec K|)\rceil{} + \textstyle\sum_{i=0}^{|\Task.\qualifier|}|Q\Tag_i|\\~~~ + \textstyle\sum_{i=0}^{|\Task.\disqualifier|}(1 + 2|D\Tag_i|)$} $\biggr)$ & & 1 & & 2 \\
        $\Pi_{\Participate, \Service}$ & \\[6pt]

        $\Pi_{\Payout, \Participant}$ & $6 + n$ &  & $4 + 2n + 2 \lceil{}log_2(\mathfrak{B} + 5n)\rceil{}$ & & $n$ & & $n$ \\
        $\Pi_{\Payout, \Service}$ & \\
        \bottomrule\\
    \end{tabular}
    \label{tab:efficiency}
\end{table*}

We summarize the efficiency of the concrete instantiation of our scheme in \Cref{tab:efficiency}.
 
\end{document}